\documentclass[12pt,preprintnumbers,superscriptaddress,amsmath,amssymb,nofootinbib]{revtex4}

\usepackage{graphicx}
\usepackage{dcolumn}
\usepackage{bm}
\usepackage{amssymb}
\usepackage{amsmath}
\usepackage{epsfig}    
\usepackage{color}
\usepackage{slashed}
\usepackage[hypertex]{hyperref}
\begin{document} 

\title{Gauge invariant one-loop corrections to Higgs boson couplings\\ in non-minimal Higgs models }

\author{Shinya Kanemura}
\email{kanemu@phys.sci.osaka-u.ac.jp }
\affiliation{Department of Physics, Osaka University, Toyonaka, Osaka 560-0043, Japan}

\author{Mariko Kikuchi}
\email{marikokikuchi@hep1.phys.ntu.edu.tw}
\affiliation{Department of Physics, National Taiwan University, Taipei 10617, Taiwan}

\author{Kodai Sakurai}
\email{sakurai@jodo.sci.u-toyama.ac.jp}
\affiliation{Department of Physics, University of Toyama, \\3190 Gofuku, Toyama 930-8555, Japan}

\author{Kei Yagyu}
\email{yagyu@fi.infn.it}
\affiliation{INFN, Sezione di Firenze, and Department of Physics and Astronomy, University of Florence, Via
G. Sansone 1, 50019 Sesto Fiorentino, Italy}

\preprint{OU-HET 930}
\preprint{UT-HET 120}

\begin{abstract}

We comprehensively evaluate renormalized Higgs boson couplings at one-loop level 
in non-minimal Higgs models such as the Higgs Singlet Model (HSM) and the four types of Two Higgs Doublet Models (THDMs) with a softly-broken $Z_2$ symmetry. 
The renormalization calculation is performed in the on-shell scheme improved by using the pinch technique to eliminate the gauge dependence in 
the renormalized couplings. 
We first review the pinch technique for scalar boson two-point functions in the Standard Model (SM), the HSM and the THDMs. 
We then discuss the difference in the results of the renormalized Higgs boson couplings between the improved on-shell scheme and the ordinal one 
with a gauge dependence appearing in mixing parameters of scalar bosons. 
Finally, we widely investigate how we can identify the HSM and the THDMs 
focusing on the pattern of deviations in the renormalized Higgs boson couplings from predictions in the SM. 

\end{abstract}
\maketitle

\newpage
\section{Introduction}

In spite of the success of the Standard Model (SM), there are many reasons to introduce
new physics beyond the SM from both experiments and theory considerations. 
At the LHC a Higgs boson has been found, but none of new particle has been found yet. 
It is expected that current and future collider experiments will find something new, namely either discovering direct evidence of new particles
or detecting deviations from the SM predictions. 

Although the Higgs boson was found, the structure of the Higgs sector remains unknown. 
The current data indicate that the observed Higgs boson behaves like the SM one~\cite{LHC1}. 
Still, there is no compelling reason for the minimal Higgs sector of the SM, and 
there are many possibilities for non-minimal structures in the Higgs sector. 

It is actually very important to clarify the structure of the Higgs sector from the view point of 
exploring new physics beyond the SM. 
The strength of the interaction, multiplet structures, symmetries of the Higgs sector are closely 
related to specific scenarios of new physics beyond the SM. 
Therefore, the Higgs sector is a probe of new physics. 

Non-minimal structures of the Higgs sector can be explored by directly discovering additional scalar particles 
at current and future experiments. 
Once we discover such a new particle, we can reconstruct the structure of the Higgs sector by measuring 
masses and couplings of these particles in details. 
However, it is not clear whether such new particles can be in the reach of direct searches in the near future. 

As a complementary way, there is a possibility to indirectly discover evidence of new physics beyond the SM by detecting deviations from the predictions in the SM. 
In particular, with the new observables after the discovery of the Higgs boson, such as the coupling constants with the Higgs boson $h(125)$,  
deviations in these coupling constants can make a specific pattern, by which we can fingerprint models with non-minimal Higgs sectors~\cite{finger}. 

Current magnitudes of the precision for the Higgs couplings measurements are typically the order of 10\% level or worse at the LHC experiments~\cite{LHC1}. 
They will be improved in the near future at future experiments such 
as LHC Run-II and High-Luminosity LHC (HL-LHC)~\cite{HLLHC_ATLAS,HLLHC_CMS}, and those at $e^+e^-$ colliders, e.g., 
the International Linear Collider (ILC)~\cite{ILC1}, the Compact Linear Collider (CLIC)~\cite{CLIC} and the Future $e^+e^-$ Circular Collider (FCCee). 
For example at the HL-LHC (at the Initial Phase of the ILC), the $hZZ$, $hb\bar{b}$ and $h\tau\tau$ couplings will be measured 
with 2--4\% (0.58\%), 4--7\% (1.5\%) and 2--5\% (1.9\%) at 1$\sigma$~\cite{Snowmass,ILC1}, respectively. 
Obviously, theory predictions for the Higgs boson couplings must be evaluated with more accuracy than those experimental errors, namely, 
we need to go beyond the tree level calculation. 
Therefore, it is important to systematically prepare the calculation of various Higgs boson couplings at loop levels. 
In addition, these calculations should be systematically performed in various kinds of extended Higgs sectors. 

So far, one-loop corrections to the Higgs boson couplings have been investigated in various models. 
In the SM, one-loop corrections to the $hVV$ ($V=W,Z$) couplings were calculated in Refs.~\cite{Fleischer,Kniehl_hzz,Kniehl_hww}. 
For the $hf\bar{f}$ couplings, one-loop QCD and electroweak corrections were respectively computed in 
Refs.~\cite{hff_QCD1,hff_QCD2,hff_QCD3,hff_QCD4} and Refs.~\cite{Fleischer,Kniehl_hff}. 
These calculations have been established in early 90's mainly based on the electroweak on-shell renormalization scheme~\cite{OS1,OS2,OS3}. 
After that, one-loop corrected Higgs boson couplings have also been calculated in various models beyond the SM. 
For example, in the minimal supersymmetric SM (MSSM), one-loop corrections to the $hf\bar{f}$ couplings have been intensively 
studied in Refs.~\cite{hff_MSSM0,hff_MSSM1,hff_MSSM2,hff_MSSM3,hff_MSSM4}, because of the sizable amount of the supersymmetric QCD corrections. 
In addition, in Refs.~\cite{hhh_MSSM1,hhh_MSSM2} the Higgs boson self-coupling $hhh$ has been calculated at one-loop level in the MSSM. 
In (non-supersymmetric) two Higgs doublet models (THDMs), one-loop corrected $hZZ$~\cite{KOSY}, $hhh$~\cite{Kiyoura,KOSY}
and $hf\bar{f}$~\cite{THDM-KKY1} couplings have been studied. 
In Ref.~\cite{THDM-KKY2},  an improved fingerprinting identification of THDMs has been discussed using the one-loop renormalized 
$hVV$ $(V=Z,W)$, $hf\bar{f}$ and $hhh$ couplings. 
In the other extended Higgs sectors, such calculations are also found in Refs.~\cite{HSM-KKY1,HSM-KKY2} for the Higgs Singlet Model (HSM), 
in Refs.~\cite{IDM1,IDM2} for the inert doublet model, and in Refs.~\cite{AKSY1,AKSY2} for the Higgs triplet model. 

However, it has been known that 
gauge dependence appears in the renormalization of mixing parameters among fields, 
e.g., fermions~\cite{Gambino:1998ec,Kniehl:2000rb,Barroso:2000is,Yamada} or scalar bosons~\cite{Yamada,Espinosa:2002cd,MSSM-gauge} based on the on-shell scheme, which 
is proven by using the Nielsen identity~\cite{NI}. 
Fortunately, it has already been known the way to remove such gauge dependence by using 
the pinch technique~\cite{Cornwall:1981zr,Cornwall:1989gv,Papavassiliou:1989zd,Papavassiliou:1994pr,Binosi:2009qm,Degrassi:1992ue}, and 
the gauge invariant scheme has been constructed in various models, e.g., in the MSSM~\cite{MSSM-gauge, Baro:2008bg, Baro:2009gn}, the HSM~\cite{HSM-gauge}, and the THDM~\cite{THDM-gauge,Denner:2016etu,Altenkamp:2017ldc}. 

In this paper, we comprehensively calculate one-loop corrections to Higgs boson couplings 
based on the on-shell renormalization scheme improved by using the pinch technique (the so-called pinched tadpole scheme~\cite{Fleischer}) to remove the gauge dependence.  
In particular, as important examples of extended Higgs sectors 
we concentrate on the HSM and the THDMs with a softly-broken $Z_2$ symmetry which is imposed to avoid Flavor Changing Neutral Currents (FCNCs) at tree level~\cite{GW}. 
For the latter models, we consider all possible four independent types of Yukawa interactions called Type-I, Type-II, Type-X and Type-Y appearing due to 
different choices of the $Z_2$ charge assignment for fermions~\cite{Barger,Grossman,typeX}. 
We first explicitly show the cancellation of the gauge dependence in Higgs boson two-point functions computed in the general $R_\xi$ gauge by adding 
pinch-terms which are extracted from vertex corrections and box diagrams of a 2--fermion to 2--fermion scattering process in the SM, the HSM and THDMs. 
We then define the gauge independent renormalized mixing angles based on the pinched tadpole scheme, and 
discuss the difference in various one-loop corrected Higgs boson couplings based on the pinched tadpole scheme 
and those based on the ordinal on-shell scheme with the gauge dependence~\cite{OS1,OS2,OS3}. 
We then investigate how we can identify the HSM and THDMs by ``fingerprinting'' various one-loop corrected Higgs boson couplings with usage of 
the gauge invariant renormalization scheme. 
Namely, these extended Higgs sectors can be disentangled by looking at the difference in the pattern of deviations in the Higgs boson couplings from the SM prediction. 
In order to concretely show how the fingerprinting works, 
we display various correlations between $\kappa_Z^{}$--$\kappa_\tau$, $\kappa_\tau$--$\kappa_b$, $\kappa_\tau$--$\kappa_c$ and $\kappa_Z$--$\kappa_\gamma$, 
where $\kappa_X^{}$ denote the normalized $hXX$ couplings by the SM prediction ($h$ being the discovered Higgs boson with the mass of 125 GeV). 
As a result, if $|\kappa_Z^{}-1|$ is found to be $\sim 1\%$ or larger, there is a possibility to distinguish these models by the combination of the measurements of 
$\kappa_\tau$, $\kappa_b$ and $\kappa_c$. 

The originality of this paper should be the following. 
We discuss how we can discriminate various Higgs sectors by focusing on the couplings of $h(125)$ with SM particles at the one-loop level without gauge dependence 
in various non-minimal Higgs sectors. 
In the previous studies~\cite{HSM-gauge,THDM-gauge}, 
one-loop corrected non-SM couplings with extra Higgs bosons have been discussed in a specific non-minimal Higgs sector. 
In addition, we provide details of calculations for the part of the pinch technique explicitly, some of which have not been shown in the literature, 
which might be useful for people who try to follow the calculation.

This paper is organized as follows. In Sec.~\ref{sec:model}, we give a brief review of the HSM and THDMs, i.e., defining their Lagrangians and giving mass formulae for 
Higgs bosons. We also discuss various constraints on parameters of these models. 
Sec.~\ref{sec:pt}, we show the cancelation of the gauge dependence in Higgs boson two-point functions using the pinch technique 
in the SM, the HSM and the THDMs in order. 
Full set of relevant Feynman diagrams giving rise to the gauge dependence of two-point functions and those to extract pinch-terms are displayed. 
Sec.~\ref{sec:inv}, we discuss the difference in the renormalized Higgs boson couplings calculated in the pinched tadpole scheme without the gauge dependence and 
in the ordinal on-shell scheme with the gauge dependence. 
Sec.~\ref{sec:num}, we numerically show predictions of various scaling factors $\kappa_X^{}$ in the HSM and the THDMs. We then discuss how we can identify these models 
by the difference in predictions of $\kappa_X^{}$. 
Conclusions are given in Sec.~\ref{sec:con}.

\section{Extended Higgs Sectors\label{sec:model}}

In order to fix notation, we give a brief review of the HSM and the THDMs with a softly-broken $Z_2$ symmetry and CP-conservation. 

\subsection{HSM}

The Higgs sector of the HSM is composed of an isospin doublet field $\Phi$ with the hypercharge $Y =1/2$ and a real isospin singlet scalar field $S$ with $Y=0$. 
The most general scalar potential is given by
\begin{align}
V(\Phi,S) =& +m_\Phi^2|\Phi|^2+\lambda |\Phi|^4  
+\mu_{\Phi S}^{}|\Phi|^2 S+ \lambda_{\Phi S} |\Phi|^2 S^2 
+t_S^{}S +m^2_SS^2+ \mu_SS^3+ \lambda_SS^4,\label{Eq:HSM_pot}
\end{align} 
where the doublet and singlet fields can be parameterized by 
\begin{align}
\Phi=\left(\begin{array}{c}
G^+\\
\frac{v+\phi+iG^0}{\sqrt{2}}
\end{array}\right),\quad
S=v_S^{} + s. \label{hsm_f}
\end{align}
In Eq.~(\ref{hsm_f}), $G^+$ and $G^0$ are the Nambu-Goldstone (NG) bosons which are absorbed into the longitudinal components of the $W^+$ and $Z$ bosons, respectively. 
The Vacuum Expectation Value (VEV) $v$ of $\Phi$ is directly related to the Fermi constant $G_F$ by $v = (\sqrt{2}G_F)^{-1/2}\simeq 246$ GeV. 
On the other hand, the singlet VEV $v_S^{}$ of $S$ contributes to neither the electroweak symmetry breaking nor generation of fermion masses. 
In addition, a shift of the singlet VEV $v_S^{} \to v_S'$ can be absorbed by the reparameterization of parameters in the potential~\cite{CDL}. 
Therefore, we simply take $v_S^{}=0$ in the following discussion. 

The mass eigenstates of the two CP-even scalar states are defined by 
\begin{align}
\begin{pmatrix}
s \\
\phi
\end{pmatrix} = R(\alpha)
\begin{pmatrix}
H \\
h
\end{pmatrix}~~\text{with}~~R(\theta) = 
\begin{pmatrix}
\cos\theta & -\sin \theta \\
\sin\theta & \cos\theta
\end{pmatrix}.   \label{mat_r}
\end{align}
Hereafter, we introduce the shorthand notation $s_\theta = \sin\theta$ and $c_\theta = \cos\theta$. 
Their masses are calculated after imposing the tree level tadpole conditions, i.e., 
\begin{align}
\left. \frac{\partial V}{\partial \phi}\right|_0 = \left. \frac{\partial V}{\partial s}\right|_0 = 0, 
\end{align}
by which we can eliminate $m_\Phi^2$ and $t_S$ parameters, where $|_0$ denotes taking all the scalar fields to be zero after the derivative. 
The squared masses ($m_H^2$ and $m_h^2$) and the mixing angle $\alpha$ are then expressed as
\begin{align}
 &m_H^2=M_{11}^2c^2_\alpha +M_{22}^2s^2_\alpha +M_{12}^2s_{2\alpha} , \label{mbh}\\
 &m_h^2=M_{11}^2s^2_\alpha +M_{22}^2c^2_\alpha -M_{12}^2s_{2\alpha} , \label{mh}\\
 &\tan 2\alpha=\frac{2M_{12}^2}{M_{11}^2-M_{22}^2}, \label{tan2a}
 \end{align} 
where $M_{ij}^2$ are the mass matrix elements for the CP-even scalar states in the basis of $(s,\phi)$:
\begin{align}
M^2_{11}= 2m_S^2+ v^2\lambda_{\Phi S}  ,\quad M^2_{22}=2\lambda v^2,\quad M^2_{12}=v\mu_{\Phi S}^{}. \label{mij}
\end{align}
We identify $h$ as the discovered Higgs boson at the LHC, so that we take $m_h = 125$ GeV. 
From Eqs.~(\ref{mbh})--(\ref{mij}), we can solve ($\lambda$, $m_S^2$, $\mu_{\Phi S}^{}$) in terms of ($m_h$, $m_H$, $\alpha$) as 
\begin{align}
\begin{split}
\lambda &= \frac{1}{2v^2}(m_h^2c^2_\alpha + m_H^2 s^2_\alpha), \\
m_S^2  &= \frac{1}{2}\left(m_h^2s_\alpha^2 +  m_H^2c_\alpha^2 -\lambda_{\Phi S}v^2\right), \\
\mu_{\Phi S}^{} &= \frac{1}{v}s_\alpha c_\alpha\left(m_H^2-m_h^2 \right).  
\end{split}\label{aaa}
\end{align}
Consequently, we can choose the following 5 free parameters as inputs:
\begin{align}
&m_H,~~\alpha,~~\lambda_S,~~\lambda_{\Phi S},~~\mu_{S}. 
\end{align}
These parameters can be constrained by taking into account the following arguments with respect to the theoretical consistency. 

First, we impose the perturbative unitarity bound~\cite{lqt} defined by 
\begin{align}
|a_0^i| \leq \frac{1}{2}, 
\end{align}
where $a_0^i$ are the eigenvalues of the $s$-wave amplitude matrix for elastic 2 body to 2 body scalar boson scatterings. 
There are 4 independent eigenvalues written in terms of dimensionless parameters in the potential in the HSM~\cite{pu_HSM}, which can be rewritten 
in terms of the physical parameters, e.g.,  $m_H^{}$ and $\alpha$ via Eq.~(\ref{aaa}). 

Second, we require that the Landau pole does not appear at a certain energy scale. 
In this paper, we impose the following condition as the triviality bound:
\begin{align}
|\lambda_i(\mu)| \leq 4\pi, ~~\text{for}~~^\forall\mu~\text{with}~m_Z^{} \leq \mu \leq \Lambda_{\text{cutoff}}, 
\end{align} 
where $\Lambda_{\text{cutoff}}$ is the cutoff of the model, and $\lambda_i(\mu)$
are the dimensionless running parameters at the scale $\mu$, which can be evaluated by solving 
the one-loop renormalization group equations. 
The one-loop beta functions in the HSM are given in Ref.~\cite{beta_HSM}. 

Third, we require the condition to guarantee the potential being bounded from below in any direction of the scalar field space. 
The sufficient condition to avoid the vacuum instability at any scale $\mu$ up to the cutoff $\Lambda_{\text{cutoff}}$
is given by~\cite{vs_HSM} 
\begin{align}
\lambda(\mu) \geq 0,~ \lambda_S(\mu) \geq 0,~ 2\sqrt{\lambda(\mu) \lambda_S(\mu)} + \lambda_{\Phi S}(\mu) \geq 0, ~~\text{for}~~^\forall\mu~\text{with}~m_Z^{} \leq \mu \leq 
\Lambda_{\text{cutoff}}.
\end{align}

Fourth, we impose the bound from conditions to avoid wrong vacua~\cite{CDL,HSM-EWBG3}. 
Because of the existence of the scalar trilinear couplings $\mu_S^{}$ and $\mu_{\Phi S}^{}$, 
non-trivial local extrema can appear in the Higgs potential.
Therefore, we need to check if the true extremum at
$(\sqrt{2}\langle \Phi \rangle, \langle S \rangle) = (v,0)$ corresponds to the minimum of the potential. 
This condition can be expressed as
 \begin{align}
 V_{\text{nor}}(v_\pm,x_\pm) > 0, \,\,\,\,\,
 V_{\text{nor}}(0,x_{1,2,3}) > 0,
 \end{align}
where $V_{\text{nor}}$ is the normalized Higgs potential satisfying $V_{\text{nor}}(v,0)=0$. 
The analytic formulae of the false VEVs for the doublet field $v_\pm$ and those for the singlet field $x_\pm$ and $x_{1,2,3}$ 
are found in Ref.~\cite{HSM-KKY2}. 

Finally, we take into account the constraint from the electroweak oblique parameters $S$ and $T$ introduced by Peskin and Takeuchi~\cite{peskin-takeuchi}. 
We define new physics contributions to the $S$ and $T$ parameters as 
$\Delta S\equiv S_{\text{NP}} - S_{\text{SM}}$ and $\Delta T \equiv T_{\text{NP}} - T_{\text{SM}}$ with 
$S_{\text{NP(SM)}}$ and $T_{\text{NP(SM)}}$ are the new physics (SM) prediction to the $S$ and $T$ parameters, respectively. 
From Ref.~\cite{stu}, the fitted values of the $\Delta S$ and $\Delta T$ are given under $\Delta U=0$ by
\begin{align}
\Delta S = 0.05 \pm 0.09,\quad \Delta T =  0.08 \pm 0.07, 
\end{align}
with the correlation factor $\rho_{\text{ST}}= +0.91$. We require that the prediction of the model is within the 95\% confidence level (CL) region, 
which is expressed by $\Delta \chi^2(\Delta S,\Delta T)\leq 5.99$. 
The analytic expressions for the new contributions $\Delta S$ and $\Delta T$ can be found in e.g., Ref.~\cite{st_HSM}. 

Before closing this subsection, let us give the trilinear interaction terms among the Higgs bosons and weak bosons or fermions. 
Because the singlet field $S$ does not couple to weak bosons and fermions, 
the singlet-like Higgs boson $H$ couples to these SM fields only through the non-zero mixing angle $\alpha$, while
the SM-like Higgs boson $h$ couplings are universally suppressed by the factor of $\cos\alpha$. 
As a result, we obtain the following interaction Lagrangian:
\begin{align}
{\cal L}_{\text{trilinear}} & = \left(\frac{h}{v}c_\alpha +\frac{H}{v}s_\alpha \right)(2m_W^2 W^+_\mu W^{-\mu} + m_Z^2Z_\mu Z^\mu - m_f\bar{f}f). 
\end{align}
In App.~\ref{sec:rhiggs}, we also give scalar trilinear couplings. 

\subsection{THDM}

\begin{table}[t]
\begin{center}
{\renewcommand\arraystretch{1.2}
\begin{tabular}{c|ccccccc|ccc}\hline\hline
&$\Phi_1$&$\Phi_2$&$Q_L$&$L_L$&$u_R$&$d_R$&$e_R$&$\zeta_u$ &$\zeta_d$&$\zeta_e$ \\\hline
Type-I &$+$&
$-$&$+$&$+$&
$-$&$-$&$-$&$\cot\beta$&$\cot\beta$&$\cot\beta$ \\\hline
Type-II&$+$&
$-$&$+$&$+$&
$-$
&$+$&$+$& $\cot\beta$&$-\tan\beta$&$-\tan\beta$ \\\hline
Type-X &$+$&
$-$&$+$&$+$&
$-$
&$-$&$+$&$\cot\beta$&$\cot\beta$&$-\tan\beta$ \\\hline
Type-Y &$+$&
$-$&$+$&$+$&
$-$
&$+$&$-$& $\cot\beta$&$-\tan\beta$&$\cot\beta$ \\\hline\hline
\end{tabular}}
\caption{Charge assignment of the $Z_2$ symmetry and the $\zeta_f$ ($f=u,d,e$) factors appearing in Eq.~(\ref{zeta-hff}) in each of four types of Yukawa interactions. 
}
\label{yukawa_tab}
\end{center}
\end{table}

The Higgs sector is composed of two isospin doublets $\Phi_1$ and $\Phi_2$ with $Y=1/2$. 
In order to avoid FCNCs at the tree level, we impose a 
$Z_2$ symmetry in the Higgs sector, which can be softly broken by a parameter in the potential. 
We fix the $Z_2$ charge assignment for two doublets and fermions as given in Table~\ref{yukawa_tab}. 
Depending on the $Z_2$ charge assignment on right-handed fermions, we can define four types of Yukawa interactions~\cite{Barger,Grossman} called as 
Type-I, Type-II, Type-X and Type-Y~\cite{typeX}. 

The Higgs potential under the $Z_2$ symmetry and the CP invariance is given by  
\begin{align}
V(\Phi_1,\Phi_2) &= +m_1^2|\Phi_1|^2+m_2^2|\Phi_2|^2-m_3^2(\Phi_1^\dagger \Phi_2 +\text{h.c.})\notag\\
& +\frac{1}{2}\lambda_1|\Phi_1|^4+\frac{1}{2}\lambda_2|\Phi_2|^4+\lambda_3|\Phi_1|^2|\Phi_2|^2+\lambda_4|\Phi_1^\dagger\Phi_2|^2
+\frac{1}{2}\lambda_5\left[(\Phi_1^\dagger\Phi_2)^2+\text{h.c.}\right], \label{pot_thdm2}
\end{align}
where the two doublet fields can be parameterized as 
\begin{align}
\Phi_i=\left(\begin{array}{c}
w_i^+\\
\frac{v_i+h_i+iz_i}{\sqrt{2}}
\end{array}\right),\hspace{3mm}(i=1,2), 
\end{align}
with $v_i$ being the VEVs for $\Phi_i$. These two VEVs can be expressed as ($v,\tan\beta$) defined by
$v = \sqrt{v_1^2+v_2^2}=(\sqrt{2}G_F)^{-1/2}$ and $\tan\beta=v_2/v_1$. 

The mass eigenstates for the scalar bosons are obtained by the following orthogonal transformations:
\begin{align}
\left(\begin{array}{c}
w_1^\pm\\
w_2^\pm
\end{array}\right)&=R(\beta)
\left(\begin{array}{c}
G^\pm\\
H^\pm
\end{array}\right),\quad 
\left(\begin{array}{c}
z_1\\
z_2
\end{array}\right)
=R(\beta)\left(\begin{array}{c}
G^0\\
A
\end{array}\right),\quad
\left(\begin{array}{c}
h_1\\
h_2
\end{array}\right)=R(\alpha)
\left(\begin{array}{c}
H\\
h
\end{array}\right), \label{mixing}
\end{align}
where $\alpha$ is the mixing angle between two CP-even scalar states. Similar to the HSM case, we regard the $h$ state as the discovered Higgs boson at the LHC. 

By imposing the tree level tadpole conditions, i.e., 
\begin{align}
\left. \frac{\partial V}{\partial h_1}\right|_0 = \left. \frac{\partial V}{\partial h_2}\right|_0 = 0, 
\end{align}
we can eliminate the $m_1^2$ and $m_2^2$ parameters. 
We then obtain the mass formulae of the physical Higgs bosons. 
First, the squared masses of $H^\pm$ and $A$ are calculated as 
\begin{align}
m_{H^\pm}^2 = M^2-\frac{v^2}{2}(\lambda_4+\lambda_5),\quad m_A^2&=M^2-v^2\lambda_5,  \label{mass1}
\end{align}
where $M$ describes the soft breaking scale of the $Z_2$ symmetry defined as: 
\begin{align}
M^2=\frac{m_3^2}{s_\beta c_\beta}. \label{bigm}
\end{align}
The masses for the CP-even Higgs bosons and the mixing angle $\alpha$ can be expressed by
\begin{align}
&m_H^2=c^2_{\beta-\alpha} M_{11}^2 + s^2_{\beta-\alpha} M_{22}^2 - s_{2(\beta-\alpha)} M_{12}^2, \label{111}\\
&m_h^2=s^2_{\beta-\alpha} M_{11}^2 + c^2_{\beta-\alpha} M_{22}^2 + s_{2(\beta-\alpha)}M_{12}^2,  \label{222}\\
&\tan 2(\beta-\alpha)= -\frac{2M_{12}^2}{M_{11}^2-M_{22}^2}, \label{333}
\end{align} 
where $M_{ij}^2$ ($i,j=1,2$) are the mass matrix elements for the CP-even scalar states in the basis of $(h_1,h_2)R(\beta)$: 
\begin{align}
M_{11}^2&=v^2(\lambda_1c^4_\beta+\lambda_2s^4_\beta)+\frac{v^2}{2}\lambda_{345}s^2_{2\beta},  \label{m11}  \\
M_{22}^2&=M^2+v^2s^2_\beta c^2_\beta(\lambda_1+\lambda_2-2\lambda_{345}), \label{m22}  \\
M_{12}^2&=\frac{v^2}{2} s_{2\beta}( \lambda_2s^2_\beta  -\lambda_1c^2_\beta)+\frac{v^2}{2}s_{2\beta} c_{2\beta} \lambda_{345},  \label{m12}
\end{align}
with $\lambda_{345}\equiv \lambda_3+\lambda_4+\lambda_5$. 
From Eqs.~(\ref{mass1}) and (\ref{111})--(\ref{m12}), 
the quartic couplings $\lambda_1$--$\lambda_5$ in the potential are rewritten in terms of the physical parameters as
\begin{align}
\begin{split}
\lambda_1v^2 &= (m_H^2\tan^2\beta + m_h^2)s^2_{\beta-\alpha} 
            +(m_H^2 + m_h^2\tan^2\beta)c^2_{\beta-\alpha} \\
             &~~~+2(m_H^2-m_h^2) s_{\beta-\alpha} c_{\beta-\alpha} \tan\beta -M^2\tan^2\beta, \\
\lambda_2v^2 &= (m_H^2\cot^2\beta + m_h^2) s^2_{\beta-\alpha}
            +(m_H^2 + m_h^2\cot^2\beta)c^2_{\beta-\alpha} \\
             &~~~-2(m_H^2-m_h^2) s_{\beta-\alpha} c_{\beta-\alpha} \tan\beta -M^2\cot^2\beta, \\
\lambda_3v^2 &= (m_H^2- m_h^2)[c^2_{\beta-\alpha} -s^2_{\beta-\alpha} +(\tan\beta-\cot\beta)s_{\beta-\alpha}c_{\beta-\alpha}]\\
             &~~~ +2m_{H^\pm}^2-M^2, \\
\lambda_4v^2&=M^2+m_A^2-2m_{H^\pm}^2,\\
\lambda_5v^2&=M^2-m_A^2. 
\end{split}\label{lams}
\end{align}
From the above discussion, we can choose the following 6 free parameters as inputs:
\begin{align}
m_H^{},~ m_A^{},~ m_{H^\pm},~ s_{\beta-\alpha},~ \tan\beta,~ M^2. 
\end{align} 

As we discussed in the previous subsection, we can constrain these parameters by taking into account 
bounds from the perturbative unitarity, the triviality, the vacuum stability and the $S$ and $T$ parameters. 
The 12 independent eigenvalues $a_0^i$ of the $s$-wave amplitude matrix are given in Refs.~\cite{pu_THDM1,pu_THDM2,pu_THDM3,pu_THDM4,pu_THDM5}. 
The sufficient condition for the vacuum stability~\cite{vs_THDM1,vs_THDM2,vs_THDM3,Klimenko} at an arbitrary scale $\mu$ is given by
\begin{align}
&\lambda_1(\mu) \geq 0, \quad \lambda_2(\mu) \geq0,~
\sqrt{\lambda_1(\mu) \lambda_2(\mu)}+\lambda_3(\mu) + 
\text{MIN}[0,\lambda_4(\mu) \pm \lambda_5(\mu)]\geq 0,\notag\\
&\text{for}~~^\forall\mu~\text{with}~m_Z^{} \leq \mu \leq \Lambda_{\text{cutoff}}. 
\end{align}
The beta functions for the 5 dimensionless couplings can be found in Ref.~\cite{beta_THDM}. 
In addition, the analytic expressions for the new contributions $\Delta S$ and $\Delta T$ are given in Refs.~\cite{st_THDM1,st_THDM2,st_THDM3,st_THDM4,st_THDM5}. 

Apart from the discussion of the potential, let us consider the Yukawa Lagrangian. 
Under the $Z_2$ symmetry~\cite{GW}, the general form of the Yukawa Lagrangian is given by 
\begin{align}
{\mathcal L}_Y =
&-Y_u{\overline Q}_Li\sigma_2\Phi^*_uu_R^{}
-Y_d{\overline Q}_L\Phi_dd_R^{}
-Y_e{\overline L}_L\Phi_e e_R^{}+\text{h.c.},
\end{align}
where $\Phi_{u,d,e}$ are either $\Phi_1$ or $\Phi_2$. 
Then, we can extract the trilinear interaction terms among the Higgs bosons and weak bosons or fermions as 
\begin{align}
\mathcal {L}_{\text{trilinear}}
&=(s_{\beta-\alpha} h+ c_{\beta-\alpha} H) \Big(\frac{2m_W^2}{v} W^{+\mu} W^-_\mu +  \frac{m_Z^2}{v} Z^\mu Z_\mu \Big)\notag\\
&-\sum_{f=u,d,e}\frac{m_f}{v}\left( \zeta_{hff}{\overline f}fh+\zeta_{Hff} {\overline f}fH-2iI_f \zeta_f{\overline f}\gamma_5fA\right)\notag\\
&-\frac{\sqrt{2}}{v}\left[V_{ud}\overline{u}
\left(m_d\zeta_d\,P_R-m_u\zeta_uP_L\right)d\,H^+
+m_e\zeta_e\overline{\nu^{}}P_Re^{}H^+ +\text{h.c.}\right],  \label{yukawa_thdm}
\end{align}
where $I_f$ represents the third component of the isospin of a fermion $f$; i.e., $I_f=+1/2$ $(-1/2)$ for $f=u~(d,e)$, and 
$\zeta_{hff}$ and $\zeta_{Hff}$ are defined by
\begin{align}
\zeta_{hff} &= s_{\beta-\alpha} + \zeta_f c_{\beta-\alpha}, \quad \zeta_{Hff} = c_{\beta-\alpha} - \zeta_f s_{\beta-\alpha}.  \label{zeta-hff}
\end{align}
In the above expression, the $\zeta_f$ factor is either $\cot\beta$ or $-\tan\beta$ depending on the fermion type and type of Yukawa interactions
as given in Table~\ref{yukawa_tab}. 
In App.~\ref{sec:rhiggs}, we also give scalar trilinear couplings.

\section{Gauge invariant scalar boson two-point functions \label{sec:pt}}

In the previous section, we gave the tree level formulae of the Higgs boson couplings with weak bosons and fermions. 
By focusing on the difference in various correlations of the deviation in the $hVV$ and $hf\bar{f}$ couplings from the SM prediction, we can discriminate
HSM and the THDM with four types of Yukawa interactions as it has been clearly shown in Ref.~\cite{finger}.  
Currently, the Higgs boson couplings, e.g., $hZZ$, $hWW$, $h\gamma\gamma$ and $hf\bar{f}$ ($f=t,b,\tau$) are measured to be typically 
order of 10\% level or even worse particularly for the Yukawa couplings at the LHC Run-I experiment~\cite{LHC1}. 
However, the accuracy of the Higgs boson coupling measurements are expected to be significantly improved at future collider experiments such as the 
HL-LHC~\cite{HLLHC_ATLAS,HLLHC_CMS} and future $e^+e^-$ colliders~\cite{ILC1,CLIC}. 
Therefore, to compare such precise measurements, we need to calculate the Higgs boson coupling at loop levels. 

In order to obtain finite predictions of one-loop corrected observables, 
renormalization of the Lagrangian parameters has to be done.   
Among various renormalization schemes, the on-shell scheme~\cite{OS1,OS2,OS3} provides clear definition of the renormalized parameters, namely, renormalized masses do not receive any corrections at their on-shell. 
By this requirement, we can determine counter terms of the Lagrangian parameters which cancel the ultra-violet (UV) divergence appearing from one-loop diagrams. 
Although the on-shell scheme has aforementioned nice feature, 
it has been known that gauge dependence appear in the renormalization of mixing parameters between scalar bosons as mentioned in Introduction. 

In this section, we discuss the cancelation of the gauge dependence in scalar boson two-point functions by using the 
pinch technique~\cite{Cornwall:1981zr,Cornwall:1989gv,Papavassiliou:1989zd,Papavassiliou:1994pr,Binosi:2009qm,Degrassi:1992ue} in the three models, i.e., the SM, the HSM and the THDM.  
We adopt the general $R_\xi$ gauge to the following calculation in order to explicitly show how the gauge dependence is canceled. 
In the $R_\xi$ gauge, a propagator of a gauge boson $V^\mu$ ($V=W,Z,\gamma$) is expressed in terms of the gauge parameter $\xi_V^{}$ as  
\begin{align}
\Delta_V^{\mu\nu} &= \frac{1}{p^2 -m_V^2}\left[g^{\mu\nu} - (1-\xi_V)\frac{p^\mu p^\nu}{p^2-\xi_V m_V^2}\right]. \label{rx}
\end{align}
We note that for $V=W\,(Z)$, $\xi_W m_W^2$ ($\xi_Z m_Z^2$) corresponds 
to the squared mass of the associated NG boson $G^\pm$ ($G^0$) and the Faddeev-Popov (FP) ghost field $c^\pm$ $(c_Z^{})$. 
In order to simply express the difference between an amplitude 
calculated in the $R_\xi$ gauge and that in the 't~Hooft-Feynman gauge, i.e., $\xi_W^{}=\xi_Z^{}=\xi_\gamma^{}=1$, we introduce the following symbol:
\begin{align}
\Delta_{\xi} {\cal M}^{} \equiv  \sum_{V = W,Z,\gamma}\Delta_{\xi_V^{}} {\cal M}_V^{}~~\text{with}~~
\Delta_{\xi_V^{}} {\cal M}_V^{} \equiv   {\cal M}_V - {\cal M}_V|_{\xi_V^{}=1}, \label{xim}
\end{align}
where ${\cal M}_V^{}$ denotes an amplitude with a dependence on $\xi_V^{}$. 
In the following, diagrams providing a $\xi_V\xi_{V'}$ ($V\neq V'$) dependence do not appear, so that 
we can separate the amplitude in the way shown in Eq.~(\ref{xim}). 
Furthermore, we introduce the following shorthand notations of the Passarino-Veltman functions\footnote{These functions given in 
Eqs.~(\ref{c01}) and (\ref{c02}) are also expressed in terms of the usual $C_0$ function as 
$C_0(p^2;A,B) = C_0(0,p^2,p^2;m_A^{},m_B^{},m_A^{}) + C_0(p^2,0,p^2;m_B^{},m_A^{},m_B^{})$ and 
$C_0(p^2;A,B,C) = C_0(0,p^2,p^2;m_A^{},m_B^{},m_C^{})$. }~\cite{pv}:
\begin{align}
C_0(p^2;A,B)& \equiv \frac{1}{m_A^2-m_B^2}[B_0(p^2;A,A) - B_0(p^2;B,B)], \label{c01}\\
C_0(p^2;A,B,C)& \equiv \frac{1}{m_A^2-m_B^2}[B_0(p^2;A,C) - B_0(p^2;B,C)]\label{c02}, 
\end{align}
where $B_0(p^2,X,Y) = B_0(p^2,m_X^{},m_Y^{})$ with $m_X^{}$ and $m_Y^{}$ being masses of $X$ and $Y$, respectively.

\subsection{SM}

\begin{figure}[!t]
\begin{center}
\includegraphics[width=150mm]{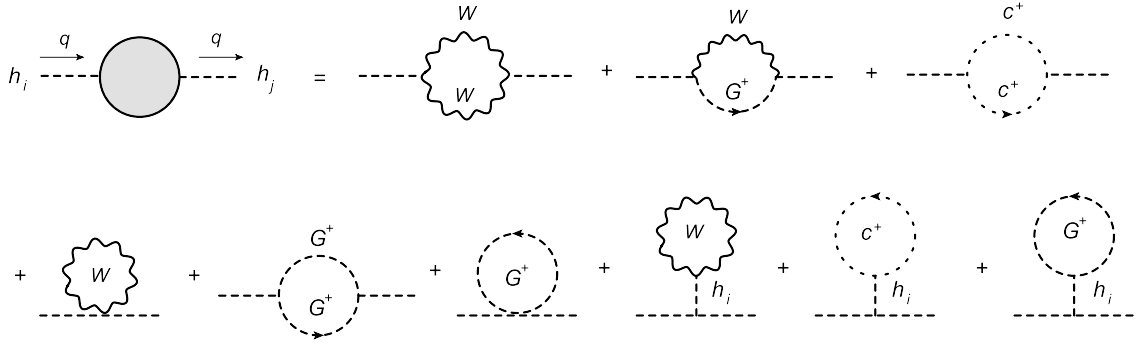}
\caption{Gauge dependent Feynman diagrams for the two-point function of CP-even Higgs bosons $h_i$ and $h_j$. 
Here, we only show the diagrams depending on $\xi_W$.  
Those depending on $\xi_Z^{}$ are obtained by replacing $(W,G^\pm,c^\pm)$ with $(Z,G^0,c_Z^{})$. 
}
\label{h2}
\end{center}
\end{figure}

\begin{figure}[t]
\begin{center}
\includegraphics[width=165mm]{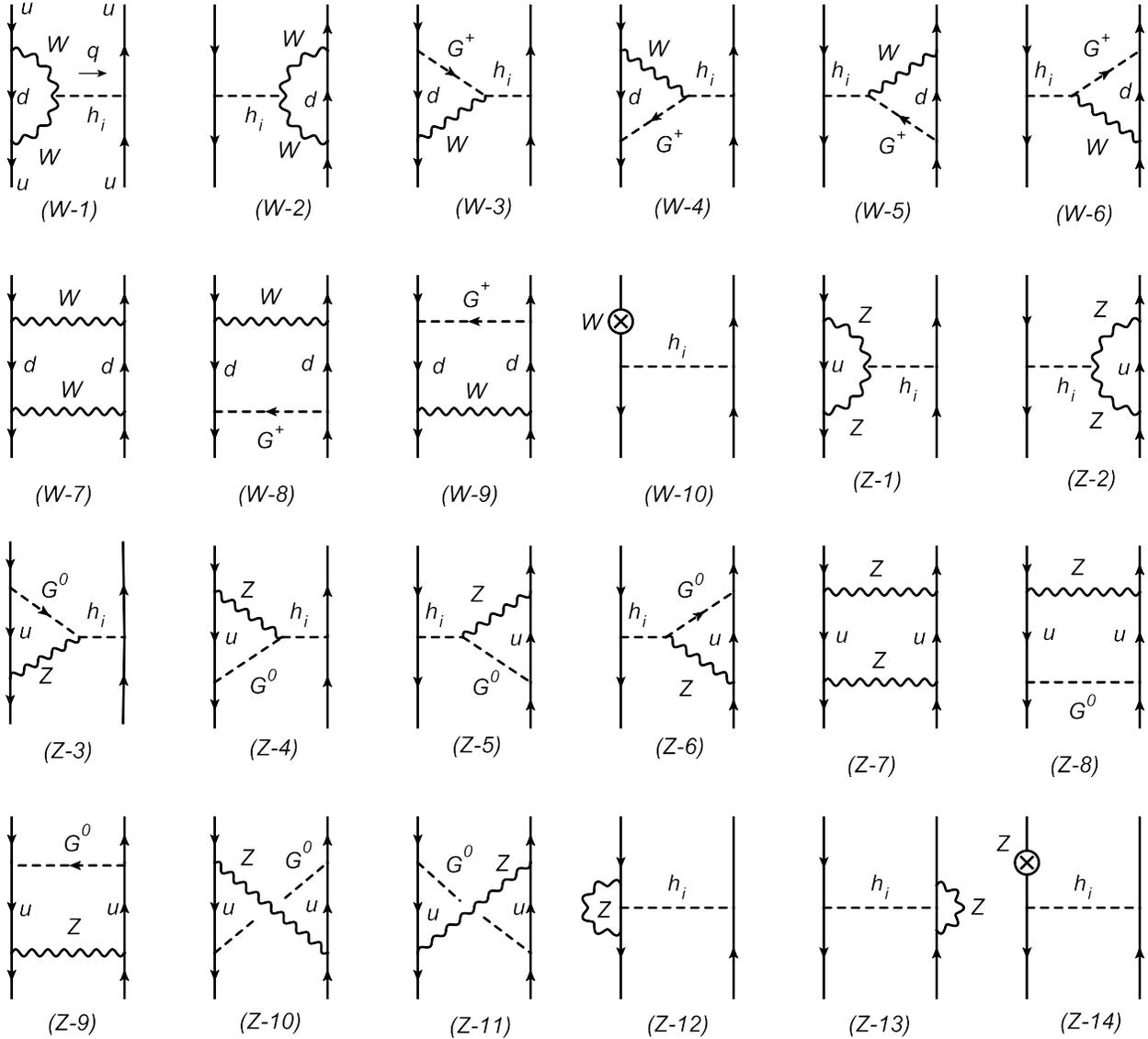}
\caption{Feynman diagrams giving pinch-terms for two-point functions of CP-even Higgs bosons in the $u\bar{u}\to u\bar{u}$ scattering, 
where $h_i$ is a CP-even Higgs boson.
The diagrams $(W$-$10)$ and $(Z$-$14)$ denote the contribution to the $\xi_W^{}$ and $\xi_Z^{}$ dependence
from the wave function renormalization of the external quark, respectively.  }
\label{gauge-dep2}
\end{center}
\end{figure}

As a first example, we review the cancelation of the gauge dependence of the Higgs boson two-point function in the SM according to Ref.~\cite{tt}. 
The Feynman diagrams for the Higgs boson two-point functions providing the gauge dependence are shown in Fig.~\ref{h2}, where $h_{i,j} = h$ in the SM. 
By summing all these diagrams, we obtain:
\begin{align}
\Delta_\xi\Pi_{hh}(q^2)
&= \frac{g^2}{64\pi^2}(1-\xi_W)(q^2-m_h^2)\left[(q^2+m_h^2)C_0(q^2;W,G^\pm)-2B_0(0;W,G^\pm) \right]\notag\\
&+\frac{g_Z^2}{128\pi^2} (1-\xi_Z)(q^2-m_h^2) \left[(q^2+m_h^2)C_0(q^2;Z,G^0)-2B_0(0;Z,G^0) \right],  \label{gd0}
\end{align}
where $g_Z^{}\equiv g/\cos\theta_W$ with $\theta_W^{}$ being the weak mixing angle and $q^\mu$ is the four momentum of the Higgs boson. 
We see that the $\xi_V$ dependence appears in front of the factor of $(q^2-m_h^2)$, which manifestly shows satisfaction of the Nielsen identity~\cite{NI}. 
Therefore, the gauge dependence in the renormalization of the Higgs boson mass vanishes in the on-shell scheme. 
We however explicitly show how this dependence can be cancelled by the pinch technique, by which we can easily extend this result to the case for the non-minimal Higgs sectors. 

In order to show the cancelation of the gauge dependence, 
we consider the $u\bar{u} \to u\bar{u}$ scattering process, where $u$ $(\bar{u})$ are an (anti) up-type quark, as a toy process. 
We note that the cancelation does not depend on the choice of the external fermions. 
In the $u\bar{u}\to u\bar{u}$ process, the contribution from the Higgs boson self-energy is calculated from Eq.~(\ref{gd0}) by 
\begin{align}
\Delta_\xi\overline{{\cal M}}_{hh}&= \frac{g^2}{64\pi^2} \frac{1-\xi_W}{q^2-m_h^2} \left[ (q^2+m_h^2)C_0(q^2;W,G^\pm)-2B_0(0;W,G^\pm) \right]\notag\\
&+\left[(g,m_W^{},\xi_W;W,G^\pm) \to (g_Z/2^{},m_Z^{},\xi_Z;Z,G^0)\right] . \label{mhh}
\end{align}
Here, we define the reduced amplitude $\overline{\cal M}$ as 
\begin{align}
{\cal M} = \overline{{\cal M}}\left(\frac{m_u}{v}\right)^2(\bar{u}u)\times (\bar{u}u). 
\end{align}
As shown in Fig.~\ref{gauge-dep2} ($h_i=h$), there are not only the contribution from the self-energy diagram but also vertex corrections, 
box diagrams and wave function renormalizations. 
The important thing is that we can extract the ``self-energy like'' contribution from these diagrams by ``pinching'' the internal fermion propagator.
This procedure can be done by using the loop momentum $k\hspace{-2mm}/$ which comes from the gauge boson propagator and/or the scalar-scalar-gauge type vertex
(after contracting with the Dirac $\gamma^\mu$ matrix).  
Such term extracted from vertex correction diagrams, box diagrams and the wave function renormalization is the so-called pinch-term. 

From the vertex correction diagrams, we can extract the pinch-term for the $\xi_W$ part as
\begin{align}
\Delta_{\xi_W^{}} ({\cal \overline{M}}_{W\text{--}1}+{\cal \overline{M}}_{W\text{--}2})
 &\to \frac{g^2}{16\pi^2}\frac{1}{q^2-m_h^2} \Bigg[ -\left(1+\frac{q^2}{2m_W^2} \right)B_0(q^2;W,W) \notag\\
&\hspace{-15mm}+\left(1 - \xi_W + \frac{q^2}{m_W^2} \right)B_0(q^2;W,G^\pm) +\left(\xi_W-\frac{q^2}{2m_W^2}\right)B_0(q^2;G^\pm,G^\pm) \Bigg], \label{pinch1}\\
\Delta_{\xi_W^{}}({\cal \overline{M}}_{W\text{--}3}+{\cal \overline{M}}_{W\text{--}4})
&\to \frac{g^2}{32\pi^2}\frac{1}{q^2-m_h^2}\Bigg[B_0(q^2;W,W)-\left( 1 - \xi_W +\frac{q^2}{m_W^2}\right)B_0(q^2;W,G^\pm)\notag\\
&-\left(\xi_W-\frac{q^2}{m_W^2}\right)B_0(q^2;G^\pm,G^\pm)  +(1-\xi_W)B_0(0;W,G^\pm) \Bigg], \\
\Delta_{\xi_W^{}}({\cal \overline{M}}_{W\text{--}5}+{\cal \overline{M}}_{W\text{--}6}) &\to \Delta_{\xi_W^{}}({\cal \overline{M}}_{W\text{--}3}+{\cal \overline{M}}_{W\text{--}4}), 
\end{align}
where $\to$ denotes the extraction of the pinched part. 
The total contribution from the vertex correction is expressed by  
\begin{align}
\Delta_{\xi_W^{}} \sum_{i=1,6}{\cal \overline{M}}_{W\text{--}i}
&\to \frac{g^2}{16\pi^2}\frac{1-\xi_W}{q^2-m_h^2} \left[B_0(0;W,G^\pm)-\frac{q^2}{2}C_0(q^2;W,G^\pm) \right].  \label{eq:vv}
\end{align}
The corresponding contribution to $\xi_Z^{}$ is obtained from the diagrams ($Z$--1)--($Z$--6), and its expression is obtained by
replacing $(g,m_W^{},\xi_W;W,G^\pm)$ with $(g_Z/2^{},m_Z^{},\xi_Z;Z,G^0)$ in Eq.~(\ref{eq:vv}). 

The box diagrams give the following pinch-terms:
\begin{align}
&\Delta_{\xi_W^{}}{\cal \overline{M}}_{W\text{--}7} \to \frac{1}{64\pi^2}\frac{g^2}{m_W^2}\left[B_0(q^2;W,W)-2B_0(q^2;W,G^\pm)+B_0(q^2;G^\pm,G^\pm) \right], \\
&\Delta_{\xi_W^{}}({\cal \overline{M}}_{W\text{--}8} + {\cal \overline{M}}_{W\text{--}9})  \to \frac{1}{32\pi^2}\frac{g^2}{m_W^2}\left[B_0(q^2;W,G^\pm)-B_0(q^2;G^\pm,G^\pm) \right]. 
\end{align}
Thus, the total contribution from the box diagrams is expressed  by
\begin{align}
\Delta_{\xi_W^{}} \sum_{i=7,9}{\cal \overline{M}}_{W\text{--}i} \to \frac{g^2}{64\pi^2}(1-\xi_W)C_0(q^2;W,G^\pm). \label{boxx}
\end{align}
The corresponding contribution to $\xi_Z$ is obtained from the diagrams ($Z$--7)--($Z$--11), and 
its expression is given by replacing $(g,m_W^{},\xi_W;W,G^\pm)$ with  $(g_Z/2^{},m_Z^{},\xi_Z;Z,G^0)$ in Eq.~(\ref{boxx}). 

Finally, the contribution from the wave function renormalization ($W$--10) is calculated from the 
fermion two-point function $\Pi_{ff}$. The pinched part of $\Delta_\xi \Pi_{ff}$, which comes from $W^\mu$, $Z^\mu$ and $\gamma^\mu$ 
loop diagrams, is expressed by 
\begin{align}
\Delta_\xi \Pi_{ff}(p\hspace{-1.8mm}/) &\to -\frac{g^2}{32\pi^2}(1-\xi_W^{})p\hspace{-1.8mm}/P_LB_0(0;W,G^\pm)\notag\\
&-\frac{g_Z^2}{16\pi^2}(1-\xi_Z)(v_f+a_f\gamma_5)(p\hspace{-1.8mm}/-m_f)(v_f-a_f\gamma_5)B_0(0;Z,G^0) \notag\\ 
&-\frac{e^2}{16\pi^2}Q_f^2(1-\xi_\gamma)(p\hspace{-1.8mm}/-m_f)B_0(0;\gamma,\gamma), \label{piff}
\end{align}
where $v_f = (I_f-2\sin^2\theta_WQ_f)/2$ and $a_f=I_f/2$ with $Q_f$ being the electric charge of a fermion $f$. 
The wave function renormalization factor $\delta Z_f$ for a fermion $f$  is then obtained by 
\begin{align}
\delta Z_f = -\frac{d}{dp\hspace{-2mm}/}\Pi_{ff}(p\hspace{-1.8mm}/). 
\end{align}
Thus, the contribution from ($W$--10) is calculated as 
\begin{align}
\Delta_{\xi_W^{}} \overline{{\cal M}}_{W\text{--}10} = 4\times \left(-\frac{1}{q^2-m_h^2}\right)\times \left(\Delta_{\xi_W^{}}\frac{\delta Z_f}{2}\right) 
\to -\frac{g^2}{32\pi^2}\frac{1-\xi_W}{q^2-m_h^2}B_0(0;W,G^\pm).  \label{eq:ww}
\end{align}
The corresponding contribution to $\xi_Z$ is obtained from the diagrams ($Z$--12)--($Z$--14), and again 
its expression is given by
replacing $(g,m_W^{},\xi_W;W,G^\pm)$ with $(g_Z/2^{},m_Z^{},\xi_Z;Z,G^0)$ in Eq.~(\ref{eq:ww}). 
We note that the $v_f$ part in Eq.~(\ref{piff}) is cancelled by the diagrams ($Z$--12) and ($Z$--13). 
In addition, the $\xi_\gamma$ dependence in Eq.~(\ref{piff}) is also canceled by the diagrams ($Z$--12) and ($Z$--13) with the replacement of $Z \to \gamma$. 
By adding Eqs.~(\ref{eq:vv}) and (\ref{eq:ww}), we obtain the following expression:
\begin{align}
\Delta_{\xi_W^{}} \left(\sum_{i=1,6}{\cal \overline{M}}_{W\text{--}i} + {\cal \overline{M}}_{W-10} \right)
\to \frac{g^2}{32\pi^2}\frac{1-\xi_W}{q^2-m_h^2}X_V(q^2;W,0),   
\end{align}
where the function $X_V$ is defined as
\begin{align}
&X_V(q^2;V,\phi) \equiv B_0(0;V,G_V)-(q^2-m_\phi^2)C_0(q^2;V,G_V,\phi),\notag\\
&\text{with}~~C_0(q^2;W,G_V,0) \equiv  C_0(q^2;W,G_V). \label{xvv}
\end{align}
In Eq.~(\ref{xvv}), $V$ and $G_V$ being a gauge boson and its associated NG boson, respectively. 

Consequently, the total contributions to the pinch-term ($\Delta_\xi \overline{{\cal M}}_{\text{PT}}$) is given by
\begin{align}
\Delta_\xi \overline{{\cal M}}_{\text{PT}} & = 
\frac{g^2}{32\pi^2} \frac{1-\xi_W}{q^2-m_h^2}
\left[B_0(0;W,G^\pm) -\frac{q^2+m_h^2}{2}C_0(q^2;W,G^\pm)  \right]\notag\\
&+\left[(g,m_W^{},\xi_W;W,G^\pm) \to (g_Z/2^{},m_Z^{},\xi_Z;Z,G^0)\right], 
\end{align}
which exactly cancels Eq.~(\ref{mhh}), i.e., $\Delta_\xi(\overline{{\cal M}}_{hh} + \overline{{\cal M}}_{\text{PT}}) = 0$. 
This means that the Higgs boson two-point function calculated with a fixed gauge parameter 
becomes gauge independent by adding the pinch-term calculated with the same fixed gauge parameter. 
In App.~\ref{sec:fg}, we present the expression of the pinch-term calculated in the 't~Hooft-Feynman gauge, in which 
the diagrams ($W$--3)--($W$--6) and ($Z$--3)--($Z$--6) give the non-zero contribution.  

\subsection{HSM}

We discuss the cancelation of the gauge dependence in two-point functions for CP-even Higgs bosons in the HSM. 
We here discuss only the $\xi_W$ dependence, since the $\xi_Z$ dependence are obtained by the simple replacement of $(g,m_W^{},\xi_W;W,G^\pm)$ with $(g_Z/2^{},m_Z^{},\xi_Z;Z,G^0)$
as we have seen it in the previous subsection. 
Similar to the SM, the diagrams which give the gauge dependence in the two-point functions of the CP-even Higgs bosons are shown in Fig.~\ref{h2}, where 
$h_i$ and $h_j$ can be either $h$ or $H$. 
The gauge dependent part of the self-energy type diagrams in the $u\bar{u}\to u\bar{u}$ process $(\Delta_{\xi_W^{}}\overline{{\cal M}}_{h_ih_j})$
is calculated by
\begin{align}
\Delta_{\xi_W^{}} \overline{{\cal M}}_{h_ih_j}
&\to \frac{g^2}{64\pi^2}\frac{(1-\xi_W)\zeta_i^2 \zeta_j^2}{(q^2-m_{h_i}^2)(q^2-m_{h_j}^2)}  \notag\\
&\times \Big[ (q^4-m_{h_i}^2m_{h_j}^2)C_0(p^2;W,G^\pm)-\left(2q^2-m_{h_i}^2 - m_{h_j}^2 \right)B_0(0,W,G^\pm)  \Big], 
\end{align}
where $i,j=1,2$ with 
\begin{align}
(h_1,\zeta_1)=(h,c_\alpha)~~\text{and}~~(h_2,\zeta_2)=(H,s_\alpha). \label{hsm_short}
\end{align}

The pinch-term can be extracted from the diagram shown in Fig.~\ref{gauge-dep2}, where $h_i=h$ or $H$. 
Similar to the case in the SM, each diagram gives the following pinch-term:
\begin{align}
\Delta_{\xi_W^{}}\left(\sum_{i=1,6} {\cal \overline{M}}_{W\text{--}i} 
+ {\cal \overline{M}}_{W\text{--}10}\right)
&\to  \frac{g^2}{32\pi^2}(1-\xi_W) X_V(q^2;W,0)\left(\frac{s_\alpha^2}{q^2-m_H^2}+\frac{c_\alpha^2}{q^2-m_h^2} \right), 
\notag\\
\Delta_{\xi_W^{}} \sum_{i=7,9} {\cal \overline{M}}_{W\text{--}i} &\to \frac{g^2}{64\pi^2}(1-\xi_W)C_0(q^2;W,G^\pm). 
\end{align}
The total pinch-term is then expressed by 
\begin{align}
\Delta_{\xi_W^{}} {\cal \overline{M}}_{\text{PT}} 
&=\frac{g^2}{32\pi^2}(1-\xi_W)\Bigg\{
\frac{c_\alpha^2}{q^2-m_h^2}\Big[B_0(0;W,G^\pm)- \frac{q^2+m_h^2}{2}C_0(q^2;W,G^\pm)   \Big]\notag\\
&\quad\quad\quad\quad +\frac{s_\alpha^2}{q^2-m_H^2}\Big[B_0(0;W,G^\pm)- \frac{q^2+m_H^2}{2}C_0(q^2;W,G^\pm) \Big] \Bigg\}. 
\end{align}
We can correctly share the above pinch-term by splitting the trigonometric functions as $c_\alpha^2 = c_\alpha^4 + c_\alpha^2s_\alpha^2$ and 
$s_\alpha^2 = s_\alpha^4 + c_\alpha^2s_\alpha^2$. 
Namely, the $c_\alpha^4$, $s_\alpha^4$ and $s_\alpha^2 c_\alpha^2$ parts exactly cancel 
$\Delta_{\xi_W^{}} \overline{{\cal M}}_{hh}$, $\Delta_{\xi_W^{}} \overline{{\cal M}}_{HH}$ and $\Delta_{\xi_W^{}} (\overline{{\cal M}}_{Hh} + \overline{{\cal M}}_{hH})$, respectively. 
After adding the $\Delta_{\xi_Z^{}}$ part, we can confirm
\begin{align}
& \Delta_\xi[\overline{{\cal M}}_{hh} + \overline{{\cal M}}_{HH} +\overline{{\cal M}}_{Hh} + \overline{{\cal M}}_{hH}+  {\cal \overline{M}}_{\text{PT}}] =0. 
\end{align}
In App.~\ref{sec:fg}, we give the expression of the pinch-term for the two-point functions of 
$h$--$h$, $H$--$H$ and $H$--$h$  in the 't~Hooft-Feynman gauge. 

\subsection{THDM}

We discuss the cancelation of the gauge dependence not only in the two-point function for  
the CP-even Higgs bosons, but also that for the CP-odd and the singly-charged scalar bosons. 
For the CP-odd (charged) scalar sector, 
we show the cancelation in the two-point function of $A$--$A$ and $A$--$G^0$ ($H^\pm$--$H^\pm$ and $H^\pm$--$G^\pm$). 
The cancellation for the NG boson two-point functions $G^0$--$G^0$ and $G^\pm$--$G^\pm$ has been 
discussed in Ref.~\cite{Papavassiliou:1994pr}, so that we do not deal with these two-point functions in this paper. 

\subsubsection{CP-even sector}

\begin{figure}[!t]
\begin{center}
\includegraphics[width=150mm]{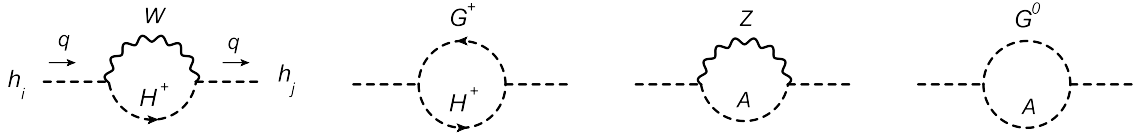}
\caption{Additional diagrams giving the $\xi_W^{}$ and $\xi_Z^{}$ dependence in 
the two-point function for the CP-even Higgs bosons $h_i$ and $h_j$ ($=h$ or $H$) in the THDM. 
}
\label{fig2}
\end{center}
\end{figure}

The contribution to the $u\bar{u}\to u\bar{u}$ process from the self-energy type diagram is calculated by the similar way to the case in the HSM. 
However, we need to add new contributions shown in Fig.~\ref{fig2} in addition to the diagrams shown in Fig.~\ref{h2}
with $h_i$ and $h_j$ being $h$ or $H$, 
in which the physical charged Higgs boson $H^\pm$ or the CP-odd Higgs boson $A$ is running in the loop. 
Again, we only show the $\xi_W$ dependent part since the $\xi_Z$ dependent part are obtained by the replacement of the 
$(g,m_W^{},\xi_W;W,G^\pm,H^\pm)$ part with $(g_Z/2^{},m_Z^{},\xi_Z;Z,G^0,A)$.
Taking into account these new contributions, the $\xi_W^{}$ dependence of the contribution to the $u\bar{u}\to u\bar{u}$ process 
from the self-energy type diagrams is calculated as follows:
\begin{align}
\Delta_{\xi_W^{}} \overline{{\cal M}}_{hh}
&=\frac{g^2}{64\pi^2} \frac{\zeta_{huu}^2}{q^2-m_h^2}(1-\xi_W) \Big\{ s_{\beta-\alpha}^2(q^2+m_h^2)C_0(q^2;W,G^\pm)-2B_0(0;W,G^\pm) \notag\\
&+ 2c_{\beta-\alpha}^2(q^2+m_h^2-2m_{H^\pm}^2)C_0(q^2;W,G^\pm,H^\pm)\Big\}, \label{41} \\
\Delta_{\xi_W^{}} \overline{{\cal M}}_{HH}
&=\frac{g^2}{64\pi^2} \frac{\zeta_{Huu}^2}{q^2-m_H^2}(1-\xi_W) \Big\{ c_{\beta-\alpha}^2(q^2+m_H^2)C_0(q^2;W,G^\pm)-2B_0(0;W,G^\pm) \notag\\
&+ 2s_{\beta-\alpha}^2(q^2+m_H^2-2m_{H^\pm}^2)C_0(q^2;W,G^\pm,H^\pm)\Big\}, \label{42} \\
\Delta_{\xi_W^{}} \overline{{\cal M}}_{Hh}
&=\frac{g^2}{64\pi^2}
 \frac{\zeta_{huu}\zeta_{Huu} s_{\beta-\alpha}c_{\beta-\alpha}}{(q^2-m_h^2)(q^2-m_H^2)}
  (1-\xi_W)\Big\{(q^4-m_h^2m_H^2)C_0(q^2;W,G^\pm)\notag\\
&- 2 [(q^2-m_{H^\pm}^2)^2-(m_{H^\pm}^2-m_h^2)(m_{H^\pm}^2-m_H^2)]C_0(q^2;H^\pm,W,G^\pm)\Big\}, 
\end{align}
where $ \Delta_{\xi_W^{}} \overline{{\cal M}}_{hH}= \Delta_{\xi_W^{}} \overline{{\cal M}}_{Hh}$, and 
$\zeta_{huu}$ and $\zeta_{Huu}$ are given in Eq.~(\ref{zeta-hff}). 

The pinch-terms can be extracted from the diagram shown in Fig.~\ref{gauge-dep2} ($h_i=h$ or $H$) with the additional diagrams which are obtained by 
the replacement $G^\pm \to H^\pm$. 
Thus, each diagram involving $G^\pm$, i.e., (W--3)--(W--6) and (W--8)--(W--9) should be understood as the sum of $G^\pm$ and $H^\pm$ loop contributions. 
We then obtain the following pinch-term contributions:
\begin{align}
\Delta_{\xi_W^{}} \sum_{i=1,6}{\cal \overline{M}}_{W\text{-}i} &\to  \frac{g^2}{16\pi^2}(1-\xi_W) \left[ B_0(0;W,G^\pm) - \frac{q^2}{2}C_0(q^2;W,G^\pm) \right]
\left(\frac{s_{\beta-\alpha}\zeta_{huu}}{q^2-m_h^2}+\frac{c_{\beta-\alpha}\zeta_{Huu}}{q^2-m_H^2} \right) \notag\\
&+\frac{g^2\zeta_u}{16\pi^2}(1-\xi_W^{}) X_V(q^2;W,H^\pm) 
\left(\frac{c_{\beta-\alpha}\zeta_{huu}}{q^2-m_h^2}-\frac{s_{\beta-\alpha}\zeta_{Huu}}{q^2-m_H^2} \right), \label{12}\\
\Delta_{\xi_W^{}} \sum_{i=7,9}{\cal \overline{M}}_{W\text{-}i} 
&\to \frac{g^2}{64\pi^2}(1-\xi_W)C_0(q^2;W,G^\pm)+\frac{g^2\zeta_u^2}{32\pi^2}(1-\xi_W)C_0(q^2;W,G^\pm,H^\pm), \label{13}\\
\Delta_{\xi_W^{}} {\cal \overline{M}}_{W\text{-}10} &\to-\frac{g^2}{32\pi^2}(1-\xi_W)\left(\frac{\zeta_{huu}^2}{q^2-m_h^2}+\frac{\zeta_{Huu}^2}{q^2-m_H^2} \right) B_0(0;W,G^\pm), 
\end{align}
where the second term of the right-hand side (RHS) in Eqs.~(\ref{12}) and (\ref{13}) is the contribution from the charged Higgs boson loop. 
The total pinch-term is then expressed by 
\begin{align}
\Delta_{\xi_W^{}} {\cal \overline{M}}_{\text{PT}}
&= \frac{g^2}{64\pi^2}\frac{1-\xi_W}{q^2-m_h^2}\Big[2\zeta_{huu}^2B_0(0;W,G^\pm) - (q^2+m_h^2)s_{\beta-\alpha}\zeta_{huu}C_0(q^2;W,G^\pm)\notag\\
&\quad\quad\quad\quad\quad\quad\quad\quad-2(q^2+m_h^2-2m_{H^\pm}^2)c_{\beta-\alpha}\zeta_u\zeta_{huu} C_0(q^2;W,G^\pm,H^\pm)  \Big]\notag\\
&+ \frac{g^2}{64\pi^2}\frac{1-\xi_W}{q^2-m_H^2}\Big[2\zeta_{Huu}^2B_0(0;W,G^\pm) - (q^2+m_H^2)c_{\beta-\alpha}\zeta_{Huu}C_0(q^2;W,G^\pm)\notag\\
&\quad\quad\quad\quad\quad\quad\quad\quad+2(q^2+m_H^2-2m_{H^\pm}^2)s_{\beta-\alpha}\zeta_u \zeta_{Huu}C_0(q^2;W,G^\pm,H^\pm)  \Big]. \label{ptt}
\end{align}
The following sum rule is useful to obtain the above expression:
\begin{align}
&s_{\beta-\alpha}\zeta_{huu} + c_{\beta-\alpha}\zeta_{Huu}= 1,\quad
c_{\beta-\alpha}\zeta_{huu} - s_{\beta-\alpha}\zeta_{Huu}= \zeta_u. \label{relations}
\end{align}
In Eq.~(\ref{ptt}), we can correctly split this expression into the pinch-term for $h$--$h$, $H$--$H$ and $H$--$h$ by the following way. 
First, we rewrite $s_{\beta-\alpha}\zeta_{huu} = s_{\beta-\alpha}^2\zeta_{huu}^2 + s_{\beta-\alpha}c_{\beta-\alpha}\zeta_{huu}\zeta_{Huu}$ 
and $c_{\beta-\alpha}\zeta_u\zeta_{huu} = c_{\beta-\alpha}^2\zeta_{huu}^2 - s_{\beta-\alpha}c_{\beta-\alpha}\zeta_{huu}\zeta_{Huu}$
in the first term of the RHS of Eq.~(\ref{ptt}).  
Second, we rewrite $c_{\beta-\alpha}\zeta_{Huu} = c_{\beta-\alpha}^2\zeta_{Huu}^2 + s_{\beta-\alpha}c_{\beta-\alpha}\zeta_{huu}\zeta_{Huu}$ 
and $s_{\beta-\alpha}\zeta_u\zeta_{Huu} = -\zeta_{Huu}^2s_{\beta-\alpha}^2 + \zeta_{huu}\zeta_{Huu}s_{\beta-\alpha}c_{\beta-\alpha}$
in the second term of the RHS of Eq.~(\ref{ptt}). 
After that, Eq.~(\ref{ptt}) is written by the terms proportional to $\zeta_{huu}^2$, $\zeta_{Huu}^2$ and $\zeta_{huu}\zeta_{Huu}$, and 
each of them respectively gives the pinch-term for the two-point functions of $h$--$h$, $H$--$H$ and $H$--$h$. 
By adding the $\Delta_{\xi_Z}$ part, we can confirm the cancellation of the gauge dependence:
\begin{align}
\Delta_\xi (\overline{{\cal M}}_{hh}+\overline{{\cal M}}_{HH}+\overline{{\cal M}}_{Hh}+\overline{{\cal M}}_{hH}+\overline{{\cal M}}_{\text{PT}})= 0. 
\end{align}

\subsubsection{CP-odd sector}

 \begin{figure}[!t]
 \begin{center}
 \includegraphics[width=150mm]{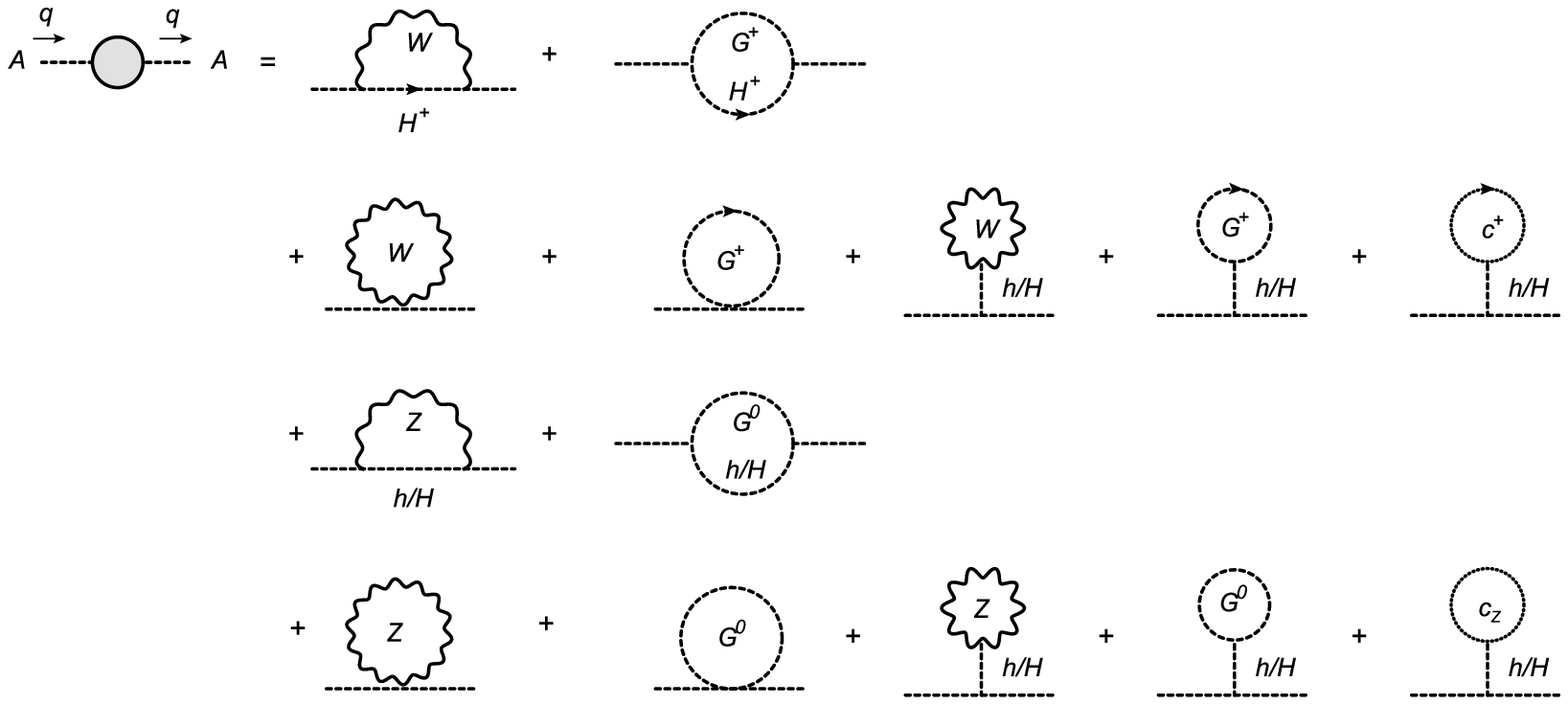}
 \caption{Gauge dependent part of the Feynman diagrams for the two-point function of $A$.  }
 \label{aaloop}
 \end{center}
 \begin{center}
 \includegraphics[width=130mm]{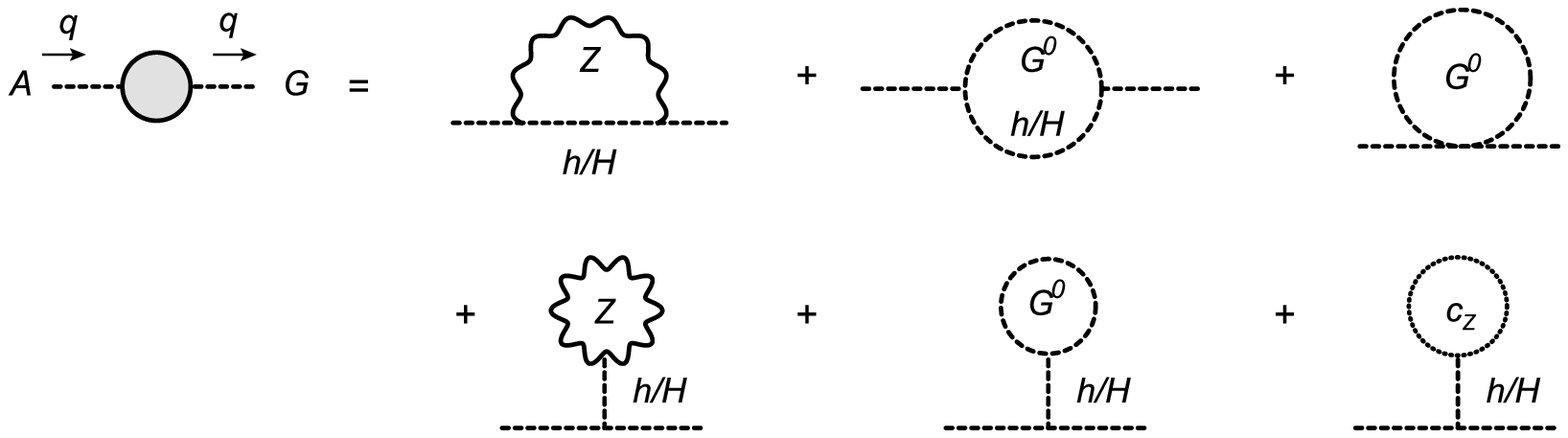}
 \caption{Gauge dependent part of the Feynman diagrams for the $A$--$G^0$ mixing.}
  \label{agloop}
 \end{center}
\end{figure}
 \begin{figure}[!t]
\begin{center}
\includegraphics[width=160mm]{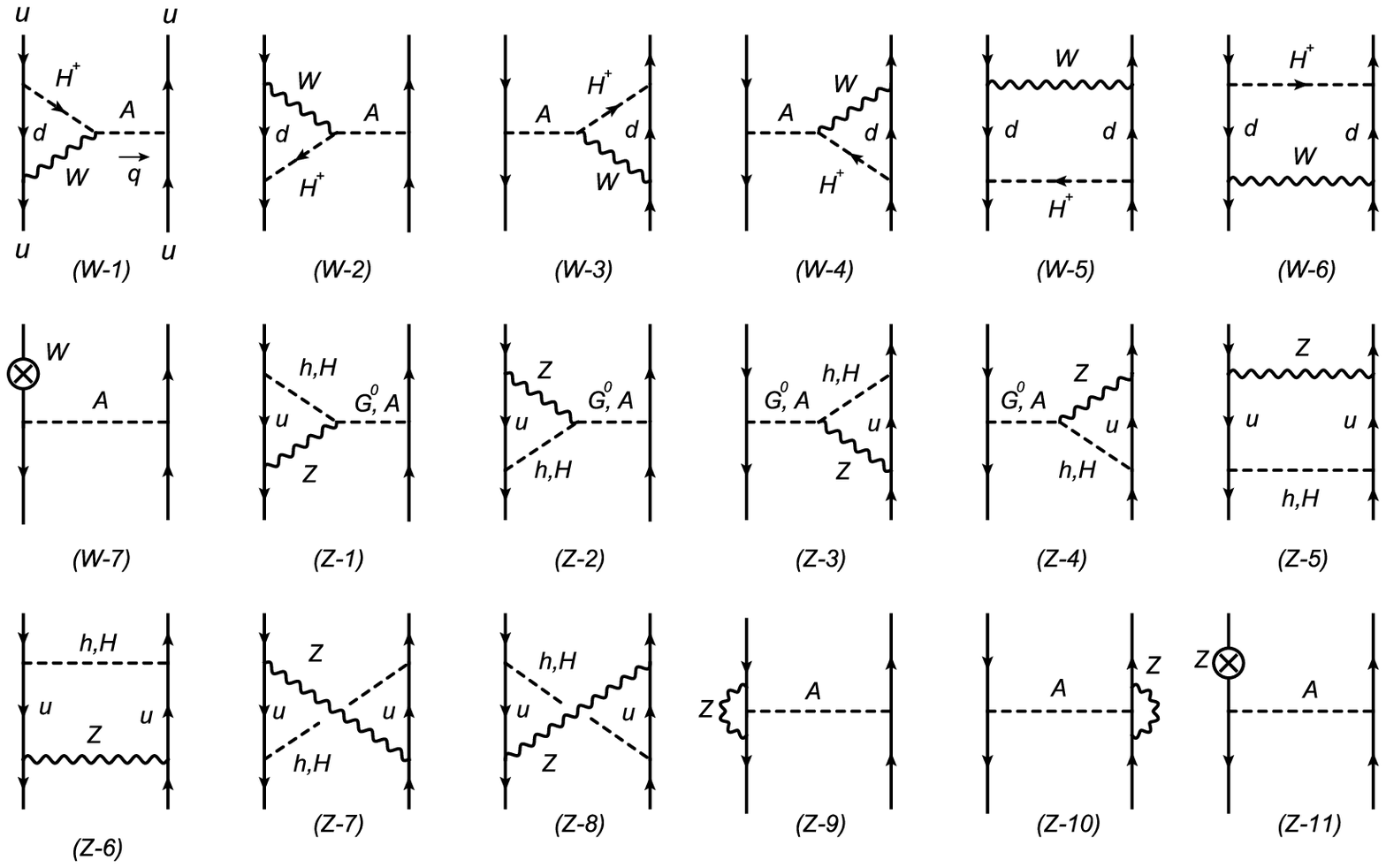}
\caption{Feynman diagrams giving pinch-terms for the two-point functions of $A$--$A$ and $A$--$G^0$ in the $u\bar{u}\to u\bar{u}$ scattering. 
The diagrams $(W$-$7)$ and $(Z$-$11)$ denote the contribution to the $\xi_W^{}$ and $\xi_Z^{}$ dependence from the wave function renormalization of the external quark, respectively.
}
\label{gauge-dep4}
\end{center}
\end{figure}

Next, we see the cancellation of the gauge dependence in the two-point functions for the CP-odd scalar bosons $A$--$A$ and $A$--$G^0$, where  
the relevant Feynman diagrams are shown in Figs.~\ref{aaloop} and \ref{agloop}, respectively. 
We note that in the $A$--$G^0$ mixing, the $\xi_W$ dependence appears from tadpole diagrams and a seagull diagram with the $G^\pm$ loop, 
but these contributions are exactly cancelled with each other. 
As a result, only the $\xi_Z^{}$ dependence remains. 
The contribution from the self-energy type diagrams to the $u\bar{u}\to u\bar{u}$ scattering is expressed by 
\begin{align}
\Delta_\xi \overline{{\cal M}}_{AA}&= 
\frac{g^2}{32\pi^2}\frac{1-\xi_W}{q^2-m_A^2}\zeta_u^2
\Big[(q^2+m_A^2-2m_{H^\pm}^2)C_0(q^2;W,G^\pm,H^\pm)-B_0(0;W,G^\pm)\Big] \notag\\
&+\frac{g_Z^2}{64\pi^2}\frac{1-\xi_Z}{q^2-m_A^2}\zeta_u^2
\Big[c_{\beta-\alpha}^2(q^2+m_A^2-2m_h^2)C_0(q^2;Z,G^0,h)\notag\\
&\quad\quad\quad+s_{\beta-\alpha}^2(q^2+m_A^2-2m_H^2)C_0(q^2;Z,G^0,H)-B_0(0;G^0,Z)\Big], \label{AA}\\
\Delta_\xi \overline{{\cal M}}_{AG^0} &= \frac{g_Z^2}{64\pi^2}
\frac{(1-\xi_Z)s_{\beta-\alpha}c_{\beta-\alpha}\zeta_u}{(q^2-m_{G^0}^2)(q^2-m_A^2)}
\Big\{\left[q^2(q^2-2m_h^2)+m_h^2m_A^2\right]C_0(q^2;Z,G^0,h)
\notag\\
&\quad\quad\quad
-\left[q^2(q^2-2m_H^2)+m_H^2m_A^2\right]C_0(q^2;Z,G^0,H)\Big\},  
\end{align}
where $m_{G^0}^2 = \xi_Z^{} m_Z^2$. 
In this subsubsection, the reduced amplitude $\overline{\cal M}$ is defined by 
\begin{align}
{\cal M} = -\overline{{\cal M}}\left(\frac{m_u}{v}\right)^2(\bar{u}\,\gamma_5\,u)\times (\bar{u}\,\gamma_5\,u). 
\end{align}

The pinch-terms are extracted from the diagrams shown in Fig.~\ref{gauge-dep4}. 
We obtain 
\begin{align}
&\Delta_\xi \sum_{i=1,4}(\overline{{\cal M}}_{W\text{-}i} + \overline{{\cal M}}_{Z\text{-}i} )
\to  \frac{g^2}{16\pi^2}\frac{1-\xi_W}{q^2-m_A^2}\zeta_u^2\, X_V(q^2;W,H^\pm)\notag\\
&\hspace{2cm}+\frac{g_Z^2}{32\pi^2}\frac{1-\xi_Z}{q^2-m_{G^0}^2}\left[s_{\beta-\alpha}\zeta_{huu}X_V(q^2;Z,h)+c_{\beta-\alpha}\zeta_{Huu}X_V(q^2;Z,H) \right]\notag\\
&\hspace{2cm} + \frac{g_Z^2}{32\pi^2}\frac{1-\xi_Z}{q^2-m_A^2}\zeta_u\left[c_{\beta-\alpha}\zeta_{huu}X_V(q^2;Z,h)-s_{\beta-\alpha}\zeta_{Huu}X_V(q^2;Z,H) \right],\\
&\Delta_\xi \left(\sum_{i=5,6}\overline{{\cal M}}_{W\text{-}i} + \sum_{i=5,8}\overline{{\cal M}}_{Z\text{-}i} \right)
\to\frac{g^2}{32\pi^2}(1-\xi_W)\zeta_u^2C_0(q^2;W,G^\pm,H^\pm)  \notag\\
&\hspace{2cm}+ \frac{g_Z^2}{64\pi^2}(1-\xi_Z)\left[\zeta_{huu}^2C_0(q^2;Z,G^0,h)+\zeta_{Huu}^2C_0(q^2;Z,G^0,H)\right],  \\
&\Delta_\xi \left(\overline{{\cal M}}_{W\text{-}7} +\sum_{i=9,11} \overline{{\cal M}}_{Z\text{-}i} \right)\notag\\
&\hspace{2cm} \to-\frac{1}{32\pi^2}\frac{\zeta_u^2}{q^2-m_A^2}\left[g^2(1-\xi_W)B_0(0;W,G^\pm)+\frac{g_Z^2}{2}(1-\xi_Z)B_0(0;Z,G^0) \right]. 
\end{align}
The total pinch-term $\Delta_\xi\overline{{\cal M}}_{\text{PT}}$
can be classified by the power of the $\zeta_u$ factor, i.e., $\zeta_u^2$, $\zeta_u^1$ and $\zeta_u^0$, where 
the terms with $\zeta_u^2$ and $\zeta_u^1$ denoting
$\Delta_\xi\overline{{\cal M}}_{\text{PT}}^{AA}$ and $\Delta_\xi\overline{{\cal M}}_{\text{PT}}^{AG^0}$
respectively give the pinch-terms for $A$--$A$ and $A$--$G^0$. 
These are expressed as 
\begin{align}
\Delta_\xi\overline{{\cal M}}_{\text{PT}}^{AA} &= -\Delta_\xi \overline{{\cal M}}_{AA}, \\
\Delta_\xi\overline{{\cal M}}_{\text{PT}}^{AG^0} &=
\frac{g_Z^2}{32\pi^2}(1-\xi_Z)s_{\beta-\alpha}c_{\beta-\alpha}\zeta_u\Big\{C_0(q^2;Z,G^0,h)-C_0(q^2;Z,G^0,H)\notag\\
&+\left(\frac{1}{q^2-m_{G^0}^2}+\frac{1}{q^2-m_A^2}\right)\Big[X_V(q^2;Z,h)-X_V(q^2;Z,H) \Big] \Big\}. \label{pt-ag}
\end{align}
For $\Delta_\xi\overline{{\cal M}}_{\text{PT}}^{AG^0}$, this pinch-term is used not only to cancel $\Delta_\xi \overline{{\cal M}}_{AG^0}$ but 
also the gauge dependence of the $A$--$Z$ mixing. 
In order to correctly share the pinch-term of $\Delta_\xi\overline{{\cal M}}_{\text{PT}}^{AG^0}$, we use the following identity:
\begin{align}
\Lambda_{G^0} = \frac{q^2\Lambda_{G^0}}{q^2-m_{G^0}^2} - im_Z^{} \Lambda_Z^\mu (\Delta_{Z})_{\mu\nu} \,q^\nu, \label{sep}
\end{align}
where  $\Lambda_{G^0}$ and $\Lambda_Z^\mu$ are the $\bar{u}uG^0$ and $\bar{u}uZ^\mu$ vertices, respectively, expressed as 
\begin{align}
\Lambda_{G^0}  = -\frac{m_u}{v}\bar{u}\gamma_5 u, \quad
\Lambda_Z^\mu &= ig_Z^{}\bar{u}\gamma^\mu (v_u - a_u\gamma_5) u. 
\end{align}
In Eq.~(\ref{sep}), the first term of the RHS can be used for the  pinch-term of the $A$--$G^0$ mixing. 
Using this identity, we can construct the correct pinch-term for the $A$--$G^0$ mixing from Eq.~(\ref{pt-ag}) as 
\begin{align}
\left(\frac{1}{q^2-m_{G^0}^2} + \frac{1}{q^2-m_{A}^2}\right)\Lambda_{G^0} &= 
\left(\frac{1}{q^2-m_{G^0}^2}\frac{q^2-m_A^2}{q^2-m_{A}^2} + \frac{1}{q^2-m_{A}^2}\frac{q^2}{q^2-m_{G^0}^2}\right)\Lambda_{G^0} + \cdots \notag\\
& = \frac{2q^2 - m_A^2}{(q^2-m_{G^0}^2)(q^2-m_{A}^2)}\Lambda_{G^0}  + \cdots, \label{sep10}
\end{align}
where the $\cdots$ part comes from the second term in Eq.~(\ref{sep}). 
We can confirm that after replacing the factor $[(q^2-m_{G^0}^2)^{-1} + (q^2-m_{A}^2)^{-1}]$ with $(2q^2 - m_A^2)[(q^2-m_{G^0}^2)(q^2-m_{A}^2)]^{-1}$ in Eq.~(\ref{pt-ag}), 
we obtain $\Delta_\xi\overline{{\cal M}}_{\text{PT}}^{AG^0} = -\Delta_\xi \overline{{\cal M}}_{AG^0}$. 

\subsubsection{Charged sector}

 \begin{figure}[!t]
 \begin{center}
 \includegraphics[width=150mm]{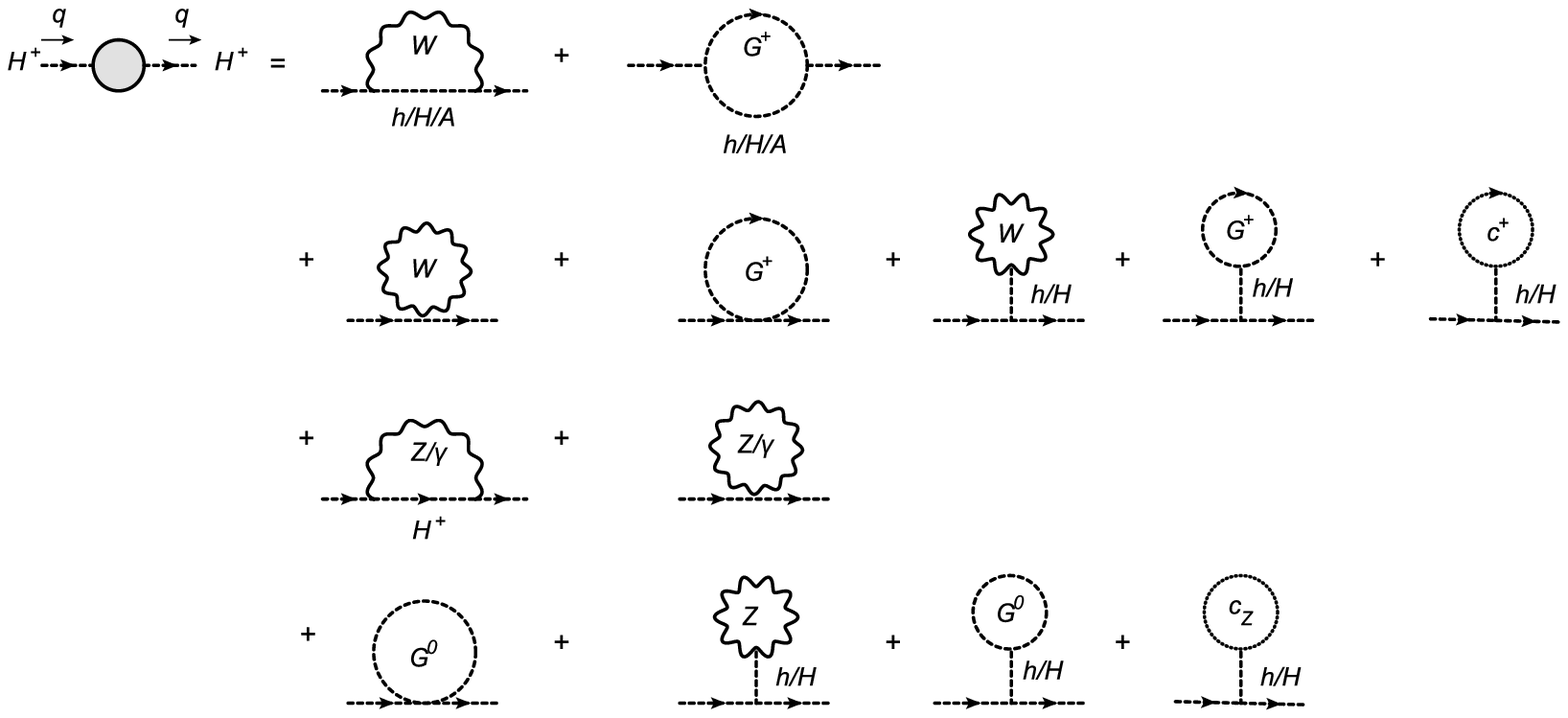}
 \caption{Gauge dependent part of the Feynman diagrams for the two-point function of $H^\pm$. }
 \label{hphm}
 \end{center}
 \begin{center}
 \includegraphics[width=140mm]{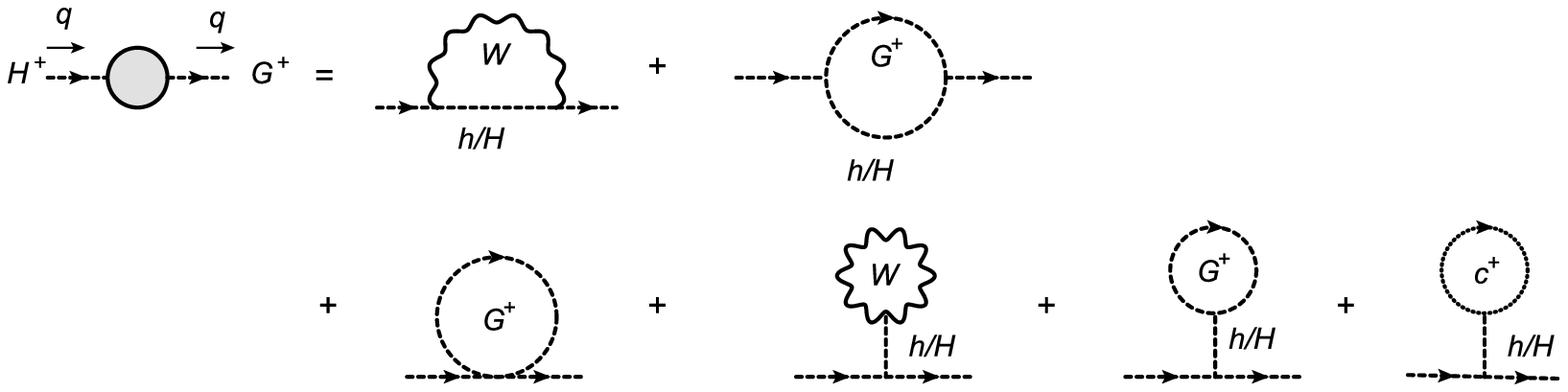}
 \caption{Gauge dependent part of the Feynman diagrams for the $H^\pm$--$G^\pm$ mixing. }
 \label{hpmix}
 \end{center}
\end{figure}
 \begin{figure}[!t]
\includegraphics[width=160mm]{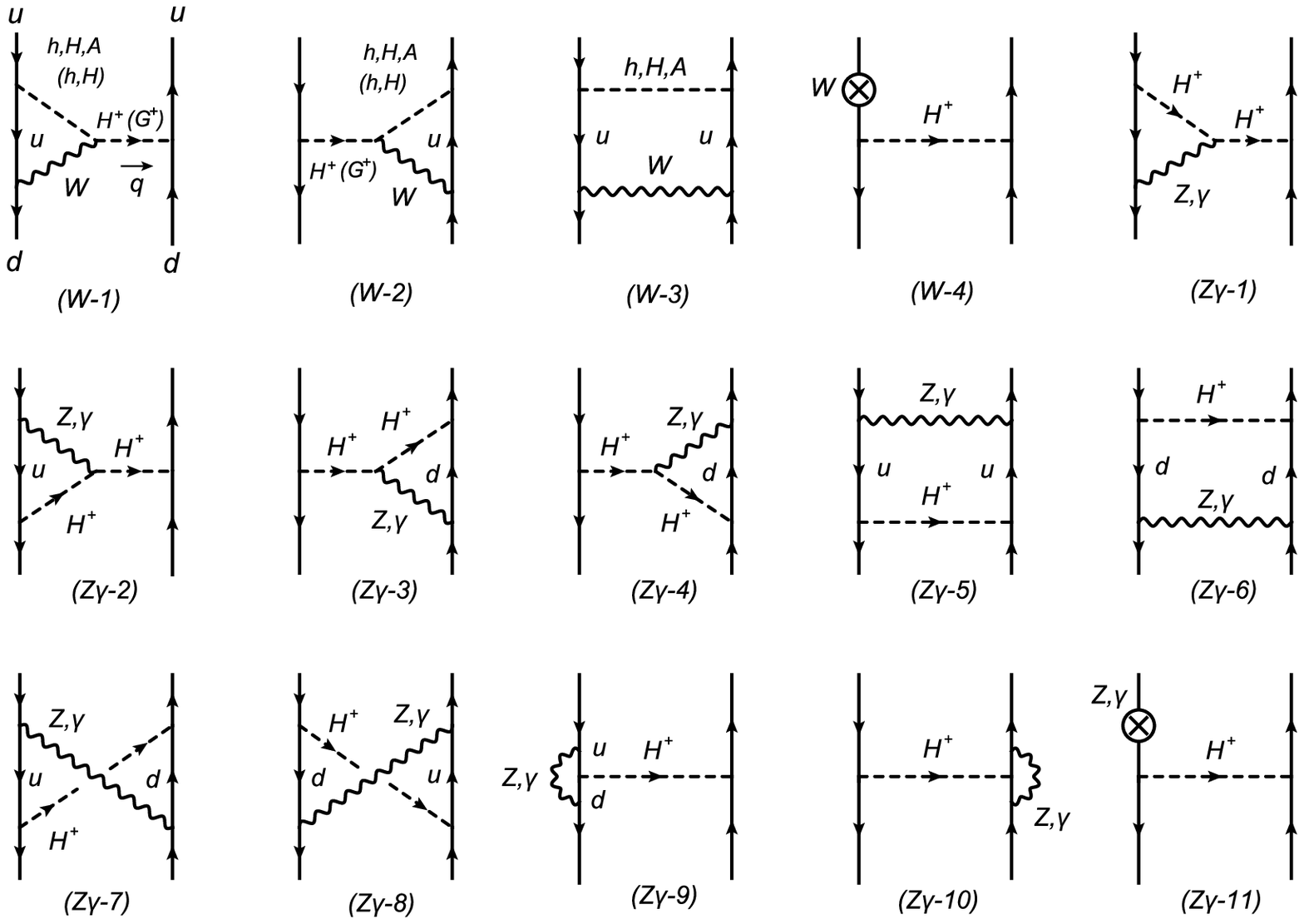}
\caption{Feynman diagrams giving pinch-terms for charged scalar two-point functions in the $u\bar{d}\to u\bar{d}$ scattering. 
The diagrams $(W$-$4)$ and $(Z\gamma$-$11)$ denote the contribution to the $\xi_W$ and $\xi_{Z,\gamma}$ dependence 
from the wave function renormalization of the external quark, respectively.}
\label{pinch-hp}
\end{figure}

The Feynman diagrams which provide gauge dependence in the two-point functions $H^\pm$--$H^\pm$ and $H^\pm$--$G^\pm$ are shown in 
Figs.~\ref{hphm} and \ref{hpmix}, respectively. 
We note that for the $H^\pm$--$G^\pm$ mixing, the $\xi_Z^{}$ dependence appears from tadpole diagrams and a seagull diagram with the $G^0$ loop, but 
these contributions are exactly cancelled with each other. As a result, only the $\xi_W^{}$ dependence remains. 

For the charged Higgs sector, we consider the $u\bar{d} \to u\bar{d}$ process instead of $u\bar{u} \to u\bar{u}$ process. 
The self-energy type diagram contributions to the $u\bar{d} \to u\bar{d}$ process are calculated~as
\begin{align}
\Delta_{\xi}\overline{{\cal M}}_{H^+H^-}&=\frac{g^2}{64\pi^2}\frac{1-\xi_W}{q^2-m^2_{H^\pm}}\Big[
(q^2+m_{H^\pm}^2-2m_A^2)C_0(q^2;W,G^\pm,A) \notag \\
&+c^2_{\beta-\alpha}(q^2+m_{H^\pm}^2-2m_h^2)C_0(q^2;W,G^\pm,h) \notag\\ 
&+s^2_{\beta-\alpha}(q^2+m_{H^\pm}^2-2m_H^2)C_0(q^2;W,G^\pm,H)
-2B_0(0,W,G^\pm) \Big] \notag \\
&-\frac{g_Z^2c^2_{2W}}{64\pi^2}\frac{1-\xi_Z}{q^2-m^2_{H^\pm}}X_V(q^2;Z,H^\pm)   \notag\\
&-\frac{e^2}{16\pi^2}\frac{1-\xi_\gamma}{q^2-m^2_{H^\pm}}\Big[B_0(0;\gamma,\gamma)-(q^2-m^2_{H^\pm})C_0(0,q^2,q^2;\gamma,\gamma,H^\pm)\Big], \label{mhphm}\\
\Delta_{\xi}\overline{{\cal M}}_{H^+G^-}&=\frac{g^2}{64\pi^2}\frac{1-\xi_W}{(q^2-m_{H^\pm}^2)(q^2-m_{G^\pm}^2)}s_{\beta-\alpha}c_{\beta-\alpha}\notag\\
&\Big[ \big(q^4-(2q^2-m_{H^\pm}^2)m_h^2\big)C_0(q^2;W,G^\pm,h) \notag\\
&+\big(q^4-(2q^2-m_{H^\pm}^2)m_H^2\big)C_0(q^2;W,G^\pm,H) 
\Big],  \label{mhpgm}
\end{align}  
where $m_{G^\pm}^2 = \xi_W^{} m_W^2$. 
In this subsubsection, the reduced amplitude $\overline{\cal M}$ is defined by 
\begin{align}
\mathcal{M} =\overline{\cal M} \frac{2m_u^2}{v^2} (\bar{d}\,P_R\, u)\times (\bar{u}\,P_L\, d), 
\label{amplitude_notation}
\end{align} 
where we neglect the down quark mass to make expressions simpler, and it does not change expressions for pinch-terms given below. 

The pinch-terms are extracted from diagrams shown in Fig.~\ref{pinch-hp} as follows:
 \begin{align}
& \Delta_\xi \sum_{i=1,4}(\overline{\mathcal{M}}_{W\text{-}i} + \overline{\mathcal{M}}_{Z\gamma\text{-}i})  \notag\\
&\to  \frac{g^2}{32\pi^2}\frac{1-\xi_W}{q^2-m^2_{H^\pm}}\zeta_u
\Big[c_{\beta-\alpha}\zeta_{huu}X_V(q^2;W,h) - s_{\beta-\alpha}\zeta_{Huu}X_V(q^2;W,H) + \zeta_uX_V(q^2;W,A)\Big] \notag\\
&+\frac{g^2}{32\pi^2}\frac{1-\xi_W}{q^2-m^2_{G^\pm}}
\Big[s_{\beta-\alpha}\zeta_{huu}X_V(q^2;W,h) + c_{\beta-\alpha}\zeta_{Huu}X_V(q^2;W,H) \Big] \notag\\
&+\frac{g_Z^2c_{2W}^2}{32\pi^2}\frac{1-\xi_Z^{}}{q^2-m^2_{H^\pm}}\zeta_u^2X_V(q^2;Z,H^\pm) \notag\\
&+\frac{e^2}{8\pi^2}\frac{1-\xi_\gamma}{q^2-m^2_{H^\pm}}\zeta_u^2\Big[B_0(q^2;\gamma,\gamma)-(q^2-m_{H^\pm}^2)C_0(0,q^2,q^2;\gamma,\gamma,H^\pm)  \Big], \\
& \Delta_\xi \sum_{i=5,8}\left(\overline{\mathcal{M}}_{W\text{-}i} + \overline{\mathcal{M}}_{Z\gamma\text{-}i}\right)\notag\\
&\to \frac{g^2}{64\pi^2}(1-\xi_W)\Big[\zeta_{huu}^2C_0(q^2;W,G^\pm,h)+\zeta_{Huu}^2C_0(q^2;W,G^\pm,H) +\zeta_u^2C_0(q^2;W,G^\pm,A)\Big] \notag\\
&+\frac{g_Z^2c_{2W}^2}{64\pi^2}(1-\xi_Z^{})\zeta_u^2C_0(q^2;Z,G^0,H^\pm) 
+\frac{e^2}{16\pi^2}(1-\xi_\gamma)\zeta_u^2C_0(0,q^2,q^2;\gamma,\gamma,H^\pm), \\
&\Delta_{\xi} \left(\overline{\mathcal{M}}_{W\text{-}9} + \sum_{i=9,11}\overline{\mathcal{M}}_{Z\gamma\text{-}i} \right)\notag\\
&\to -\frac{1}{32\pi^2}\frac{\zeta_u^2}{q^2-m^2_{H^\pm}}\Big[g^2(1-\xi_W)B_0(0;W,G^\pm) + \frac{g_Z^2c_{2W}^2}{2}(1-\xi_Z)B_0(0;Z,G^0)\notag\\
&\hspace{1cm} +2e^2(1-\xi_\gamma)B_0(0;\gamma,\gamma)\Big]. 
 \end{align}  
Similar to the case for the CP-odd sector, we can separate the total pinch-term contribution $\Delta_\xi\overline{\mathcal{M}}_{\text{PT}}$ into the three parts 
by the power of $\zeta_u$ factor. 
The term proportional to $\zeta_u^2$ ($\Delta_\xi\overline{{\cal M}}_{\text{PT}}^{H^+H^-}$) and $\zeta_u^1$ ($\Delta_\xi\overline{{\cal M}}_{\text{PT}}^{H^+G^-}$) can be 
used as the pinch-terms for $H^\pm$--$H^\pm$ and $H^\pm$--$G^\pm$, respectively. 
These are expressed as 
\begin{align}
\Delta_{\xi}\overline{\mathcal{M}}^{\text{PT}}_{H^+H^-}&= -\Delta_{\xi}\overline{\mathcal{M}}_{H^+H^-}, \\
\Delta_{\xi}\overline{\mathcal{M}}^{\text{PT}}_{H^+G^-}&= 
\frac{g^2}{32\pi^2}(1-\xi_W)s_{\beta-\alpha}c_{\beta-\alpha}\zeta_u \Bigg\{
\Big[C_0(q^2;W,G^\pm,h) - C_0(q^2;W,G^\pm,H) \Big]\notag\\
& + \left(\frac{1}{q^2-m^2_{H^\pm}} + \frac{1}{q^2-m^2_{G^\pm}}\right)
\Big[X_V(q^2;W,h) - X_V(q^2;W,H) \Big] \Bigg\}.  \label{abc}
\end{align} 
As in the $A$--$G^0$ mixing, we need to correctly share the pinch-term for the $G^\pm$--$H^\pm$ mixing and 
the $W^\pm$--$H^\pm$ mixing. Similar to Eq.~(\ref{sep}), we have the following identity:
\begin{align}
\Lambda_{G^+} = \frac{q^2\Lambda_{G^+}}{q^2-m_{G^\pm}^2} -  m_W^{}\Lambda_W^\mu (\Delta_{W})_{\mu\nu}\,q^\nu, \label{sep2}
\end{align}
where $\Lambda_{G^+}$ and $\Lambda_W^\mu$ are the $\bar{u}dG^+$ and $\bar{u}dW^{+\mu}$ vertex, respectively. These are given by  
\begin{align}
\Lambda_{G^+} &= i\frac{\sqrt{2}}{v}\bar{u}m_uP_L d, \quad
\Lambda_W^\mu = i\frac{g}{\sqrt{2}}\bar{u}\gamma^\mu P_L d. 
\end{align}
In Eq.~(\ref{sep2}), the first term of the RHS can be used for the pinch-term of the $G^\pm$--$H^\pm$ mixing. 
From this identity, we can construct the correct pinch-term for the $G^\pm$--$H^\pm$ mixing by repeating the similar procedure done in Eq.~(\ref{sep10}). 

\section{Renormalized Higgs boson couplings with gauge invariance \label{sec:inv}}

\begin{figure}[!t]
\begin{center}
\includegraphics[width=150mm]{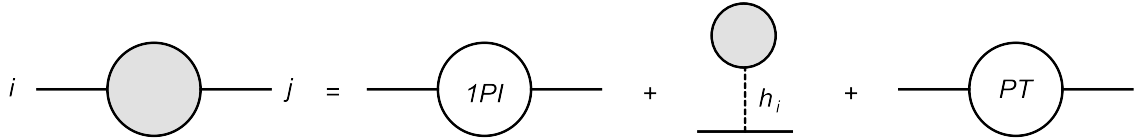}
\caption{Non-renormalized two-point functions in the pinched tadpole scheme. }
\label{tad}
\end{center}
\end{figure}

We compute the renormalized Higgs boson couplings at the one-loop level based on the 
pinched tadpole scheme~\cite{Fleischer}, 
in which the gauge dependence in the scalar boson mixing is successfully removed by using the pinch technique as discussed in the previous section. 
We then clarify the difference in the renormalized Higgs boson couplings 
calculated in the pinched tadpole scheme and those calculated in the ordinal on-shell scheme with the gauge dependence. 
For the latter, we adopt the scheme defined in Ref.~\cite{KOSY}, and we call this the KOSY scheme.  
In this section, all the calculations will be done in the 't~Hooft-Feynman gauge. 

In the pinched tadpole scheme, non-renormalized two-point functions for particles $i$ and $j$ which can be a scalar boson, a gauge boson or a fermion 
are defined as follows:
\begin{align}
\Pi_{ij}(p^2) = \Pi_{ij}^{\text{1PI}}(p^2) + \Pi_{ij}^{\text{Tad}} + \Pi_{ij}^{\text{PT}}(p^2), \label{sse}
\end{align}
where $\Pi_{ij}^{\text{1PI}}$ denotes the contribution from conventional 1-particle irreducible (1PI) diagrams (the first diagram of the RHS in Fig.~\ref{tad}), 
$\Pi_{ij}^{\text{Tad}}$ represents the contribution from the tadpole graph (the second diagram of the RHS in Fig.~\ref{tad}), 
and $\Pi_{ij}^{\text{PT}}$ shows the pinch-term contribution (the third diagram of the RHS in Fig.~\ref{tad}). 
In the 't~Hooft-Feynman gauge, all the analytic expressions of the pinch-terms for 
scalar boson two-point functions are presented in App.~\ref{sec:fg} in the SM, the HSM and the THDM.
Thanks to adding the pinch-terms, the two-point function defined in Eq.~(\ref{sse}) is gauge invariant. 
We note that tadpole diagrams should be added not only to two-point functions but also to three point functions such as $hVV$ and $hhh$, so that 
we further introduce $\Gamma_{ijk}^{\text{Tad}}$ which denote tadpole inserted diagrams to the tree level vertices $\Gamma_{ijk}$. 
We also note that the wave function renormalization factors are not changed from the KOSY scheme, because 
$\Pi_{ij}^{\text{Tad}}$ do not depend on the external momentum, and the pinch-term corrections are not applied to the wave function renormalization factors. 

At one-loop level, the renormalized $\phi V^\mu V^\nu$ ($V=W,Z$) and $\phi f\bar{f}'$ vertices with $\phi$ being a scalar field 
are expressed in terms of the following form factors:
\begin{align}
\hat{\Gamma}_{\phi VV}^{\mu\nu}(p_1^2,p_2^2,q^2)&=g^{\mu\nu}\hat{\Gamma}_{\phi VV}^1
+\frac{p_1^\mu p_2^\nu}{m_V^2}\hat{\Gamma}_{\phi VV}^2
+i\epsilon^{\mu\nu\rho\sigma}\frac{p_{1\rho} p_{2\sigma}}{m_V^2}\hat{\Gamma}_{\phi VV}^3,  \label{form_factor} \\
\hat{\Gamma}_{\phi ff'}(p_1^2,p_2^2,q^2)&=
\hat{\Gamma}_{\phi ff'}^S+\gamma_5 \hat{\Gamma}_{\phi ff'}^P+p_1\hspace{-3.5mm}/\hspace{2mm}\hat{\Gamma}_{\phi ff'}^{V_1}
+p_2\hspace{-3.5mm}/\hspace{2mm}\hat{\Gamma}_{\phi ff'}^{V_2}\notag\\
&\quad +p_1\hspace{-3.5mm}/\hspace{2mm}\gamma_5 \hat{\Gamma}_{\phi ff'}^{A_1}
+p_2\hspace{-3.5mm}/\hspace{2mm}\gamma_5\hat{\Gamma}_{\phi ff'}^{A_2}
+p_1\hspace{-3.5mm}/\hspace{2mm}p_2\hspace{-3.5mm}/\hspace{2mm}\hat{\Gamma}_{\phi ff'}^{T}
+p_1\hspace{-3.5mm}/\hspace{2mm}p_2\hspace{-3.5mm}/\hspace{2mm}\gamma_5\hat{\Gamma}_{\phi ff'}^{PT}, 
\end{align}
where $p_1^\mu$ and $p_2^\mu$ ($q^\mu$) are incoming momenta for gauge bosons or fermions (the Higgs boson). 
Each of the above form factors is also the function of $(p_1^2,p_2^2,q^2)$, but we here do not explicitly denote it. 
In the THDM, Higgs-Higgs-gauge type vertices also appear, i.e.,  $hH^\pm W_\mu^\mp$ and $hAZ_\mu$ in addition to the above vertices. 
Their renormalized vertices can be expressed by 
\begin{align}
\hat{\Gamma}_{\phi_1^{} \phi_2^{} V}^\mu (p_1^2,p_2^2,q^2)  &=-i(p_1^{} - p_2^{})^\mu \hat{\Gamma}_{\phi_1^{} \phi_2^{} V}, 
\end{align}
where $p_{1}^\mu$ and $p_{2}^\mu$ are the incoming momenta for $\phi_1$ and $\phi_2$, respectively, and $q^\mu$ is that for a gauge boson $V^\mu$. 
In App.~\ref{sec:rhiggs} and App.~\ref{sec:counter}, 
we present all the relevant renormalized Higgs boson couplings and counter terms calculated in the pinched tadpole scheme, respectively. 

For the later convenience, we introduce the following symbol:
\begin{align}
\Delta_{\text{SC}}\,[\cdots] = [\cdots]_{\text{TP}} - [\cdots]_{\text{KOSY}}, 
\end{align}
where the first (second) term of the RHS denotes the quantity calculated in the pinched tadpole (KOSY) scheme. 

\subsection{SM}

We calculate the difference in the renormalized  gauge ($hVV$), Yukawa ($hf\bar{f}$) and Higgs-self ($hhh$) couplings 
calculated in the pinched tadpole scheme and those in the KOSY scheme in the SM. 
As it is shown below, there is no difference between the two schemes in the three couplings:
\begin{align}
\Delta_{\text{SC}}\,\hat{\Gamma}_{hVV}^{1}
&= \frac{2m_V^2}{v}\Delta_{\text{SC}}\,\left(\frac{\delta m_V^2}{m_V^2} -\frac{\delta v}{v} \right) + \Gamma_{hVV}^{\text{Tad}} 
= \frac{\Pi_{VV}^{\text{Tad}}}{v}  + \Gamma_{hVV}^{\text{Tad}} =0, \\
\Delta_{\text{SC}}\,\hat{\Gamma}_{hff}^S
&=-\frac{m_f}{v}\Delta_{\text{SC}}\,\left(\frac{\delta m_f}{m_f} -\frac{\delta v}{v} \right)
= -\frac{m_f}{v}\left(\frac{\Pi_{ff}^{\text{Tad}}}{m_f}-\frac{\Pi_{WW}^{\text{Tad}}}{2m_W^2} \right)=0, \\
\Delta_{\text{SC}}\,\hat{\Gamma}_{hhh}&=
  - \frac{3m_h^2}{v}\Delta_{\text{SC}}\,\left (
  \frac{\delta m_h^2}{m_h^2} - \frac{\delta v}{v}
  \right)  + \Gamma_{hhh}^\text{Tad} \notag\\
&= 
  - \frac{3m_h^2}{v}\left (
  \frac{\Pi_{hh}^{\text{Tad}}}{m_h^2}+\frac{T_h^{\text{1PI}}}{vm_h^2} - \frac{\Pi_{WW}^{\text{Tad}}}{2m_W^2}
  \right)  + \Gamma_{hhh}^\text{Tad} = 0, 
  \end{align}
where $T_{h}^\text{1PI}$ is the 1PI tadpole diagram for $h$.  
In the following, we use the generic symbol $T_{h_i}^\text{1PI}$ to express the 1PI tadpole diagram for a CP-even Higgs boson $h_i$. 
We note that there are following relations among $T_{h}^{\text{1PI}}$ and $\Pi_{ij}^{\text{Tad}}$:
\begin{align}
\frac{\Pi_{VV}^{\text{Tad}}}{2m_V^2} = -\frac{\Pi_{hh}^{\text{Tad}}}{3m_h^2} = \frac{\Pi_{ff}^{\text{Tad}}}{m_f} = -\frac{T_h^{\text{1PI}}}{vm_h^2}. 
\end{align}
Thus, in the SM
the tadpole contribution in a two-point function $\Pi_{ij}^{\text{Tad}}$ is cancelled 
by that from the other two-point function and/or the tadpole inserted contribution in the three point function $\Gamma_{ijk}^{\text{Tad}}$. 

\subsection{HSM}

In the HSM, the difference in the renormalized $hVV$ and $hf\bar{f}$ coupling is calculated by 
\begin{align}
\Delta_{\text{SC}}\,\hat{\Gamma}_{hVV}^{1} &= 
 \frac{2m_V^2}{v}c_\alpha\Delta_{\text{SC}}
\left(
\frac{\delta m_V^2}{m_V^2} -\frac{\delta v}{v} \right)
 + \Gamma_{hVV}^{\text{Tad}} =
 c_\alpha\frac{\Pi_{VV}^{\text{Tad}}}{v} + \Gamma_{hVV}^{\text{Tad}} = 0, \\
\Delta_{\text{SC}}\,\hat{\Gamma}_{hff}^{S} &= 
 -\frac{m_f}{v}c_\alpha\Delta_{\text{SC}}\left(
\frac{\delta m_f}{m_f} -\frac{\delta v}{v} \right)
 = 
 -\frac{m_f}{v}c_\alpha\left(
\frac{\Pi_{ff}^{\text{Tad}}}{m_f} -\frac{\Pi_{VV}^{\text{Tad}}}{2m_V^2} \right)=0.  
\end{align}
Similarly, we can show that there is no difference in the $HVV$ and $Hf\bar{f}$ couplings. 

In contrast to the Higgs boson couplings with weak bosons or fermions, 
we find non-zero differences in the $hhh$ and $Hhh$ couplings as follows:
\begin{align}
 \Delta_{\text{SC}}\,\hat{\Gamma}_{hhh}
  &= 6\Delta_{\text{SC}}(\delta \lambda_{hhh} +\lambda_{Hhh}\delta\alpha) + \Gamma_{hhh}^\text{Tad}\notag\\
  &=4 ! \lambda_S s_\alpha^3 \left(
    \frac{c_\alpha}{m_H^2}T_H^\text{1PI}
    - \frac{s_\alpha}{m_h^2}T_h^\text{1PI}\right)_{\text{Fin}}
    -\frac{s_\alpha c_\alpha^2}{4v}\left[\Pi_{Hh}^\text{PT}(m_h^2) + \Pi_{Hh}^\text{PT}(m_H^2) \right], \\
\Delta_{\text{SC}}\,\hat{\Gamma}_{Hhh} 
    &= \Delta_{\text{SC}}[2\delta\lambda_{Hhh}+(4\lambda_{HHh}-6\lambda_{hhh})\delta \alpha ]
 + \Gamma_{Hhh}^\text{Tad} \notag\\
    &\hspace{-5mm} =- 4 !\lambda_{S}s_\alpha^2 c_\alpha \left(
 \frac{c_\alpha}{m_H^2}T_H^\text{1PI} -\frac{s_\alpha}{m_h^2}T_h^\text{1PI}  \right)_{\text{Fin}}
    + \frac{c_{3\alpha} - 5c_\alpha}{8v}\left[\Pi_{Hh}^\text{PT}(m_h^2) + \Pi_{Hh}^\text{PT}(m_H^2)\right],  
  \end{align}
where $(\cdots)_\text{Fin}$ shows the finite part of the quantity ($\cdots$). 
These differences vanish when we take the no mixing limit, i.e., $\alpha \to 0$. 

\subsection{THDM}

In the THDM, the difference in the renormalized $hVV$ coupling is calculated by 
 \begin{align}
 \Delta_{\text{SC}}\hat{\Gamma}_{hVV}^1&=
 \frac{2m_V^2}{v} \Delta_{\text{SC}}\left[s_{\beta-\alpha}  
\left(\frac{\delta m_V^2}{m_V^2}-\frac{\delta v}{v}\right) +c_{\beta-\alpha}\delta \beta \right] + \Gamma_{hVV}^{\text{Tad}}. 
\end{align}
Differently from the previous two models, the counter term $\delta \beta$ also contributes to the difference. 
We can calculate $\Delta_{\text{SC}}\,\delta \beta$ as follows:
\begin{align}
\Delta_{\text{SC}}\,\delta \beta
&=\frac{T_H^{ \text{1PI}}}{vm_H^2}s_{\beta-\alpha}-\frac{T_h^\text{1PI} }{vm_h^2}c_{\beta-\alpha} -\frac{1}{2m_A^2}\left[\Pi_{AG}^{\text{PT}}(m_A^2)+\Pi_{AG}^{\text{PT}}(0)   \right]. 
\end{align}
Using the above result, we obtain 
\begin{align}
\Delta_{\text{SC}}\hat{\Gamma}_{hVV}^1 
&=-\frac{m_V^2}{m_A^2v}c_{\beta-\alpha} \left[\Pi_{AG}^{\text{PT}}(m_A^2)+\Pi_{AG}^{\text{PT}}(0)   \right]. \label{dsc_hvv}
\end{align}
Similar to the case in the SM and the HSM, the dependence of $T_{h_i}^{\text{1PI}}$ is exactly cancelled among the counter terms and $\Gamma_{hVV}^{\text{Tad}}$, but the non-vanishing 
contribution comes from $\delta \beta$. 
This effect, however, vanishes when we take the alignment limit $s_{\beta-\alpha} \to 1$. 
All the differences in the other gauge and Yukawa couplings also come from $\Delta_{\text{SC}}\,\delta \beta$ as follows:
\begin{align}
\Delta_{\text{SC}}\hat{\Gamma}_{HVV}^1 &= +\frac{m_V^2}{vm_A^2}s_{\beta-\alpha}\left[\Pi_{AG}^{\text{PT}}(m_A^2)+\Pi_{AG}^{\text{PT}}(0) \right], \\
\Delta_{\text{SC}}\hat{\Gamma}_{hff}^S &= -\frac{m_f}{2vm_A^2}\zeta_{hff}\zeta_f\left[\Pi_{AG}^{\text{PT}}(m_A^2)+\Pi_{AG}^{\text{PT}}(0) \right], \\
\Delta_{\text{SC}}\hat{\Gamma}_{Hff}^S &= -\frac{m_f}{2vm_A^2}\zeta_{Hff}\zeta_f\left[\Pi_{AG}^{\text{PT}}(m_A^2)+\Pi_{AG}^{\text{PT}}(0) \right], \\
\Delta_{\text{SC}}\hat{\Gamma}_{Aff}^P &= +i\frac{I_fm_f}{vm_A^2}\zeta_f^2\left[\Pi_{AG}^{\text{PT}}(m_A^2)+\Pi_{AG}^{\text{PT}}(0) \right], \\
\Delta_{\text{SC}}\hat{\Gamma}_{H^+\bar{u}_Ld_R}^R &= -\frac{m_d}{\sqrt{2}vm_A^2}\zeta_d^2\left[\Pi_{AG}^{\text{PT}}(m_A^2)+\Pi_{AG}^{\text{PT}}(0) \right], \\
\Delta_{\text{SC}}\hat{\Gamma}_{H^+\bar{u}_Rd_L}^L &= \frac{m_u}{\sqrt{2}vm_A^2}\zeta_u^2\left[\Pi_{AG}^{\text{PT}}(m_A^2)+\Pi_{AG}^{\text{PT}}(0) \right]. 
\end{align}
We note that in the Yukawa couplings for $A$ and $H^\pm$, we extract the different 
form factor with respect to those for the CP-even Higgs bosons, because of the difference in the tree level coupling structure (see App.~\ref{sec:rhiggs}). 

For the $hH^\pm W^\mp_\mu$ and $hAZ_\mu$ couplings, we have 
\begin{align}
\Delta_{\text{SC}}\hat{\Gamma}_{hH^\pm W^\mp} = \Delta_{\text{SC}}\hat{\Gamma}_{hA Z} = 0. 
\end{align}
This simply follows $\Delta_{\text{SC}}[\delta m_V^2/(2m_V^2) -\delta v/v]=0 $. 

Finally, the difference in the renormalized $hhh$ and $Hhh$ vertices is calculated as
\begin{align}
  \Delta_{\text{SC}}\hat{\Gamma}_{hhh}
  &=
  - \frac{12M^2}{v^2}\frac{c_{2\beta}^{} c_{\alpha + \beta}^{} c_{\beta - \alpha}^2}{s_{2\beta}^2}
  \left( \frac{T_h^\textrm{1PI}}{m_h^2}c_{\beta - \alpha}
  - \frac{T_H^\textrm{1PI}}{m_H^2} s_{\beta - \alpha}\right)_{\text{Fin}} \notag\\
  &-\frac{3c_{\beta-\alpha}s_{2\alpha}}{4vs_\beta c_\beta}\left[ \Pi_{hH}^\text{PT}(m_h^2) + \Pi_{hH}^\text{PT}(m_H^2) \right]
  - \frac{3F_\beta}{m_A^2}\left[\Pi_{AG}^\text{PT}(m_A^2) +   \Pi_{AG}^\text{PT}(0) \right],  \label{dsc_hhh2} \\
  \Delta_{\text{SC}}\hat{\Gamma}_{Hhh} & =
 -\frac{4M^2}{v^2}\frac{c_{2\beta}^{} c_{\beta-\alpha}^{}}{s_{2\beta}^2}
  (3s_\alpha^{}c_\alpha^{} - s_\beta^{} c_\beta^{})
  \left(\frac{T_h^\textrm{1PI}}{m_h^2}c_{\beta-\alpha}^{} 
  -\frac{T_H^\textrm{1PI}}{m_H^2}s_{\beta - \alpha} \right)_{\text{Fin}}
  \notag\\
  & +\frac{c_{3\alpha - \beta} - 5c_{\alpha + \beta}}{4v s_{2\beta}}
  \left[ \Pi_{Hh}^\text{PT}(m_h^2) + \Pi_{Hh}^\text{PT}(m_H^2)\right]
  - \frac{G_\beta}{m_A^2}\left[\Pi_{AG}^\text{PT}(m_A^2) + \Pi_{AG}^\text{PT}(0)\right], 
\end{align}
where  
\begin{align}
   F_{\beta} & = \frac{c_{\beta-\alpha}}{2v s_{2\beta}^2}
   \left[(2 + 2c_{2\alpha}c_{2\beta} - s_{2\alpha} s_{2\beta})(m_h^2 -M^2)   - s_{2\beta}^2 M^2   \right], \label{F_beta}\\
   G_{\beta} & =
   \frac{s_{2\alpha}^{}}{vs_{2\beta}^2}\left(
   c_\alpha^{}c_\beta^3 - s_\alpha^{}s_\beta^3 \right) (2m_h^2 + m_H^2)
   + \frac{1}{2v s_{2\beta}^{2}}
   \left[ s_{2\beta}^{2} s_{\beta - \alpha}^{}   - 6s_{2\alpha}^{} \left( c_\alpha^{}c_\beta^3 - s_\alpha^{}s_\beta^3 \right) \right]M^2. \label{G_beta}
\end{align}

\begin{figure}[t]
\begin{center}
\includegraphics[width=70mm]{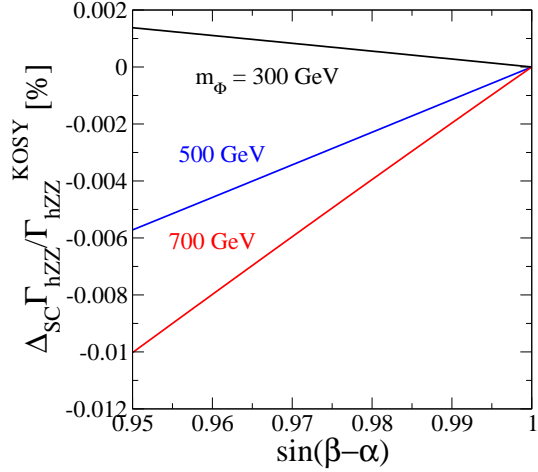}
\caption{Difference in the renormalized $hZZ$ coupling between
the pinched tadpole scheme and the KOSY scheme as a function of $s_{\beta-\alpha}$ in the THDM. 
The black, blue and red curve show the case for $m_{\Phi}^{}(=m_{H^\pm} = m_A^{}=m_H^{})=300$, 500 and 700 GeV, respectively. 
We take $\tan\beta = 1.5$, $M/m_\Phi = 0.8$ and $c_{\beta-\alpha}>0$. }
\label{scheme_hvv}
\end{center}
\end{figure}

\begin{figure}[t]
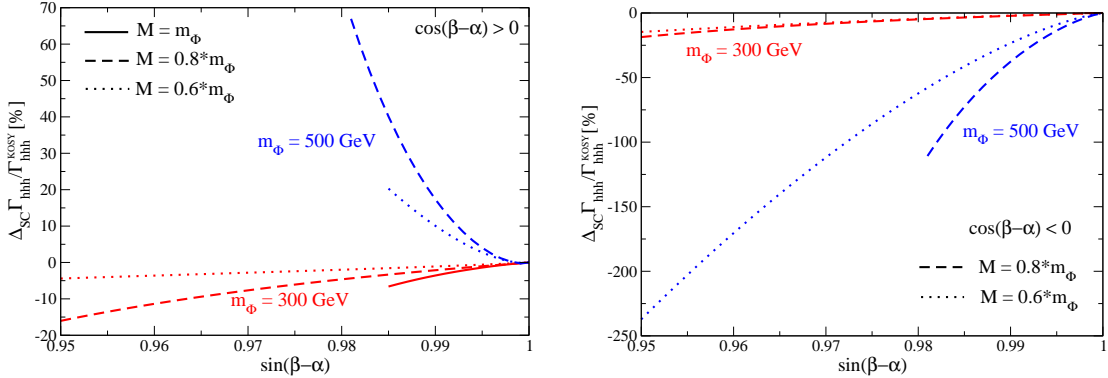

\begin{center}
\includegraphics[width=70mm]{scheme_dep_hhh_p.eps}\hspace{5mm}
\includegraphics[width=70mm]{scheme_dep_hhh_m.eps}
\caption{Difference in the renormalized $hhh$ coupling between
the pinched tadpole scheme and the KOSY scheme as a function of $s_{\beta-\alpha}$ in the THDM with $\tan\beta = 1.5$. 
The left and right panels show the case for $c_{\beta-\alpha}>0$ and $c_{\beta-\alpha}<0$, respectively. 
We only show results allowed by bounds from the perturbative unitarity and the vacuum stability. 
}
\label{scheme_hhh}
\end{center}
\end{figure}

In Fig.~\ref{scheme_hvv}, we show the scheme difference in the renormalized $hZZ$ coupling as a function of $s_{\beta-\alpha}$ in the THDM. 
Here, we take $\tan\beta=1.5$, $M/m_\Phi^{}=0.8$ ($m_{\Phi}^{}=m_{H^\pm} = m_A^{}=m_H^{})$ and $c_{\beta-\alpha}>0$, but the result does not depend on these parameters so much in this plot. 
The typical magnitude of the difference is seen to be ${\cal O}(0.01)\%$. 

In Fig.~\ref{scheme_hhh}, we show the scheme difference in the renormalized $hhh$ coupling as a function of $s_{\beta-\alpha}$ in the THDM with $\tan\beta=1.5$ 
and $c_{\beta-\alpha}>0$ (left panel) or $c_{\beta-\alpha}<0$ (right panel). 
We only show results allowed by bounds from the perturbative unitarity~\cite{pu_THDM1,pu_THDM2,pu_THDM3,pu_THDM4,pu_THDM5} 
and the vacuum stability~\cite{vs_THDM1,Klimenko,vs_THDM2,vs_THDM3}, which were discussed in Sec.~\ref{sec:model}. 
The typical magnitude of the difference is found to be ${\cal O}(10\text{--}100)\%$. 
Such large difference comes from the non-vanishing tadpole contribution $T_{h,H}^{\text{1PI}}$ in Eq.~(\ref{dsc_hhh2}). 

\begin{figure}[t]
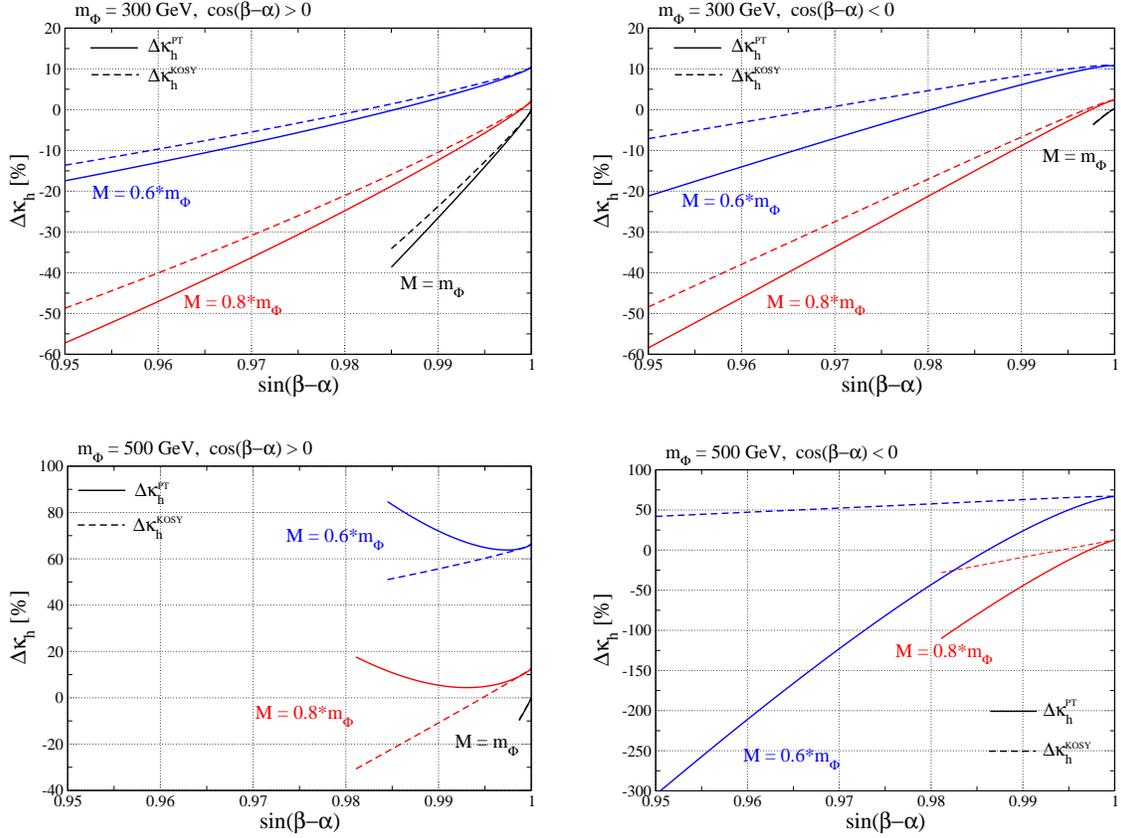

\begin{center}
\includegraphics[width=70mm]{dkh_2scheme_300_p.eps} \hspace{5mm}
\includegraphics[width=70mm]{dkh_2scheme_300_m.eps} \\ \vspace{5mm}
\includegraphics[width=70mm]{dkh_2scheme_500_p.eps} \hspace{5mm}
\includegraphics[width=70mm]{dkh_2scheme_500_m.eps} 
\caption{$\Delta \kappa_h^{}$ calculated in the pinched tadpole scheme (solid lines) and the KOSY scheme (dashed lines) in the THDM with $\tan\beta = 1.5$.
Top-left, top-right, bottom-left and bottom-right panels are results for
($m_\Phi^{}=300$ GeV, $c_{\beta-\alpha}>0$), 
($m_\Phi^{}=300$ GeV, $c_{\beta-\alpha}<0$), 
($m_\Phi^{}=500$ GeV, $c_{\beta-\alpha}>0$) and
($m_\Phi^{}=500$ GeV, $c_{\beta-\alpha}<0$), respectively.
We only show results allowed by bounds from the perturbative unitarity and the vacuum stability. 
}
\label{kh}
\end{center}
\end{figure}

In Fig.~\ref{kh}, 
we also evaluate the value of $\Delta\kappa_h^{}$ defined in Eq.~(\ref{kappas}) calculated in the two different schemes. 
The solid and dashed curves show the results in the pinched tadpole scheme and in the KOSY scheme, respectively. 
The upper-left, upper-right, lower-left and lower-right panels are the results in cases with 
($m_\Phi^{}=300$ GeV, $c_{\beta-\alpha}>0$), 
($m_\Phi^{}=300$ GeV, $c_{\beta-\alpha}<0$), 
($m_\Phi^{}=500$ GeV, $c_{\beta-\alpha}>0$) and
($m_\Phi^{}=500$ GeV, $c_{\beta-\alpha}<0$), respectively.  
We here take $\tan\beta = 1.5$ and $M/m_\Phi^{}=1$ (black), $0.8$ (red) and $0.6$ (blue). 
Similar to Fig.~\ref{scheme_hhh}, we only show results allowed by bounds from the perturbative unitarity and the vacuum stability. 
As we saw in the previous figure, a larger difference is given in the case with a large value of $1-s_{\beta-\alpha}$ and/or $m_\Phi^{}$. 
In addition, a larger value of $\Delta \kappa_h$ is obtained when we take a larger (smaller) value of $m_\Phi^{}$ $(m_\Phi^{}/M)$. 
A large value of $\Delta \kappa_h$ also is given in the alignment limit $s_{\beta-\alpha}\to 1$, e.g., $\Delta \kappa_h \sim +10 (70)\%$
in the case of $c_{\beta-\alpha} < 0$, $M/m_\Phi^{} = 0.6$ and $m_\Phi^{} = 300 (500)$ GeV. 

\section{Numerical results\label{sec:num}}

In this section, we numerically show the one-loop corrected Higgs boson couplings
based on the pinched tadpole scheme discussed in the previous section. 
We discuss how we can discriminate the HSM and the THDMs with four different types of Yukawa interactions by looking at the pattern of the deviation 
in the Higgs boson couplings. 
In addition, we clarify how the tree level results can be changed by taking into account their one-loop corrections. 

In order to discuss the deviation in the Higgs boson couplings from the SM prediction, 
we introduce the renormalized scaling factors $\kappa_X^{}$ for the $hXX$ couplings as follows:
\begin{align}
&\kappa_V^{} \equiv \frac{\hat{\Gamma}_{hVV}^1(m_V^2,(m_V^{}+m_h)^2,m_h^2)_{\text{NP}} }{ \hat{\Gamma}_{hVV}^{1}(m_V^2,(m_V^{}+m_h)^2,m_h^2)_{\text{SM}}}, 
\quad \kappa_f^{}  \equiv \frac{\hat{\Gamma}_{hff}^S(m_f^2,m_f^2,m_h^2)_{\text{NP}} }{ \hat{\Gamma}_{hff}^S(m_f^2,m_f^2,m_h^2)_{\text{SM}}}, \notag\\
&\kappa_h^{}  \equiv \frac{\hat{\Gamma}_{hhh}(m_h^2,m_h^2,4m_h^2)_{\text{NP}} }{ \hat{\Gamma}_{hhh}(m_h^2,m_h^2,4m_h^2)_{\text{SM}}},  
\quad \quad \quad \quad \quad 
\kappa_\gamma^{}  \equiv \sqrt{
                              \frac{ \Gamma(h\to \gamma\gamma)_\text{NP} }{ \Gamma(h\to \gamma\gamma)_\text{SM} } 
                             }, 
\label{kappas}
\end{align}
where $\Gamma(h\to \gamma\gamma)$ is the decay rate of the $h \to \gamma\gamma$ mode. 
We also define $\Delta \kappa_X^{} \equiv \kappa_X^{} -1$. 

For the one-loop level calculation, we scan the parameters in the HSM as
\begin{align}
 m_H^{} \geq 300~\text{GeV}, \quad -0.44 \leq \sin\alpha \leq 0.44, \quad |\lambda_{\Phi S}| \leq 3, 
\end{align}
with $\mu_S^{}=\lambda_S=0$.
In the THDMs, we scan the parameters as 
\begin{align}
 m_\Phi^{} \geq 300~\text{GeV}, \quad 0.90 \leq s_{\beta-\alpha} \leq 1, \quad |\lambda_{\Phi \Phi h}| \leq 3,\quad 1\leq \tan\beta \leq 10, \label{scan_thdm}
\end{align}
where $\lambda_{\Phi\Phi h} \equiv (m_\Phi^2 -M^2)/v^2$ and $m_\Phi^{} = m_{H^\pm}^{}(=m_A^{}=m_H^{})$. 
For the both models, we require $\Lambda_{\text{cutoff}}\geq 3$ TeV for the triviality and vacuum stability bounds (see Sec.~\ref{sec:model}). 

\begin{figure}[t]
\begin{center}\hspace{5mm}
\includegraphics[width=85mm]{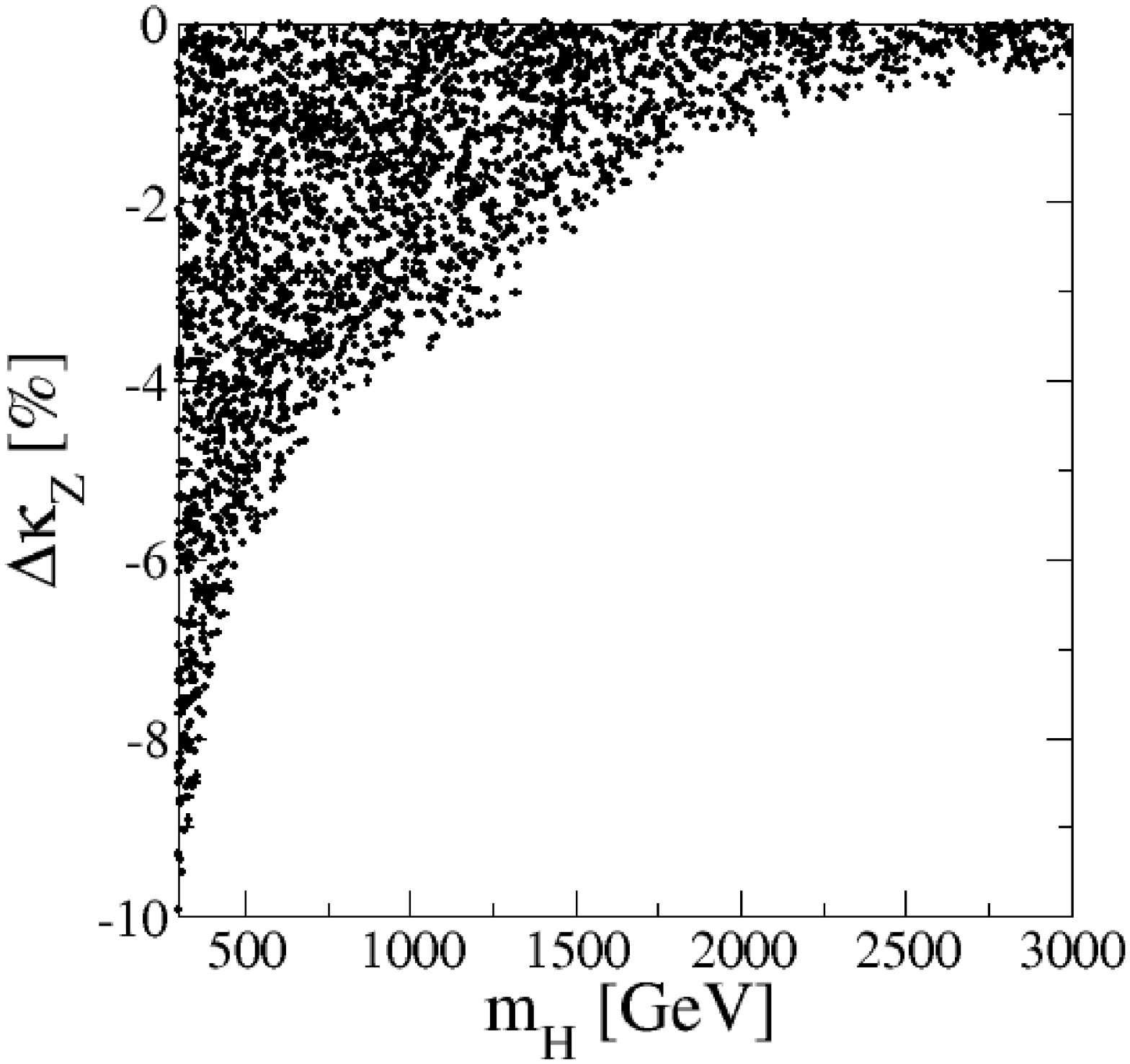} \hspace{-15mm}
\includegraphics[width=85mm]{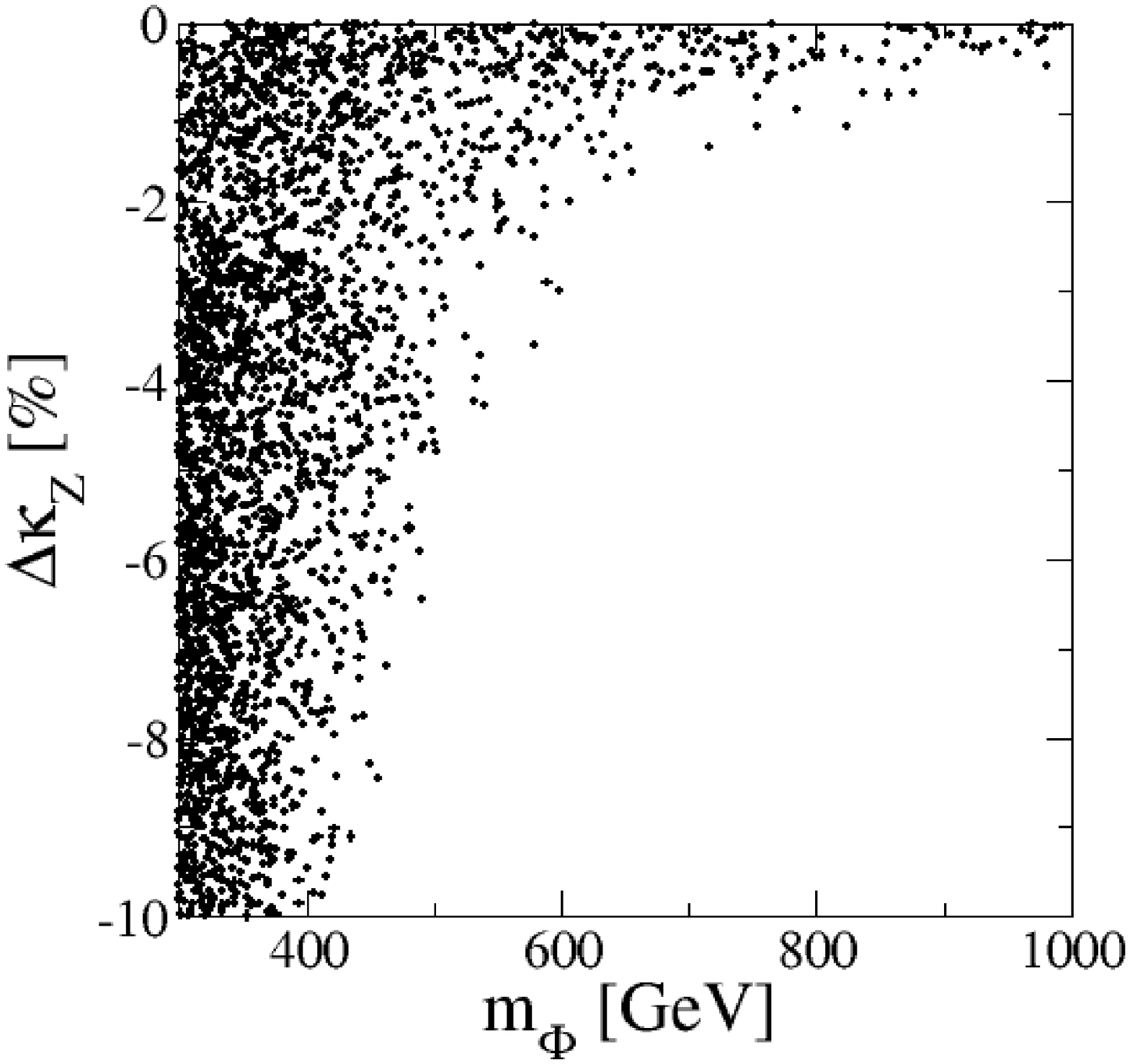} 
\caption{Allowed parameter region under the constraints from the perturbative unitarity, the vacuum stability, the triviality and the $S$, $T$ parameters
on the $\Delta\kappa_Z^{}$--$m_H^{}$ plane and the $\Delta\kappa_Z^{}$--$m_\Phi^{}$ plane 
in the HSM (left) and in the THDM (right), respectively.  }
\label{mH-kv}
\end{center}
\end{figure}

First of all in Fig.~\ref{mH-kv}, we show the allowed region on the $m_H^{}$--$\Delta \kappa_Z^{}$ plane in the HSM and that on the 
$m_\Phi^{}$--$\Delta \kappa_Z^{}$ plane in the THDMs. 
We note that the dependence on the type of Yukawa interactions in the THDM is negligible in this plot. 
In the both models, we can see the decoupling behavior, namely
the large mass limit can be taken in the limit of $\Delta \kappa_Z^{} \to 0$. 
It is also seen that the speed of the decoupling is quite different between these two models. 
For example, in the HSM the mass of $H$ can be larger than 1 TeV even if  $|\Delta \kappa_Z| \lesssim 4\%$, while 
in the THDMs $m_\Phi^{} > 1$ TeV is allowed only when $|\Delta \kappa_Z| \lesssim 0.5\%$. 
This result suggests us the existence of the upper limit on the mass of the extra Higgs bosons 
once a non-zero deviation in the $hVV$ couplings is measured at future collider experiments, and the upper limit quite depends on the structure of the Higgs sector.

\begin{figure}[t]
\begin{center} \hspace{5mm}
\includegraphics[width=90mm]{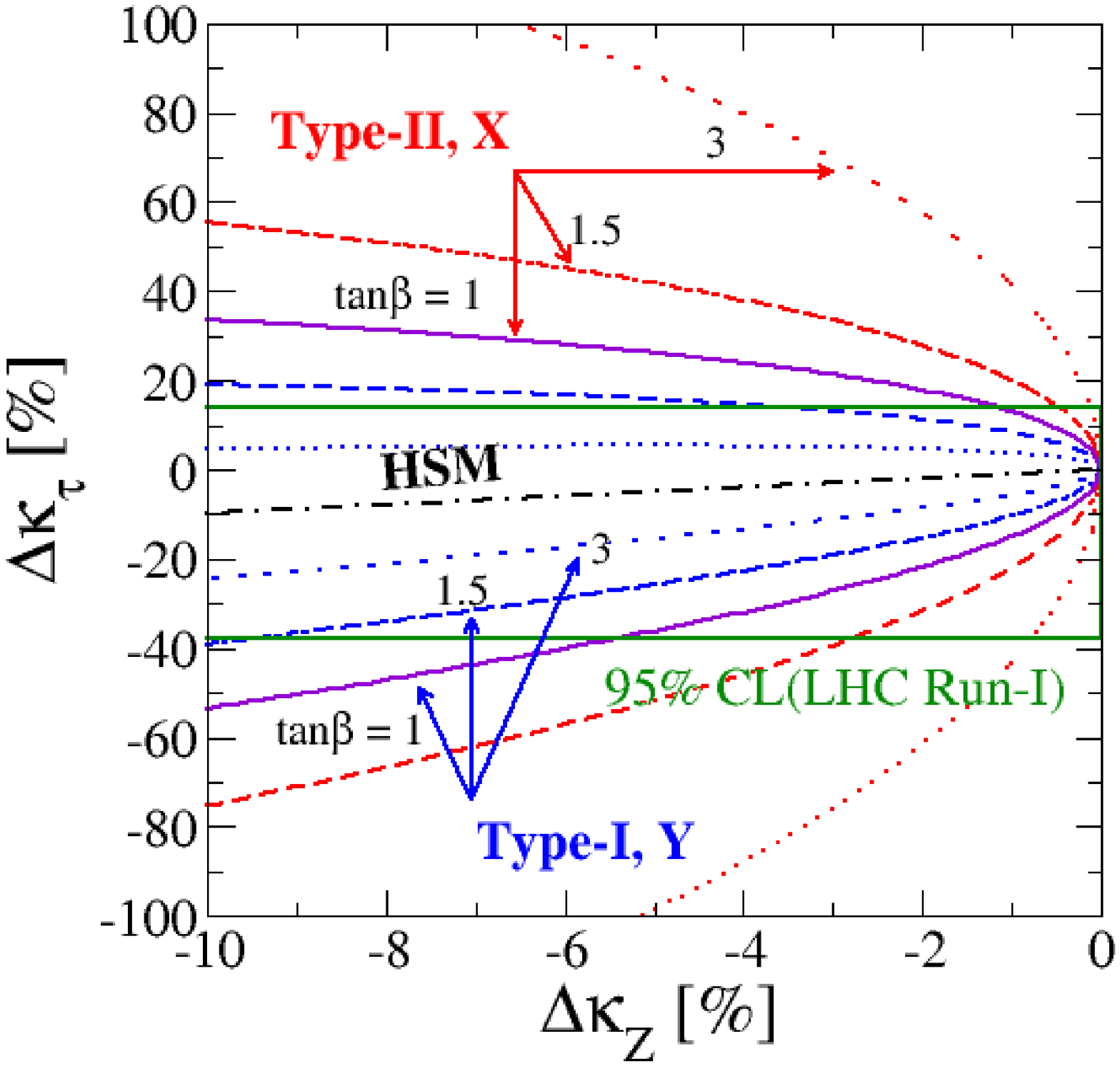} \hspace{-25mm}
\includegraphics[width=90mm]{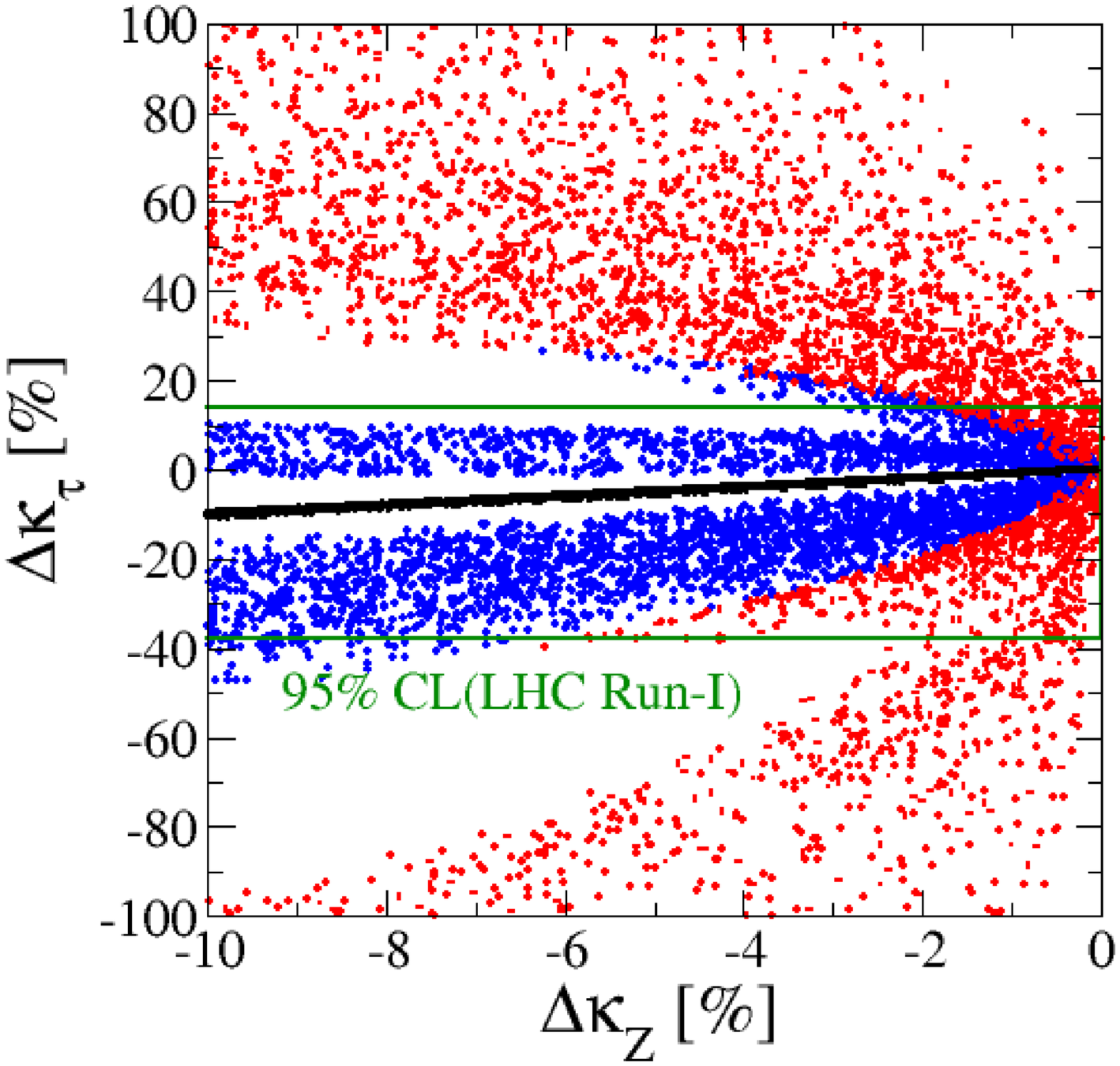} 
\caption{Correlation between $\Delta \kappa_Z^{}$--$\Delta\kappa_\tau$ in the HSM and in the THDMs. 
The left (right) panel shows the result at the tree (one-loop) level.
In the left panel, the solid, dashed and dotted curves are the results in the THDM with $\tan\beta = 1$, 1.5 and 3, respectively.  
The black dot-dashed curve is the result in the HSM.  
In the right panel, the blue, red and black dots are the results in the Type-I (Y) THDM, Type-II (X) THDM and the HSM, respectively. 
The region inside the green box is allowed with the 95\% CL from the measurement of the Higgs boson coupling at the LHC Run-I experiment. }
\label{kv-ktau}
\end{center}
\end{figure}

Next, we discuss various correlations among deviations in the Higgs boson couplings. 
In Fig.~\ref{kv-ktau}, we show the correlation between $\Delta \kappa_Z^{}$--$\Delta \kappa_\tau^{}$ in the THDMs and in the HSM. 
The left and right panels show the results at the tree level and at the one-loop level, respectively. 
Here, we also display the current 95\% CL limit\footnote{This limit is simply given by taking 2 times error bar from each measured central value of
$\Delta\kappa_Z$ and $\Delta \kappa_\tau$ without taking into account a chi-square fit nor a correlation factor. } 
on the values of $\Delta \kappa_Z^{}$ and $\Delta \kappa_\tau^{}$ from combined ATLAS and CMS analyses using the data at the LHC Run-I experiment~\cite{LHC1}. 
In the left panel, predictions of the Type-I and Type-Y THDMs are shown by the blue curves, while 
those of the Type-II and Type-X THDMs are shown by the red curves. 
The dashed and dotted curves show the cases with $\tan\beta = 1.5$ and 3, respectively. 
For $\tan\beta = 1$, all the THDMs have the same prediction 
denoted by the purple solid curve 
after scanning the sign of $c_{\beta-\alpha}$ and the value of $s_{\beta-\alpha}$ (see Eq.~(\ref{zeta-hff})). 
The black dot-dashed curve denotes the prediction of the HSM. 
From the result shown in the left panel, we can see that 
the value of $\Delta \kappa_\tau$ approaches to 0 
in the limit of $\Delta \kappa_Z^{} \to 0$ in all the 5 models, which corresponds to $s_{\beta-\alpha}\to 1$ in the THDMs and $s_\alpha\to 0$ in the HSM at the tree level. 
Thus, in this limit it is difficult to distinguish these models by looking at the correlation between $\Delta \kappa_Z^{}$ and $\Delta \kappa_\tau^{}$. 
In contrast, once $\Delta \kappa_Z^{}\neq 0$ is given, the 5 models can be separated into the 3 categories assuming $\tan\beta>1$. 
Namely, models belonging to the first (Type-I and Type-Y THDMs), the second (Type-II and Type-X THDMs) and the third (HSM) categories give the prediction 
inside the purple curve, outside the purple curve and of $\Delta \kappa_Z^{} \simeq \Delta \kappa_\tau$, respectively. 

In the right panel, we show the prediction allowed by the constraints explained in Sec.~\ref{sec:model} at the one-loop level. 
The black and blue (red) dots denote the prediction in the HSM and the Type-I and Type-Y (Type-II and Type-X) THDMs, respectively. 
We note that the white region, e.g., $20\% \lesssim \Delta \kappa_\tau^{}\lesssim 30\%$ and 
$-95\% \lesssim \Delta \kappa_\tau^{}\lesssim -50\%$ at $\Delta \kappa_Z^{} = -10\%$ is excluded by either the vacuum stability bound or 
the triviality bound. 
Although the behavior is quite similar to the tree level result after scanning the value of $\tan\beta$, the important difference is seen in 
the region with $|\Delta \kappa_Z^{}| \lesssim 1\%$, in which predictions of all the 5 models are overlapping with each other. 
This is mainly due to the fact that ${\cal O}(-1)\%$ of $\Delta \kappa_Z^{}$ can be explained by the loop effects of 
the extra Higgs bosons with $s_{\beta-\alpha}\simeq 1$. 
Therefore, taking into account the one-loop result, we can conclude that the 5 models 
can be distinguished into the 3 categories in the case of $|\Delta \kappa_Z^{}|\gtrsim 1\%$. 

\begin{figure}[t]
\begin{center} \hspace{5mm}
\includegraphics[width=90mm]{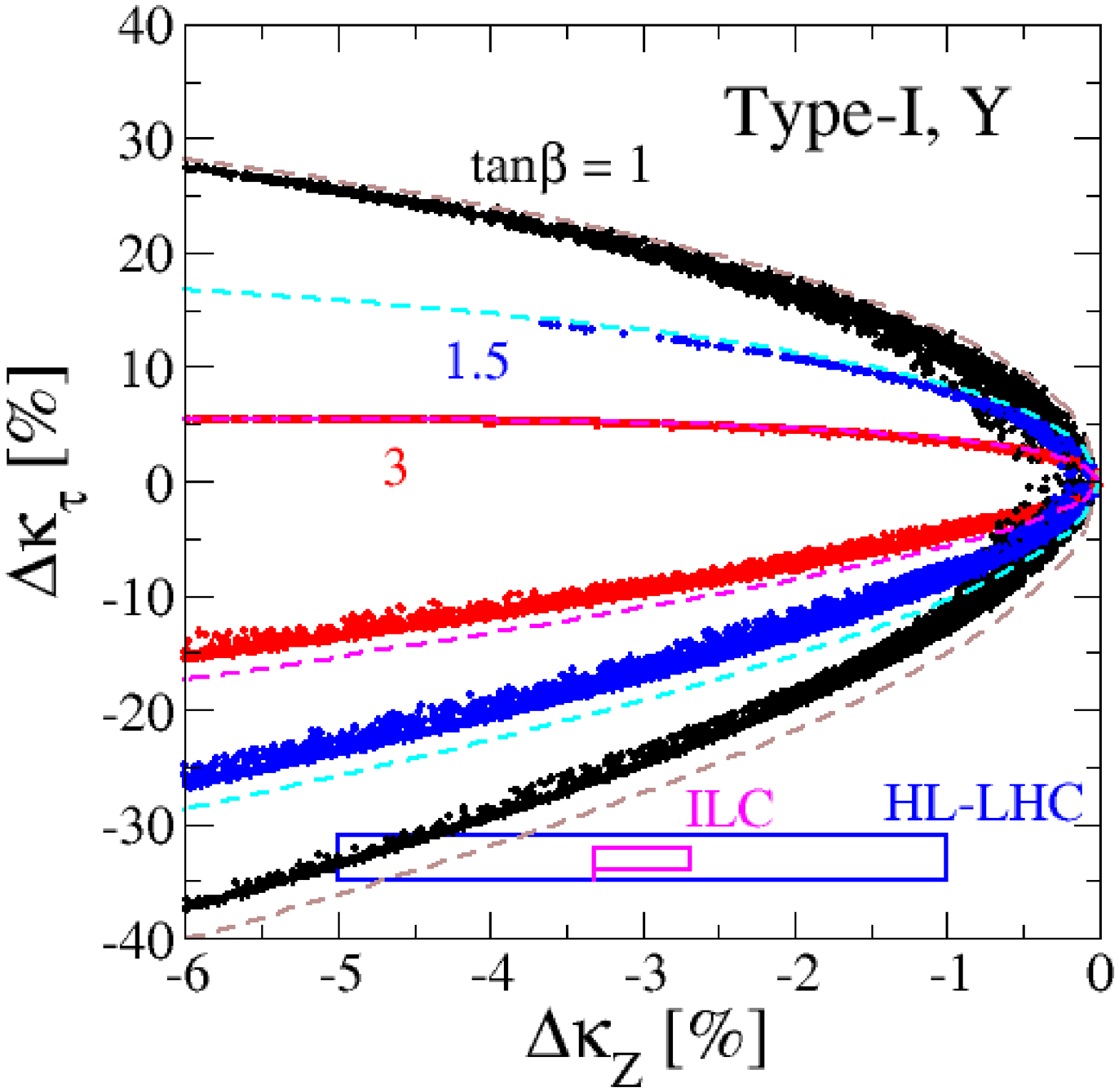} \hspace{-25mm}
\includegraphics[width=90mm]{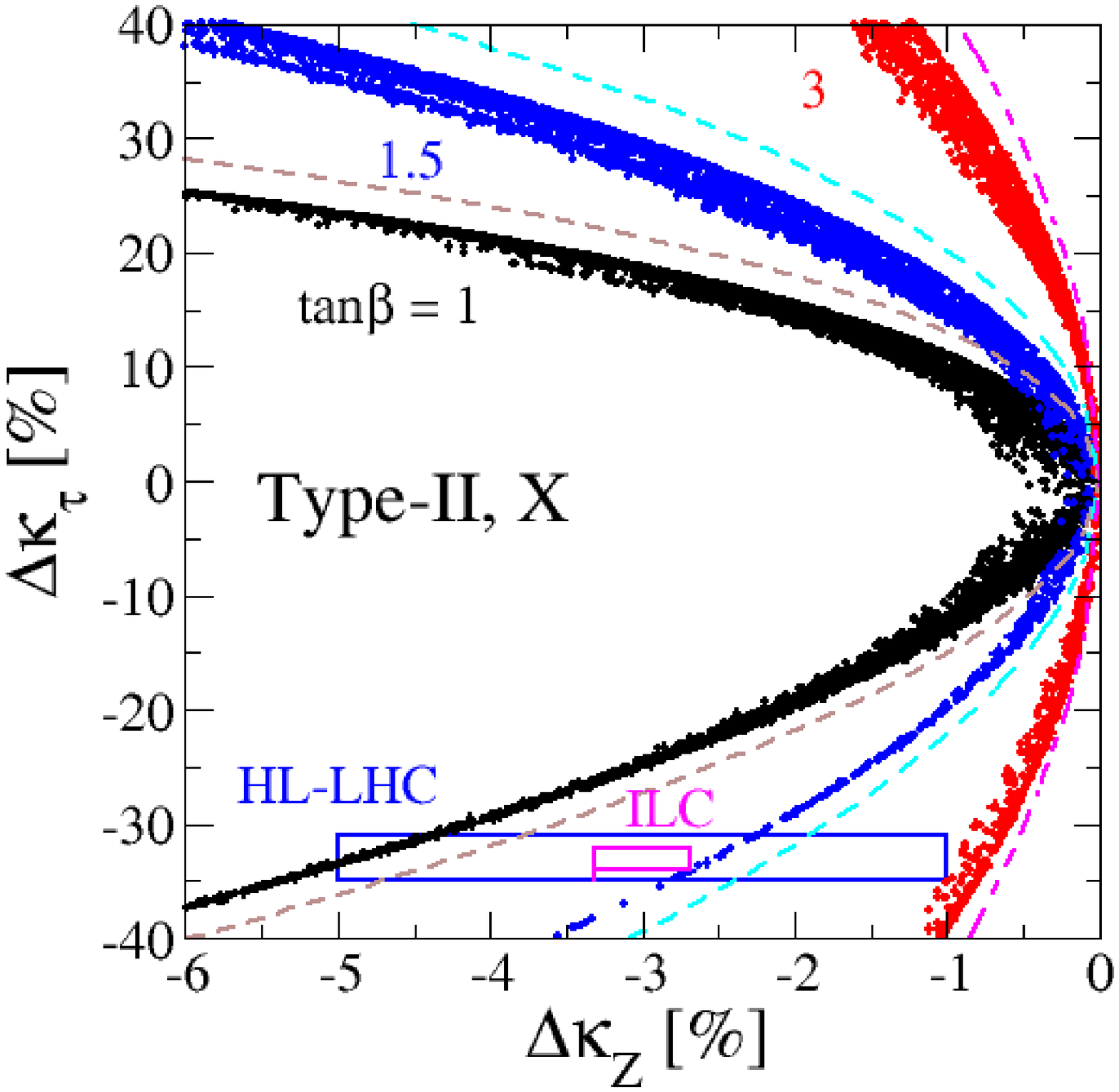} 
\caption{Correlation between $\Delta \kappa_Z^{}$--$\Delta\kappa_\tau$ in the Type-I,-Y THDM (left) and Type-II,-X THDM (right) at one-loop level. 
The black, blue and red dots show the cases for $\tan\beta = 1$, 1.5 and 3, respectively.
The tree level predictions are also shown as the dashed curves. 
The blue (magenta) box denotes the expected 1$\sigma$ accuracies for the measurement of 
$\Delta \kappa_Z^{}$ and $\Delta\kappa_\tau$ at the HL-LHC (ILC), where their central values are not reflected in the current measurements at the LHC. }
\label{kv-ktau2}
\end{center}
\end{figure}

In Fig.~\ref{kv-ktau2}, we show the correlation between $\Delta\kappa_Z^{}$--$\Delta \kappa_\tau$ in the THDMs
for a fixed value of $\tan\beta$. 
Here, we show the expected 1$\sigma$ accuracies for the measurement of $(\Delta \kappa_Z^{},\Delta\kappa_\tau)$ at the HL-LHC (2\%,2\%)~\cite{Snowmass} and at the ILC
with the full data set (0.31\%,0.9\%)~\cite{ILC1}. 
We see that the one-loop results tend to be inside the tree level curve with a small width (a few percent level).
Such a small width can be detected by using the accuracy at the ILC.

\begin{figure}[!t]
\begin{center}
\hspace{5mm}
\includegraphics[width=70mm]{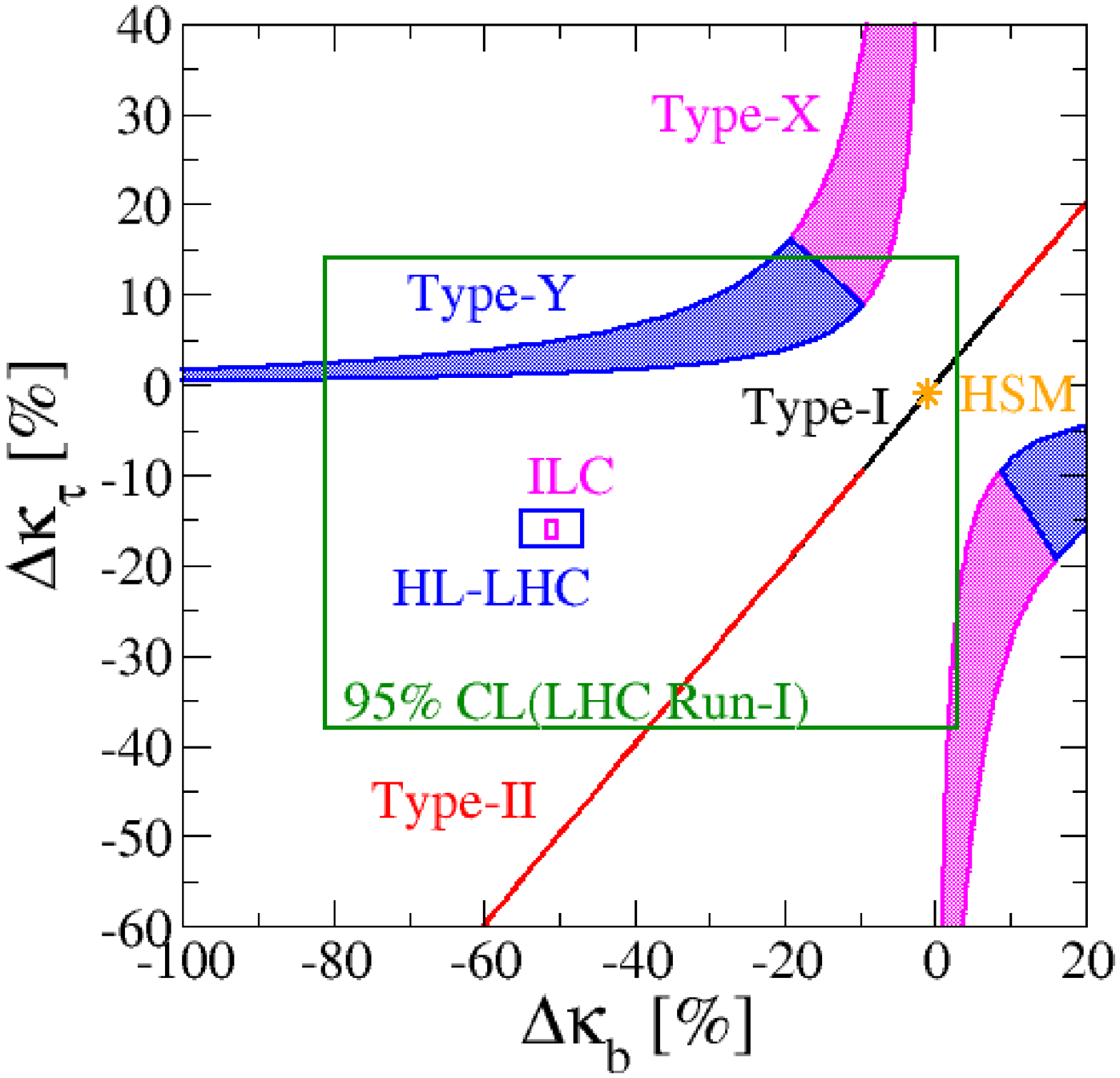} \hspace{-15mm}
\includegraphics[width=70mm]{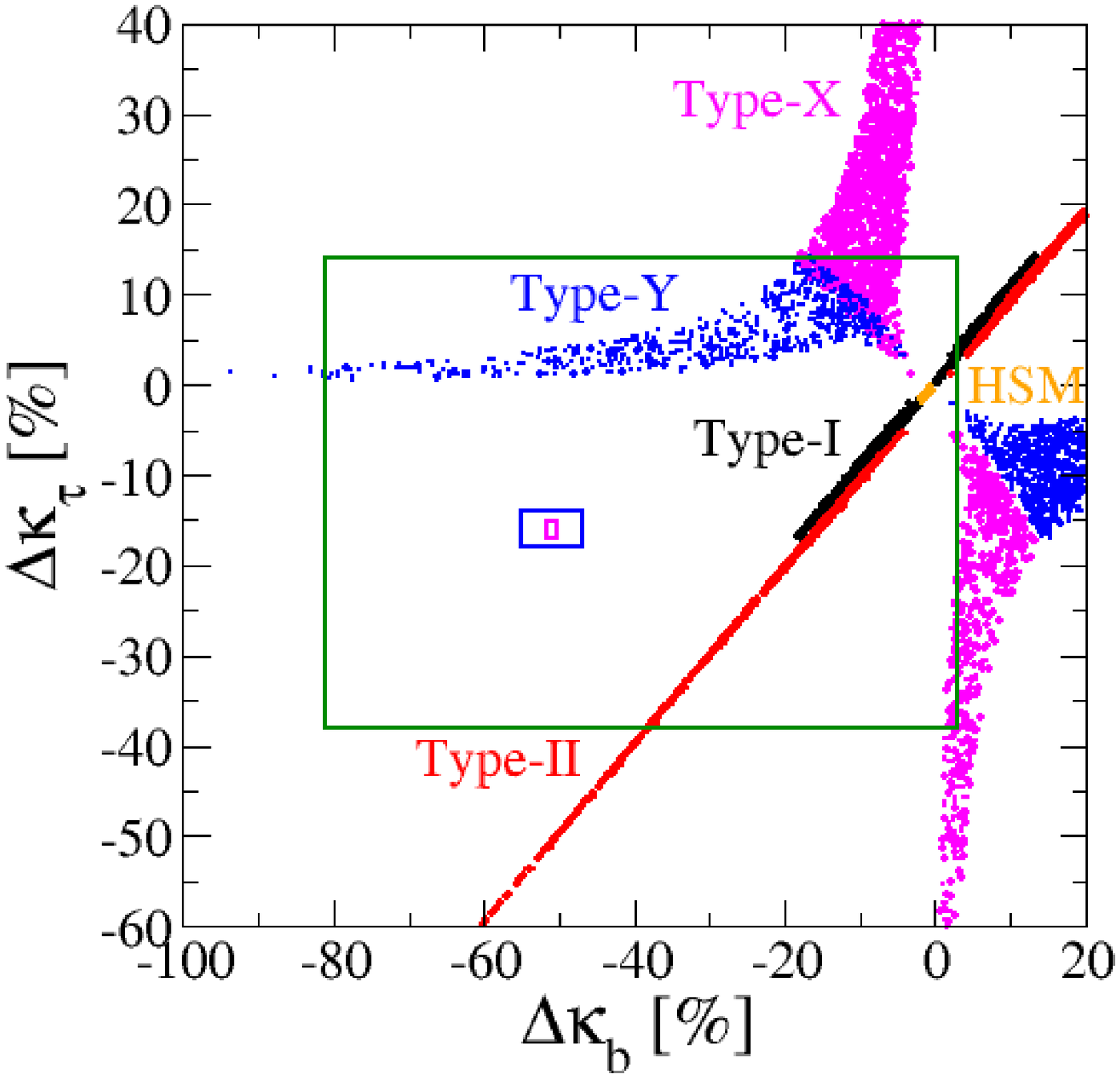} \\ 
\hspace{5mm}
\includegraphics[width=70mm]{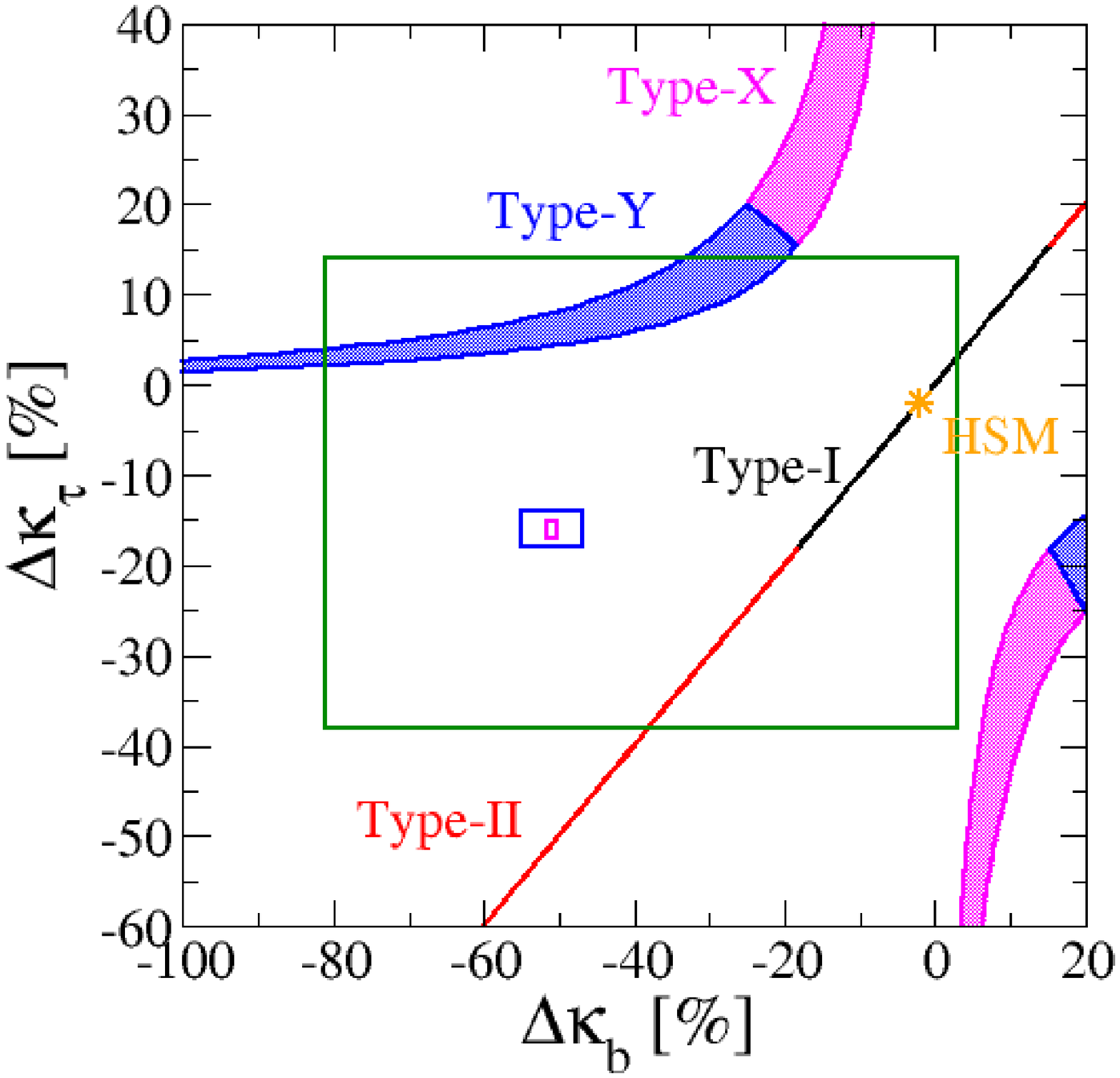} \hspace{-15mm}
\includegraphics[width=70mm]{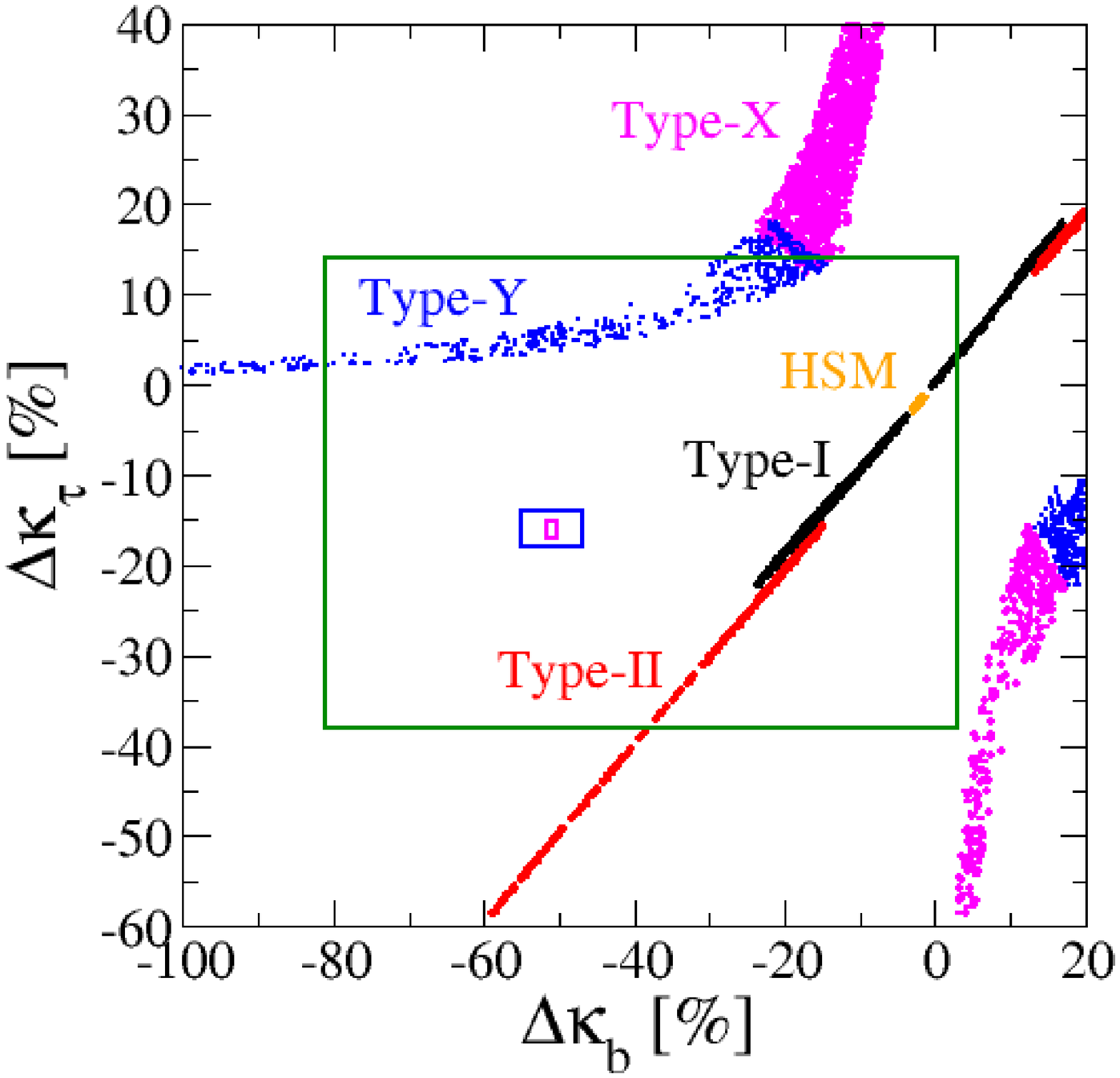} \\
\hspace{5mm}
\includegraphics[width=70mm]{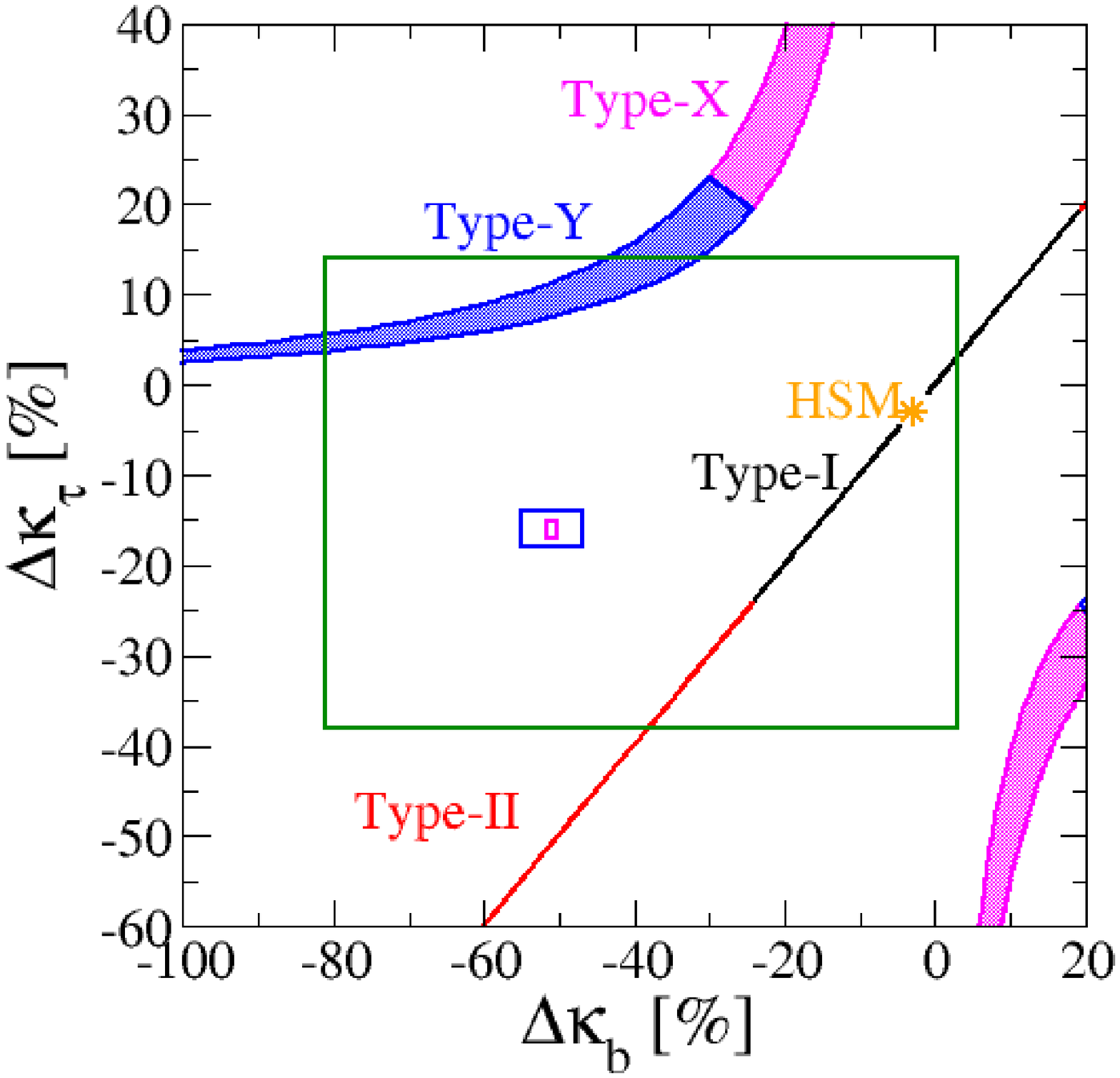} \hspace{-15mm}
\includegraphics[width=70mm]{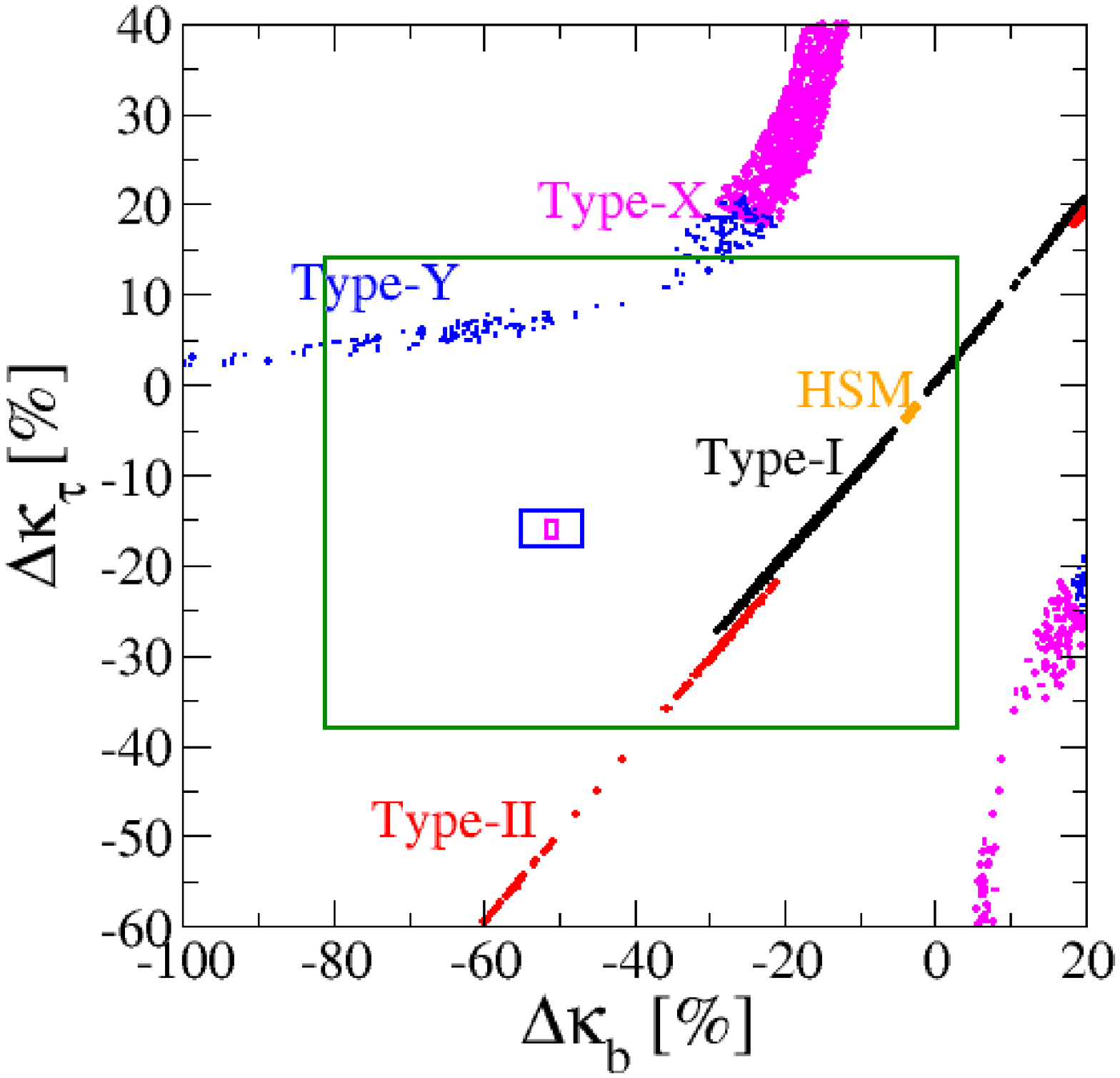} 
\caption{
Correlation between $\Delta \kappa_b^{}$--$\Delta\kappa_\tau$ in the HSM and in the THDMs. 
The left (right) panels show the tree (one-loop) level results. 
The upper, middle and lower panels respectively show the case with $\Delta\kappa_Z^{} = -1\pm 0.58\%$, 
$-2\pm 0.58\%$ and $-3\pm 0.58\%$. 
The region inside the green box is allowed with the 95\% CL from the measurement at the LHC Run-I experiment.
The blue (magenta) box denotes the expected 1$\sigma$ accuracies for the measurement of 
$\Delta \kappa_b^{}$ and $\Delta\kappa_\tau$ at the HL-LHC (ILC), where 
their central values are fixed to be those measured at the LHC Run-I experiment.  
}
\label{kb-ktau2}
\end{center}
\end{figure}

In order to further distinguish models belonging to the same category explained in the above, we need to use other observables such as $\Delta \kappa_b$. 
In Fig.~\ref{kb-ktau2}, we show the correlation between $\Delta \kappa_b$ and $\Delta \kappa_\tau$ in the 5 models. 
The left (right) panels show the tree (one-loop) level results. 
The top, middle and bottom panels display the cases with $\Delta\kappa_Z^{}=-1\pm 0.58\%$, $-2\pm 0.58\%$ and $-3\pm 0.58\%$, respectively, where 
0.58\% corresponds to the expected 1$\sigma$ uncertainty for the measurement of $\Delta \kappa_Z^{}$ by the 
Initial Phase of the ILC program~\cite{ILC1}. We here display the expected 1$\sigma$ accuracies 
for the measurements of $(\Delta \kappa_b^{},\Delta \kappa_\tau^{})$ at the HL-LHC (4\%,2\%)~\cite{Snowmass} denoted by the blue box and 
those at the ILC  with the full data set (0.7\%,0.9\%)~\cite{ILC1} denoted by the magenta box.  

Let us first discuss the tree level results (left panels). 
The predictions of the Type-I and Type-II THDMs are given on the line with $\Delta \kappa_b = \Delta \kappa_\tau$. 
On the other hand, those of the Type-X and Type-Y THDMs are given as a region filled by magenta and blue color, respectively. 
Furthermore, the point denoted by $\ast$ is the prediction of the HSM\footnote{Strictly speaking, the prediction of the HSM is not the point-like shown as $\ast$ in this figure, 
but it is a line segment with the length of $2\sqrt{2}\times 0.58$. }. 
We note that there is no overlapping region between Type-I and Type-II THDMs and that between Type-X and Type-Y THDMs, because we take $\tan\beta > 1$. 
For the case with larger $|\Delta \kappa_Z^{}|$, predictions of four THDMs tend to go more away from the SM prediction, 
i.e., ($\Delta \kappa_b$, $\Delta \kappa_\tau$)=(0,0). 

Next, by looking at the right panels, we can see how the one-loop correction changes the prediction at the tree level. 
The biggest difference can be seen by comparing the top-left and top-right panels. 
Namely, at the tree level
the predictions of the four THDMs are well separated, but 
at the one-loop level there appear overlapping regions at around $(\Delta \kappa_b,\Delta \kappa_\tau$)=(0,0). 
Such behavior happens when $s_{\beta-\alpha} \simeq 1$, in which the tree level difference in the pattern of 
($\Delta \kappa_b$,$\Delta \kappa_\tau$) among four THDMs becomes very small. 
In contrast, for the case with larger $|\Delta \kappa_Z^{}|$, the area of the overlapping region is reduced
as we can see it from the middle-right and bottom-right panels.

Therefore, combining the results given in Figs.~\ref{kv-ktau} and \ref{kb-ktau2}, 
we conclude that the 5 models can be well distinguished by measuring $\Delta \kappa_Z^{}$,  $\Delta \kappa_\tau^{}$ and $\Delta \kappa_b^{}$
as long as $|\Delta \kappa_Z^{}|\gtrsim 1\%$. 

\begin{figure}[!t]
\begin{center}
\hspace{5mm}
\includegraphics[width=75mm]{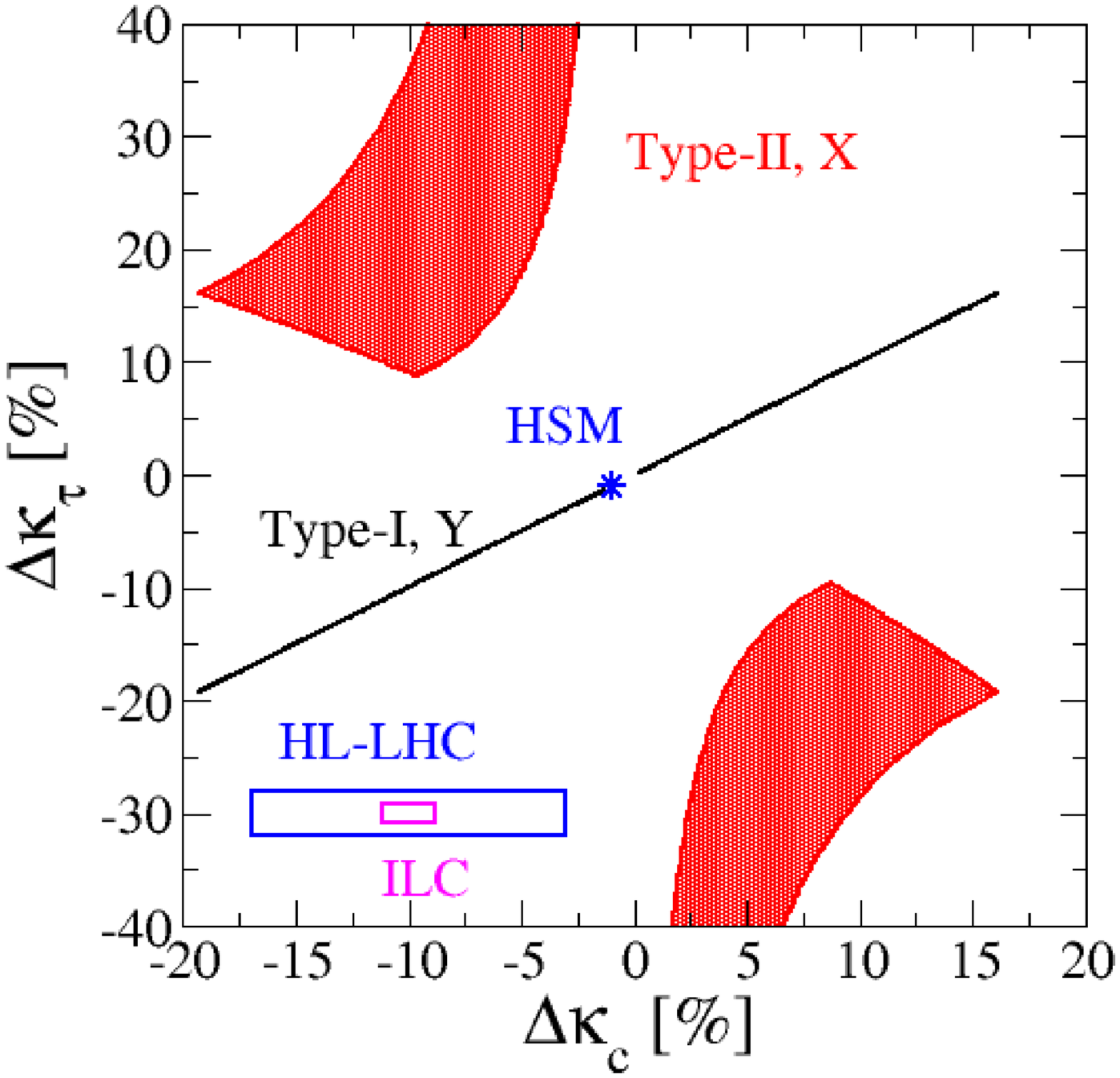} \hspace{-15mm}
\includegraphics[width=75mm]{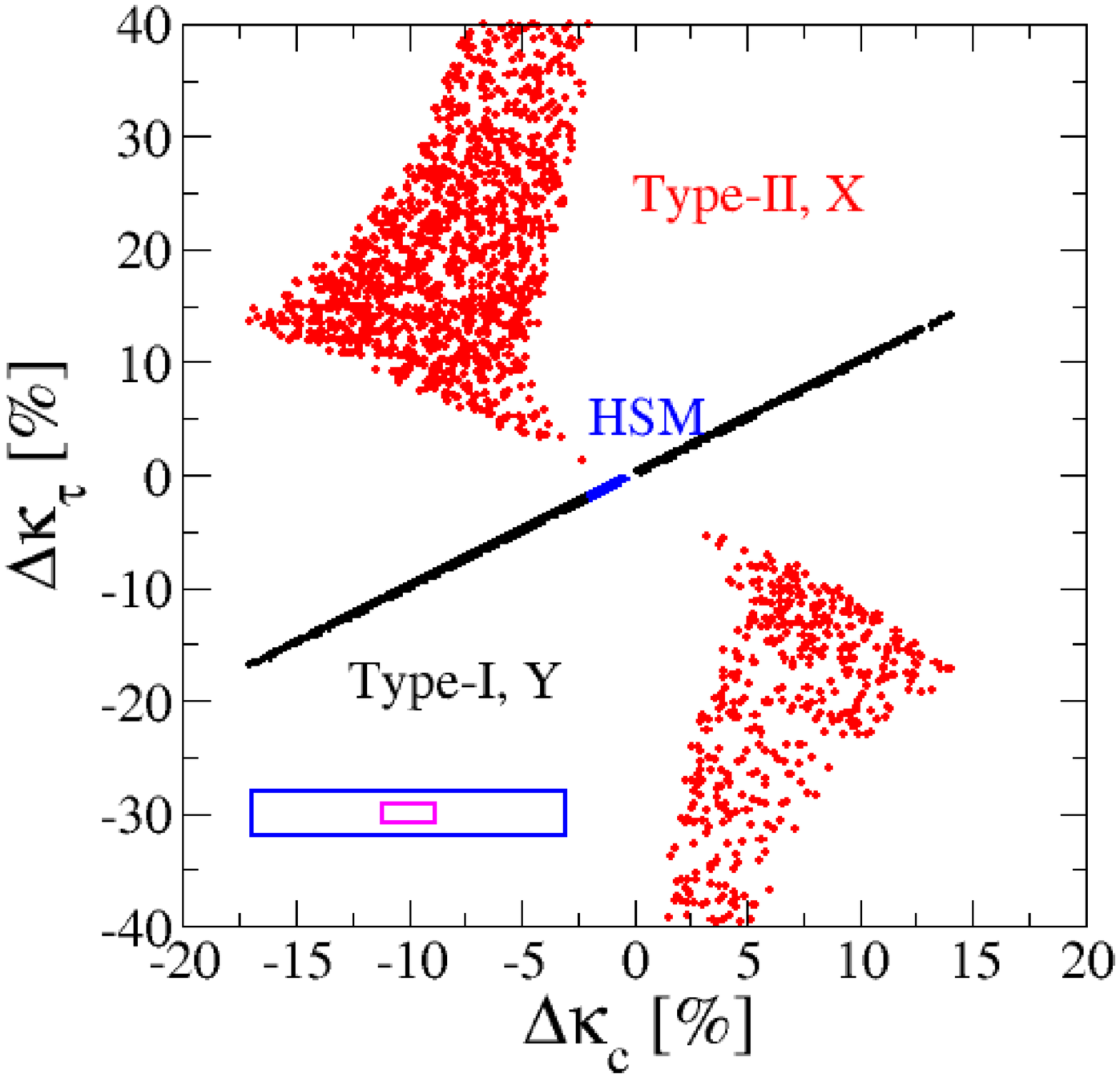} \\ 
\hspace{5mm}
\includegraphics[width=75mm]{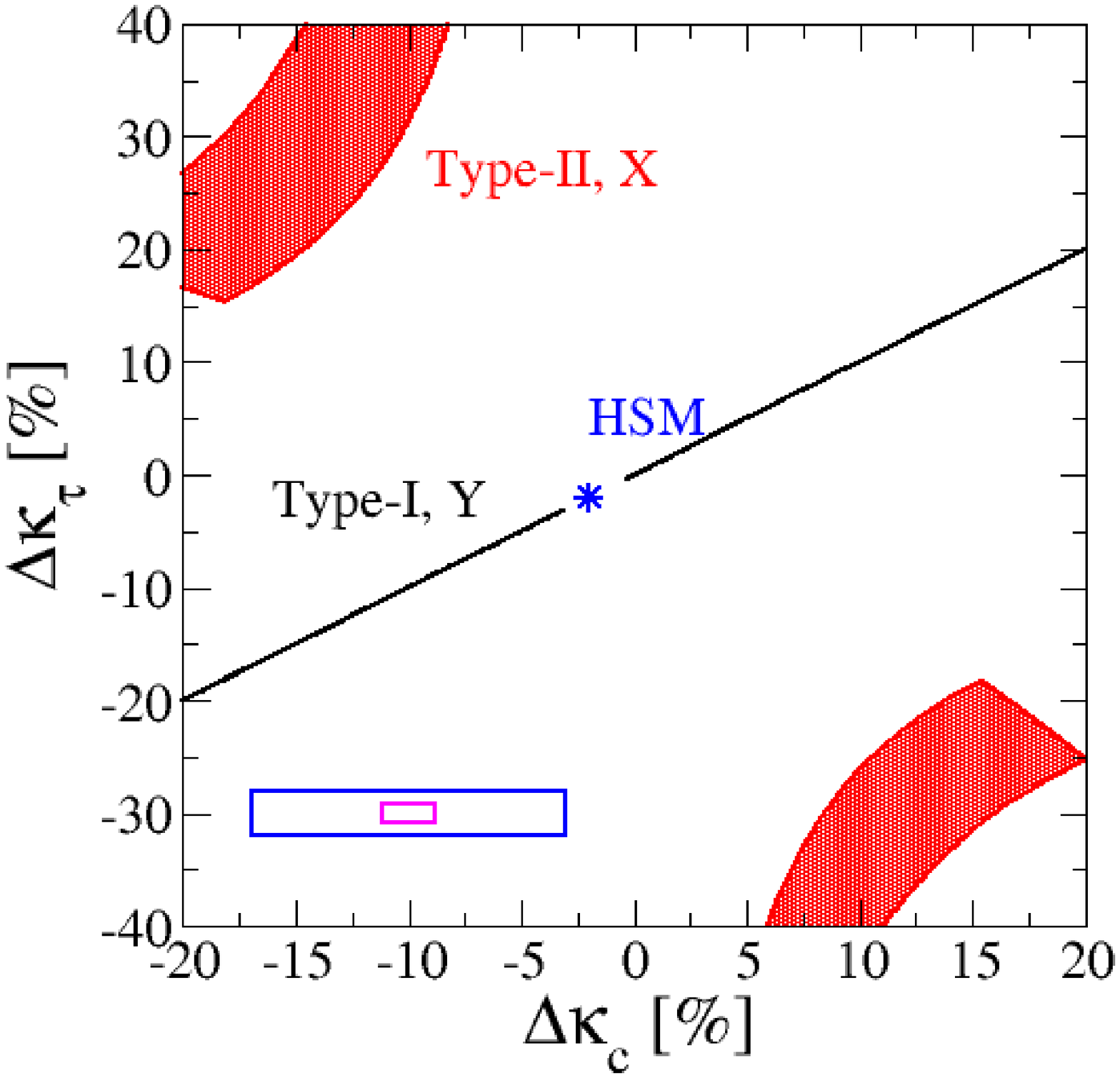} \hspace{-15mm}
\includegraphics[width=75mm]{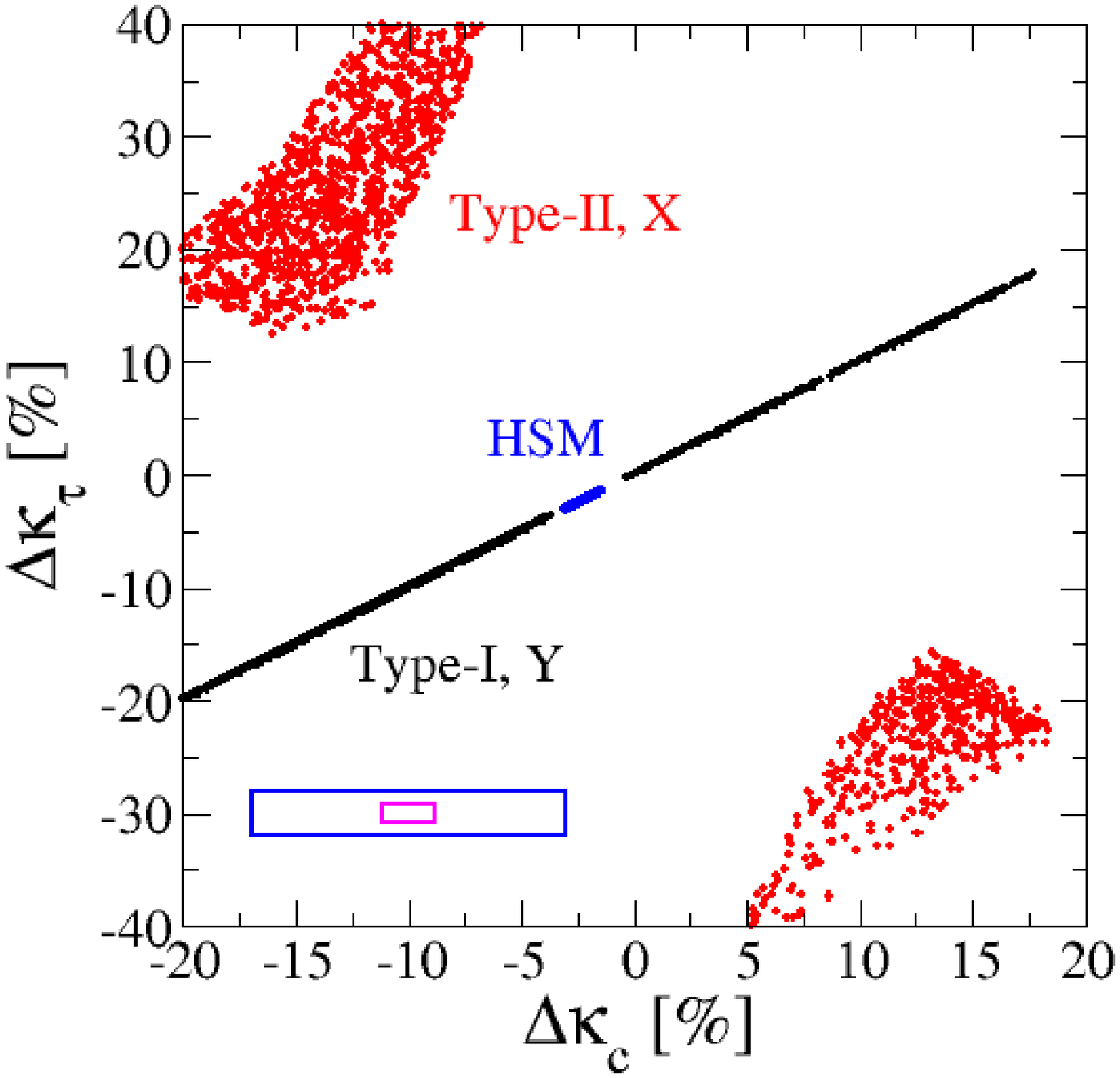} \\
\hspace{5mm}
\includegraphics[width=75mm]{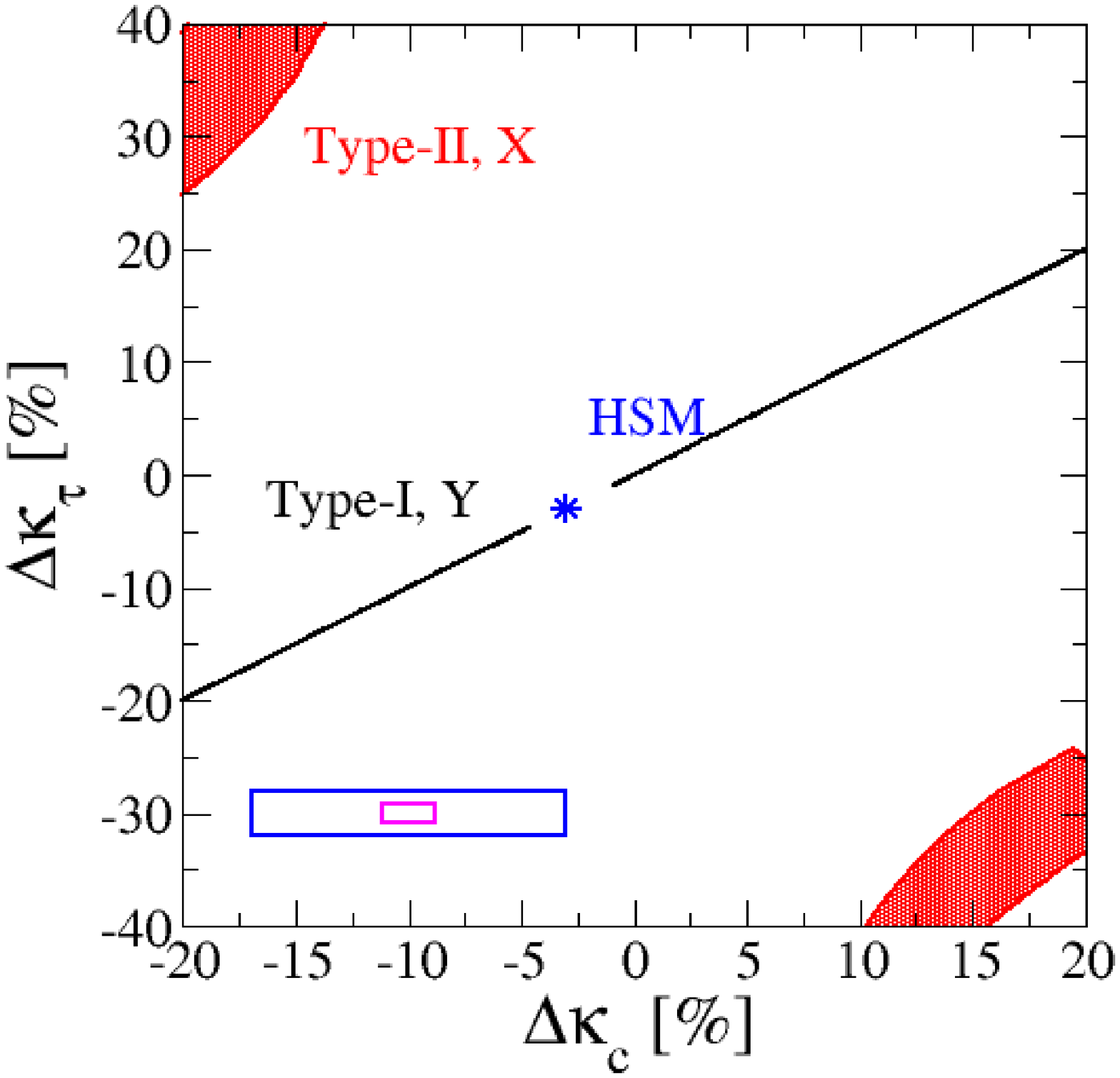} \hspace{-15mm}
\includegraphics[width=75mm]{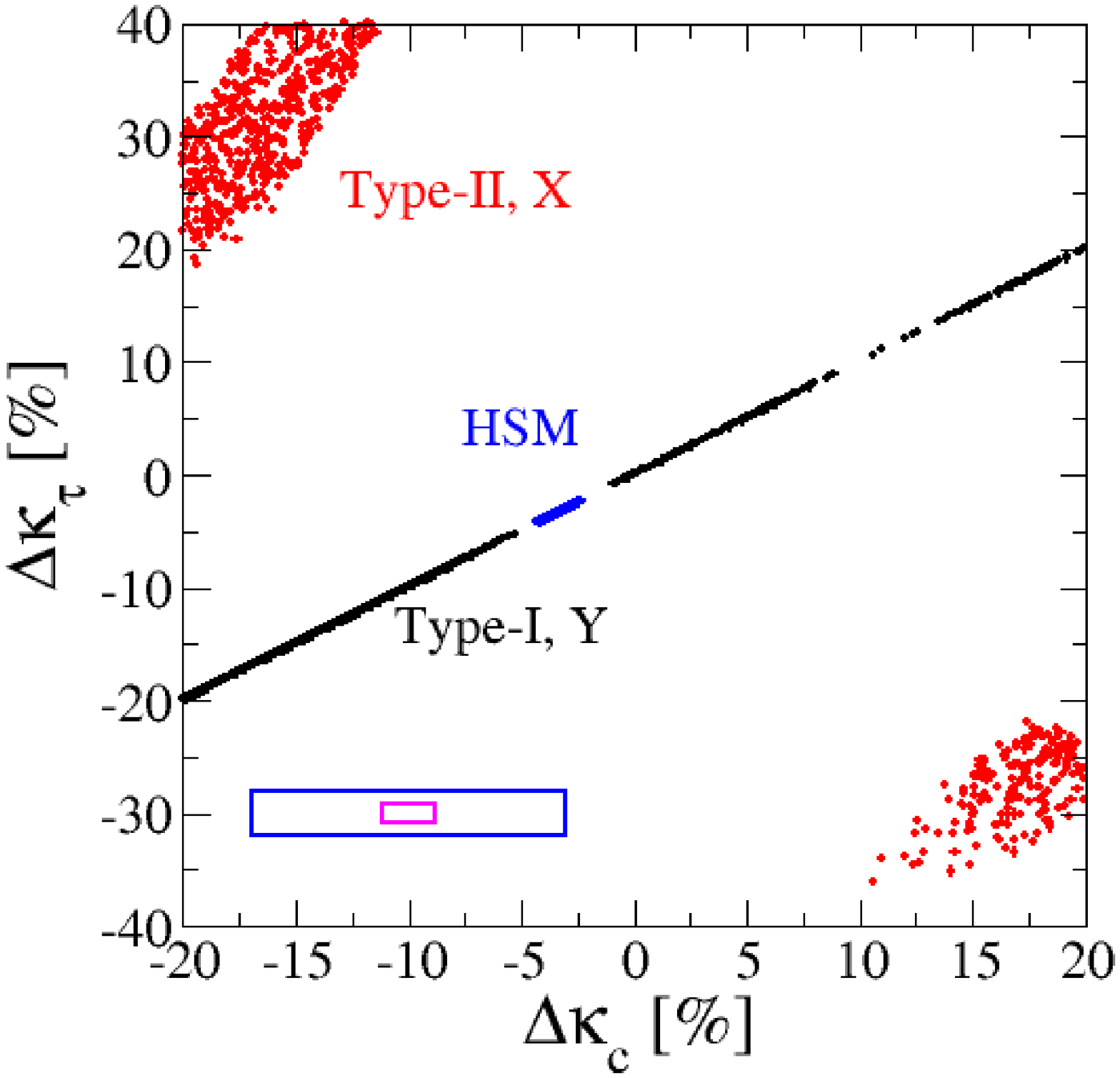} 
\caption{
Correlation between $\Delta \kappa_c^{}$--$\Delta\kappa_\tau$ in the HSM and in the THDMs. 
The left (right) panels show the tree (one-loop) level results. 
The upper, middle and lower panels respectively show the case with $\Delta\kappa_Z^{} = -1\pm 0.58\%$, 
$-2\pm 0.58\%$ and $-3\pm 0.58\%$. 
The blue (magenta) box denotes the expected 1$\sigma$ accuracies for the measurement of 
$\Delta \kappa_c^{}$ and $\Delta\kappa_\tau$ at the HL-LHC (ILC), where 
their central values are not reflected in the current measurements at the LHC.  
}
\label{kc-ktau}
\end{center}
\end{figure}

In Fig.~\ref{kc-ktau}, we show the correlation between $\Delta \kappa_c$ and $\Delta\kappa_\tau$ in the similar way to Fig.~\ref{kb-ktau2}. 
Here, we display the expected 1$\sigma$ accuracies 
for the measurements of $(\Delta \kappa_c^{},\Delta \kappa_\tau^{})$ at the HL-LHC (7\%,2\%)~\cite{Snowmass} denoted by the blue box and 
those at the ILC  with the full data set (1.2\%,0.9\%)~\cite{ILC1} denoted by the magenta box.  
In this plane, the predictions of the Type-I and Type-Y (Type-II and Type-X) THDMs are the same with each other. 

\begin{figure}[t]
\begin{center}\hspace{10mm}
\includegraphics[width=120mm]{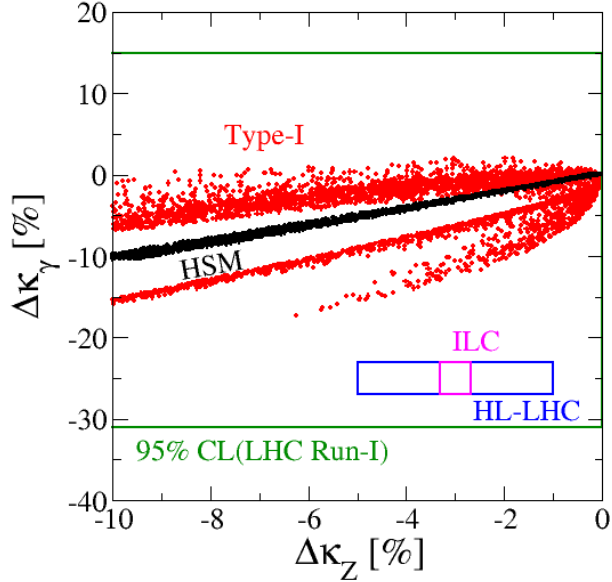} 
\caption{Correlation between $\Delta\kappa_Z^{}$ and $\Delta\kappa_\gamma$ is expressed by black (red) in the HSM (Type-I THDM).
The region inside two green lines is allowed with the 95\% CL from the measurement at the LHC Run-I experiment. 
The blue (magenta) box denotes the expected 1$\sigma$ accuracies for the measurement of 
$\Delta \kappa_Z^{}$ and $\Delta\kappa_\gamma$ at the HL-LHC (ILC), where their central values are not reflected in the current measurements at the LHC.  
}
\label{kv-kgam}
\end{center}
\end{figure}

Finally, we show the correlation between $\Delta \kappa_Z^{}$ and $\Delta\kappa_\gamma$ in Fig.~\ref{kv-kgam}. 
We here only display the results of the Type-I THDM and the HSM. 
The results of the other 3 types of THDMs are almost the same as the result of the Type-I THDM. 
The green lines denotes the current 95\% limit on the $\Delta\kappa_\gamma$ measured by the LHC Run-I experiment~\cite{LHC1}. 
The blue and magenta boxes denote the expected 1$\sigma$ accuracies for the measurement of 
$(\Delta \kappa_Z^{},\Delta\kappa_\gamma)$ at the HL-LHC (2\%,2\%)~\cite{Snowmass} and at the ILC with the full data set (0.31\%,2\%)~\cite{ILC1}, 
where the accuracy of $\Delta\kappa_\gamma$ at the ILC is referred to that given at the HL-LHC, because of its better accuracy. 

We can see that even in the region with $|\Delta \kappa_Z^{}| \lesssim 1\%$, predictions in the THDMs can be largely different from those in the HSM. 
This is because of the fact that the charged Higgs boson loop effect on the $h\gamma\gamma$ vertex in the THDM can be significant, which does not appear in the HSM. 
In addition, the tree level values of $\kappa_t$ and $\kappa_Z^{}$ are generally different in the THDMs, while 
these are common to be $c_\alpha$ in the HSM. 
As a result, in the HSM $\Delta \kappa_\gamma$ is simply given by $c_\alpha - 1$, and the prediction is given around the line with $\Delta\kappa_Z^{}= \Delta \kappa_\gamma$. 
Thus, this result is quite useful to distinguish the THDMs and the HSM even in  the case with $|\Delta \kappa_Z^{}| \lesssim 1\%$, in which 
it is difficult to separate these models only by using $\Delta \kappa_V^{}$ and $\Delta \kappa_{f}^{}$. 

\section{Conclusions \label{sec:con}}

We have computed one-loop corrected Higgs boson couplings based on the improved on-shell renormalization scheme without gauge dependence 
in the non-minimal Higgs sectors, i.e., the HSM and the THDMs with the softly-broken $Z_2$ symmetry. 
The pinch technique is adopted to remove gauge dependence in Higgs boson two-point functions, which give rise to the gauge dependence in the renormalized 
mixing parameters between Higgs bosons. 
We have explicitly shown the cancelation of the gauge dependence in the general $R_\xi$ gauge in the non-minimal Higgs sectors. 
We then have calculated the difference in various renormalized Higgs boson couplings calculated in the previous on-shell scheme with gauge dependence and those 
calculated in the improved scheme without gauge dependence. 

Having the gauge invariant one-loop corrected coupling constants,
we have investigated how we can identify the HSM and the THDMs by looking at the difference in the pattern of deviations in the renormalized Higgs boson couplings from predictions 
in the SM. 
We have shown correlations between $\Delta\kappa_Z^{}$--$\Delta\kappa_\tau$, $\Delta\kappa_\tau$--$\Delta\kappa_b$, $\Delta\kappa_\tau$--$\Delta\kappa_c$ and $\Delta\kappa_Z$--$\Delta\kappa_\gamma$. 
We can distinguish these models by the combination of the measurements of $\kappa_\tau$, $\kappa_b$ and $\kappa_c$ if 
$|\Delta\kappa_Z^{}|$ is measured to be $\sim 1\%$ or larger at future collider experiments.

\vspace*{4mm}
\noindent
\section*{Acknowledgments}
\noindent

SK's work was supported, in part, by Grant-in-Aid for Scientific Research on Innovative Areas, the Ministry of Education, Culture, Sports, Science and Technology 
No. 16H06492, Grant H2020-MSCA-RISE-2014 no. 645722 (Non Minimal Higgs). 
MK was supported by MOST 106-2811-M-002-010. 

\begin{appendix}

\section{Pinch-term in the 't~Hooft-Feynman gauge \label{sec:fg}}

We present the analytic expressions for the pinch-term contribution to the scalar boson two-point functions in the 't~Hooft-Feynman gauge. 
As expressed in Eq.~(\ref{sse}), 
the gauge invariant two-point function for scalar bosons $\varphi_{i,j}$ is obtained in the pinched tadpole scheme as follows:
\begin{align}
\Pi_{\varphi_i \varphi_j}(q^2) = \Pi_{\varphi_i \varphi_j}^{\text{1PI}}(q^2)\Big|_{\xi_V^{} = 1} 
+\Pi_{\varphi_i \varphi_j}^{\text{Tad}}\Big|_{\xi_V^{} = 1} + \Pi_{\varphi_i \varphi_j}^{\text{PT}}(q^2)\Big|_{\xi_V^{} = 1}. 
\end{align}
In the following subsections, we give the explicit formulae for $\Pi_{\varphi_i \varphi_j}^{\text{PT}}(q^2)\Big|_{\xi_V^{} = 1}$ 
in the SM, the HSM and the THDM in order. Hereafter, we do not explicitly write the symbol $\big|_{\xi_V^{} = 1}$. 

\subsection{SM}

The pinch-term for the Higgs boson $h$ two-point function is given as  
\begin{align}
\Pi_{hh}^{\text{PT}}(q^2) = -\frac{g^2}{16\pi^2}(q^2-m_h^2)B_0(q^2;W,W)
-\frac{g_Z^2}{32\pi^2}(q^2-m_h^2)B_0(q^2;Z,Z). 
\end{align}

\subsection{HSM}

The pinch-terms for the two-point functions for the CP-even Higgs bosons  $h_i$--$h_j$ are given~as 
\begin{align}
\Pi_{h_ih_j}^{\text{PT}}(q^2)&= -\frac{g^2}{32\pi^2}(2q^2-m_{h_i}^2-m_{h_j}^2)\zeta_i \zeta_j B_0(q^2;W,W)\notag\\
&-\frac{g_Z^2}{64\pi^2}(2q^2-m_{h_i}^2-m_{h_j}^2)\zeta_i \zeta_j B_0(q^2;Z,Z), 
\end{align}
where $\zeta_{i,j}$ and $h_{i,j}$ ($i,j=1,2$) are defined in Eq.~(\ref{hsm_short}). 

\subsection{THDM}

The pinch-terms for the two-point functions for the CP-even Higgs bosons $h$--$h$, $H$--$H$ and $H$--$h$ are given as 
\begin{align}
\Pi_{hh}^{\text{PT}}(q^2) &= -\frac{g^2}{16\pi^2}(q^2-m_h^2)[s_{\beta-\alpha}^2 B_0(q^2;W,W)+c_{\beta-\alpha}^2 B_0(q^2;H^\pm,W) ] \notag\\
&-\frac{g_Z^2}{32\pi^2}(q^2-m_h^2)[s_{\beta-\alpha}^2 B_0(q^2;Z,Z)+c_{\beta-\alpha}^2 B_0(q^2;A,Z) ], \\
\Pi_{HH}^{\text{PT}}(q^2) &= -\frac{g^2}{16\pi^2}(q^2-m_H^2)[c_{\beta-\alpha}^2 B_0(q^2;W,W)+s_{\beta-\alpha}^2 B_0(q^2;H^\pm,W) ] \notag\\
&-\frac{g_Z^2}{32\pi^2}(q^2-m_H^2)[c_{\beta-\alpha}^2 B_0(q^2;Z,Z)+s_{\beta-\alpha}^2 B_0(q^2;A,Z) ] , \\
\Pi_{Hh}^{\text{PT}}(q^2) &= \frac{g^2}{32\pi^2}(2q^2-m_h^2-m_H^2)s_{\beta-\alpha}c_{\beta-\alpha} [B_0(q^2;H^\pm,W)-B_0(q^2;W,W) ] \notag\\
&+\frac{g_Z^2}{64\pi^2}(2q^2-m_h^2-m_H^2)s_{\beta-\alpha}c_{\beta-\alpha} [B_0(q^2;A,Z)-B_0(q^2;Z,Z) ], 
\end{align}
where $\Pi_{hH}^{\text{PT}}(q^2)=\Pi_{Hh}^{\text{PT}}(q^2)$. 
Those for the CP-odd scalar bosons $A$--$A$ and $A$--$G^0$, we obtain: 
\begin{align}
\Pi_{AA}^{\text{PT}}(q^2) &= -\frac{g^2}{16\pi^2}(q^2-m_A^2) B_0(q^2;W,H^\pm)\notag\\
&-\frac{g_Z^2}{32\pi^2}(q^2-m_A^2)[c_{\beta-\alpha}^2 B_0(q^2;Z,h)+s_{\beta-\alpha}^2 B_0(q^2;Z,H) ], \\ 
\Pi_{AG^0}^{\text{PT}}(q^2) &= \frac{g_Z^2}{64\pi^2}(2q^2-m_A^2)s_{\beta-\alpha}c_{\beta-\alpha}[B_0(q^2;Z,H)- B_0(q^2;Z,h) ]. \label{pt_ag}
\end{align}
where $\Pi_{G^0A}^{\text{PT}}(q^2)=\Pi_{AG^0}^{\text{PT}}(q^2)$. 
Those for the charged scalar bosons $H^+H^-$ and $H^+G^-$, we obtain:  
\begin{align}
\Pi_{H^+H^-}^{\text{PT}}(q^2) &= -\frac{g^2}{32\pi^2}\left(q^2-m_{H^\pm}^2\right)\left[s_{\beta-\alpha}^2B_0(q^2;W,H)+c_{\beta-\alpha}^2B_0(q^2;W,h) \right]\notag\\
&-\frac{g^2}{32\pi^2}\left(q^2-m_{H^\pm}^2\right)B_0(q^2;W,A) \notag\\
&-\frac{g_Z^2}{32\pi^2}(1-2s_W^2)^2\left(q^2-m_{H^\pm}^2\right)B_0(q^2;Z,H^\pm)-\frac{e^2}{8\pi^2}\left(q^2-m_{H^\pm}^2\right)B_0(q^2;\gamma,H^\pm), \\
\Pi_{H^+G^-}^{\text{PT}}(q^2) &= \frac{g^2}{32\pi^2}s_{\beta-\alpha}c_{\beta-\alpha}\left(2q^2-m_{H^\pm}^2\right)\left[B_0(q^2;W,H)-B_0(q^2;W,h) \right], 
\end{align}
where $\Pi_{G^+H^-}^{\text{PT}}(q^2)=\Pi_{H^+G^-}^{\text{PT}}(q^2)$. 

\section{Renormalized Higgs boson vertices in the pinched tadpole scheme \label{sec:rhiggs}}

In this Appendix, we give the expressions for the 
renormalized Higgs boson vertices in the pinched tadpole scheme in the SM, the HSM and the THDM in order. 
The expressions for the counter terms $\delta X$ appearing in these vertices are given in App.~\ref{sec:counter}.

\subsection{SM}

The renormalized $hVV$, $hf\bar{f}$ and $hhh$ vertices are given by
\begin{align}
\hat{\Gamma}_{hVV}^{1} &=  \frac{2m_V^2}{v}\left[1+\left(\frac{\delta m_V^2}{m_V^2} -\frac{\delta v}{v} +\delta Z_V +\frac{1}{2}\delta Z_h \right)\right]
+\Gamma_{hVV}^{\text{1PI}} + \Gamma_{hVV}^{\text{Tad}}, \\
\hat{\Gamma}_{hff}^{S} &=
 -\frac{m_f}{v}\left[1+\left(
\frac{\delta m_f}{m_f} -\frac{\delta v}{v}+  \delta Z_V^f + \frac{1}{2}\delta Z_h\right) \right]+ \Gamma_{hff}^{\text{1PI}}, \\
\hat{\Gamma}_{hhh} &=   - \frac{3m_h^2}{v}\left[1+\left(
  \frac{\delta m_h^2}{m_h^2} - \frac{\delta v}{v}
  + \frac{3}{2}\delta Z_h 
  \right)\right] + \Gamma_{hhh}^\textrm{1PI}  + \Gamma_{hhh}^\text{Tad}. 
  \end{align}

\subsection{HSM}

The renormalized $h_i VV$, $h_i f\bar{f}$ and $h_ihh$ ($h_1 = h$ and $h_2 = H$) vertices are given by 
\begin{align}
\hat{\Gamma}_{h_iVV}^{1} &= 
\frac{2m_V^2}{v}\zeta_i\left[1+
\left(
\frac{\delta m_V^2}{m_V^2} -\frac{\delta v}{v} + \tan\alpha\, \delta C_h + \delta Z_V + \frac{\delta Z_{h_i}}{2} \right) \right] 
+ \Gamma_{h_i VV}^{\text{1PI}} + \Gamma_{h_iVV}^{\text{Tad}}, \\
\hat{\Gamma}_{h_iff}^{S}& = 
 -\frac{m_f}{v}\zeta_i\left[
1+\left(
\frac{\delta m_f}{m_f} -\frac{\delta v}{v} + \tan\alpha\, \delta C_h + \delta Z_V^f + \frac{\delta Z_{h_i}}{2}\right)\right] + \Gamma_{h_iff}^{\text{1PI}}, \\
 \hat{\Gamma}_{hhh} &=
  6\lambda_{hhh}\left[1 +\frac{\delta \lambda_{hhh}}{\lambda_{hhh}}  + \frac{3}{2} \delta Z_h
    + \frac{\lambda_{Hhh}}{\lambda_{hhh}} (\delta C_h +\delta \alpha )    \right]
  + \Gamma_{hhh}^\text{1PI} + \Gamma_{hhh}^{\text{Tad}}, \label{hhh-hsm2}\\
  \hat{\Gamma}_{Hhh} &=
 2\lambda_{Hhh}\left[1 + \delta Z_h + \frac{\delta Z_H}{2} 
    +\frac{ 3\lambda_{hhh}}{\lambda_{Hhh}}(\delta C_h-\delta\alpha) + \frac{2\lambda_{HHh}}{\lambda_{Hhh}}(\delta C_h+\delta\alpha) +\frac{\delta \lambda_{Hhh}}{\lambda_{Hhh}}
    \right] \notag\\
&+ \Gamma_{Hhh}^\text{1PI} + \Gamma_{Hhh}^{\text{Tad}}, \label{bHhh-hsm2} 
\end{align}
where $\zeta_{i,j}$ ($i,j=1,2$) are defined in Eq.~(\ref{hsm_short}). 
For the $hhh$ and $Hhh$ vertices, 
the relevant scalar boson trilinear couplings defined as ${\cal L} = +\lambda_{\phi_i\phi_j\phi_k}\phi_i\phi_j\phi_k \cdots$ 
are given by
\begin{align}
   \lambda_{hhh} & = -\frac{c_\alpha^3}{2v}m_h^2
                 - s_\alpha^2( c_\alpha^{}\lambda_{\Phi S}v
                            - s_\alpha^{} \mu_S^{} ), \\
 \lambda_{Hhh} &= 
 - (2m_h^2 + m_H^2 )\frac{s_\alpha c_\alpha^2}{2v}
 + \frac{s_\alpha \lambda_{\Phi S}^{} v}{2}(1 + 3c_{2\alpha}^{})
 - 3s_\alpha^2 c_\alpha^{} \mu_S^{}, \\
 \lambda_{HHh} &= 
 - (m_h^2 + 2m_H^2 )\frac{c_\alpha s_\alpha^2}{2v}
 - \frac{\lambda_{\Phi S}^{} v}{4}(c_\alpha + 3c_{3\alpha}^{})
 + 3c_\alpha^2 s_\alpha^{} \mu_S^{}. 
  \end{align}
The explicit formulae for the 1PI diagram contributions are given in Refs.~\cite{HSM-KKY1,HSM-KKY2}.

\subsection{THDM\label{sec:bthdm}}

First, we give the renormalized $hVV$ and $HVV$ vertices:
\begin{align}
\hat{\Gamma}_{hVV}^{1} &= 
\frac{2m_V^2}{v}
\left[1 + s_{\beta-\alpha}\left(
\frac{\delta m_V^2}{m_V^2} -\frac{\delta v}{v}  + \delta Z_V + \frac{\delta Z_h}{2}\right) 
+c_{\beta-\alpha}\left(\delta C_h+ \delta \beta \right)
\right]  + \Gamma_{hVV}^{\text{1PI}} + \Gamma_{hVV}^{\text{Tad}}, \\
\hat{\Gamma}_{HVV}^{1} &= 
\frac{2m_V^2}{v}
\left[1 + c_{\beta-\alpha}\left(
\frac{\delta m_V^2}{m_V^2} -\frac{\delta v}{v}  + \delta Z_V + \frac{\delta Z_H}{2}\right) 
+s_{\beta-\alpha}\left(\delta C_h -\delta \beta \right)
\right] + \Gamma_{HVV}^{\text{1PI}} + \Gamma_{HVV}^{\text{Tad}}, 
\end{align}
where the explicit formulae for $\Gamma_{hVV}^{\text{1PI}}$ are given in Ref.~\cite{HSM-KKY2}. 

Second, the renormalized Yukawa couplings are given by 
\begin{align}
\hat{\Gamma}_{hff}^S& = 
-\frac{m_f}{v}\zeta_{hff}
\left(1 +
\frac{\delta m_f}{m_f} -\frac{\delta v}{v} -\zeta_f \delta \beta+\delta Z_V^f + \frac{\delta Z_h}{2}  +\frac{\zeta_{Hff}}{\zeta_{hff}}\delta C_h \right) + \Gamma_{hff}^{\text{1PI}}, \\
\hat{\Gamma}_{Hff}^{S}& = 
-\frac{m_f}{v}\zeta_{Hff}
\left(1+
\frac{\delta m_f}{m_f} -\frac{\delta v}{v} -\zeta_f \delta \beta+\delta Z_V^f + \frac{\delta Z_H}{2}  +\frac{\zeta_{hff}}{\zeta_{Hff}}\delta C_h \right) + \Gamma_{Hff}^{\text{1PI}},\\
\hat{\Gamma}_{Aff}^{P} & = 
2i\frac{m_f}{v}\zeta_fI_f
\left(1+ 
\frac{\delta m_f}{m_f} -\frac{\delta v}{v} -\zeta_f\delta \beta+\delta Z_A^f + \frac{\delta Z_A}{2}  +\frac{\delta C_A}{\zeta_f} \right) + \Gamma_{Aff}^{\text{1PI}}, \\
\hat{\Gamma}_{H^+\bar{u}_Ld_R}^{R}& = 
-\frac{\sqrt{2}m_d}{v}\zeta_d
\left(1+
\frac{\delta m_d}{m_d} -\frac{\delta v}{v} -\zeta_d\delta \beta+\frac{\delta Z_{dR}+\delta Z_{uL}}{2} + \frac{\delta Z_{H^\pm}}{2}  +\frac{\delta C_{H^\pm}}{\zeta_d} \right) \notag\\
&+ \Gamma_{{H^+\bar{u}_Ld_R}}^{\text{1PI}},\\
\hat{\Gamma}_{H^+\bar{u}_Rd_L}^{L}& = 
\frac{\sqrt{2}m_u}{v}\zeta_u
\left(1+
\frac{\delta m_u}{m_u} -\frac{\delta v}{v} -\zeta_u\delta \beta+\frac{\delta Z_{dL}+\delta Z_{uR}}{2} + \frac{\delta Z_{H^\pm}}{2}  +\frac{\delta C_{H^\pm}}{\zeta_u} \right) \notag\\
&+ \Gamma_{{H^+\bar{u}_Rd_L}}^{\text{1PI}},
\end{align}
where $\hat{\Gamma}_{H^+ ff'}^{R,L} =\hat{\Gamma}_{H^+ ff'}^{S} \pm \hat{\Gamma}_{H^+ ff'}^{P}$. 
The $\zeta_{\phi ff}$ ($\phi=h,H$) and $\zeta_f$ factors are respectively given in Eq.~(\ref{zeta-hff}) and in Tab.~\ref{yukawa_tab}. 
The explicit formulae for $\Gamma_{hff}^{\text{1PI}}$ are given in Ref.~\cite{HSM-KKY2}. 

Third, the renormalized $hH^\pm W_\mu^\mp$ and $hA Z_\mu$ vertices are given by 
\begin{align}
\hat{\Gamma}_{hH^\pm W } & = \mp i\frac{m_W^{}}{v}c_{\beta-\alpha}\left[1+\frac{\delta m_W^2}{2m_W^2}-\frac{\delta v}{v} +\frac{1}{2}\left(\delta Z_h +\delta Z_{H^\pm}+\delta Z_W\right)
+\tan(\beta-\alpha)(\delta C_{H^\pm}-\delta C_h)\right], \\
\hat{\Gamma}_{hA Z } & = -\frac{m_Z^{}}{v}c_{\beta-\alpha}\left[1+\frac{\delta m_Z^2}{2m_Z^2}-\frac{\delta v}{v} +\frac{1}{2}\left(\delta Z_h +\delta Z_{A}+\delta Z_Z\right)
+\tan(\beta-\alpha)(\delta C_{A}-\delta C_h)\right]. 
\end{align}

Finally, the renormalized scalar trilinear vertices $\hat{\Gamma}_{hhh}$ and $\hat{\Gamma}_{Hhh}$ are 
expressed by the same form as those given in Eqs.~(\ref{hhh-hsm2}) and (\ref{bHhh-hsm2}) in the HSM, where 
the explicit expression for $\Gamma_{hhh}^{\text{1PI}}$ is given in Ref.~\cite{THDM-KKY2}. 
In addition, the relevant scalar trilinear couplings are given by 
\begin{align}
  \lambda_{hhh}
&=-\frac{m_h^2}{2v}s_{\beta-\alpha} +\frac{M^2-m_h^2}{v}s_{\beta-\alpha}c_{\beta-\alpha}^2+\frac{M^2-m_h^2}{2v}c_{\beta-\alpha}^3(\cot\beta-\tan\beta),\label{lam_hhh}\\
\lambda_{Hhh}&=-\frac{c_{\beta-\alpha}}{2v s_{2\beta}^{}}\left[(2m_h^2+m_H^2)s_{2\alpha}^{}+M^2(s_{2\beta}^{}-3s_{2\alpha}^{}) \right], \\  \label{lam_bhhh}
\lambda_{HHh}&=-\frac{s_{\beta-\alpha}}{2vs_{2\beta}}[(m_h^2 + 2m_H^2)s_{2\alpha} - M^2(s_{2\beta} + 3s_{2\alpha})].  
  \end{align}

\section{Counter terms \label{sec:counter}}

We present the explicit formulae for the relevant counter terms appearing in the previous subsection, which 
are determined in the pinched tadpole scheme~\cite{Fleischer}. 
We also explain the way to obtain the counter terms determined in the KOSY scheme~\cite{KOSY}. 

\subsection{SM}

Counter terms for the gauge boson masses $\delta m_V^2$, the VEV $\delta v$ and the wave function renormalizations of weak gauge bosons $\delta Z_V$ are given by 
\begin{align}
 \delta m_V^2 & = \Pi_{VV}^\textrm{}(m_V^2), \\
  \frac{\delta v}{v}&= \frac{1}{2}
\left[
\frac{s_W^2-c_W^2}{s_W^2}\frac{\Pi_{WW}^{\text{}}(m_W^2)}{m_W^2}+\frac{c_W^2}{s_W^2}\frac{\Pi_{ZZ}^{\text{}}(m_Z^2)}{m_Z^2}
-\frac{d}{dp^2}\Pi_{\gamma\gamma}^{\text{}}(p^2)\big|_{p^2=0}^{} \right],  \label{del_vev} \\
\delta Z_Z &= -\frac{d}{dp^2}\Pi_{\gamma\gamma}^\text{1PI}(p^2)\big|_{p^2=0}^{}-\frac{2(c_W^2-s_W^2)}{c_Ws_W}
\frac{\Pi_{Z\gamma}^{\text{1PI}}(0)}{m_Z^2} 
 +\frac{c_W^2-s_W^2}{s_W^2} \left[\frac{\Pi_{ZZ}^{\text{1PI}}(m_Z^2)}{m_Z^2}-\frac{\Pi_{WW}^{\text{1PI}}(m_W^2)}{m_W^2}\right], \label{delZzg}\\
\delta Z_W &=  -\frac{d}{dp^2}\Pi_{\gamma\gamma}^{\text{1PI}}(p^2)\big|_{p^2=0}
-\frac{2c_W}{s_W}\frac{\Pi_{Z\gamma}^{\text{1PI}}(0)}{m_Z^2} 
+\frac{c_W^2}{s_W^2}\left[\frac{\Pi_{ZZ}^{\text{1PI}}(m_Z^2)}{m_Z^2}-\frac{\Pi_{WW}^{\text{1PI}}(m_W^2)}{m_W^2}\right],
\end{align}
where $\Pi_{ij}^{}$ are the gauge invariant two-point functions defined in Eq.~(\ref{sse}), and 
$\Pi_{ij}^{\text{1PI}}$ are the part of the 1PI diagram contribution to the two-point functions. 
Counter terms for fermion masses $\delta m_f$ and the wave function renormalization of fermions ($\delta Z_V^f$ and $\delta Z_A^f$) are given by   
\begin{align}
  \delta m_f^{} & = m_f \left[
  \Pi_{ff,V}^\textrm{}(m_f^2) + \Pi_{ff,S}^\textrm{}(m_f^2)
    \right],\\
  \delta Z_V^f &=-\Pi_{ff,V}^{\text{1PI}}(m_f^2) -\Pi_{ff,V}^{\text{Tad}}
  -2m_f^2\left[\frac{d}{dp^2}\Pi_{ff,V}^{\text{1PI}}(p^2)\big|_{p^2=m_f^2}+\frac{d}{dp^2}\Pi_{ff,S}^{\text{1PI}}(p^2)\Big|_{p^2=m_f^2}\right],       \\
  \delta Z_A^f & =- \Pi_{ff,A  }^\textrm{1PI}(m_f^2)
  - \Pi_{ff,A  }^\textrm{Tad}
  + 2m_f^2 \frac{d}{dp^2}\Pi_{ff,A}^\textrm{1PI}(p^2)\big|_{p^2=m_f^2},
\end{align}
where $\Pi_{ff,V}$,~$\Pi_{ff,A}^{}$ and $\Pi_{ff,S}$ are the vector, the axial vector 
and the scalar parts of the fermion two-point functions:
\begin{align}
  \Pi_{ff} = \slashed{p}\Pi_{ff,V} - \slashed{p}\gamma_5 \Pi_{ff,A} + m_f^{} \Pi_{ff,S}.
  \end{align}
We note that the wave function renormalizations for left-handed ($\delta Z_L^f$) and right-handed ($\delta Z_R^f$) fermions are
related to $\delta Z_V^f$ and $\delta Z_A^f$ as follows:
\begin{align}
  \delta Z_L^f = \delta Z_V^f + \delta Z_A^f, \quad
  \delta Z_R^f = \delta Z_V^f - \delta Z_A^f.
  \end{align}

Counter terms for the Higgs boson mass $\delta m_h^2$ and the wave function renormalization for the Higgs boson $\delta Z_h$ are expressed as
\begin{align}
  \delta m_h^2 & = \Pi_{hh}^\textrm{}(m_h^2),  \,\,\,
  \delta Z_h^{} 
  = - \frac{d}{dp^2}\Pi_{hh}^\textrm{1PI}(p^2)
  \big|_{p^2= m_h^2}. 
\end{align}

In the following, we also present the expressions for the counter terms in the KOSY scheme~\cite{KOSY}, which are necessary to calculate the 
scheme difference discussed in Sec.~\ref{sec:inv}. 
In the KOSY scheme, two-point functions for gauge bosons and fermions are obtained by subtracting the tadpole diagram contribution as 
\begin{align}
\Pi_{ij}(p^2)\big|_{\text{KOSY}} = \Pi_{ij}^\textrm{1PI}(p^2) + \Pi_{ij}^\textrm{PT}(p^2),~~(i,j)~~\text{for}~~\text{gauge bosons or fermions}. 
\end{align}
Those for scalar bosons are obtained by subtracting the tadpole diagram and the pinch-term contributions and adding the non-vanishing 
counter terms for tadpoles  $(\delta T_{ij})$ as follows:
\begin{align}
  \Pi_{ij}(p^2)\big|_{\text{KOSY}} = \Pi_{ij}^\textrm{1PI}(p^2) + \delta T_{ij}\big|_{\text{KOSY}},~~(i,j)~~\text{for}~~\text{scalar bosons} \label{Pi_replace_SM}. 
\end{align}
In the SM, we have
\begin{align}
  \delta T_{hh}^{}\big|_{\text{KOSY}} &=   - \frac{1}{v}  T_h^\textrm{1PI}. 
\end{align}

\subsection{HSM}

We give the expressions for the counter terms appearing in Sec.~B--2. 
The explicit formulae for the relevant 1PI diagram contributions to 1-point and 2-ponint functions  are given in Refs.~\cite{HSM-KKY1,HSM-KKY2}. 

Counter terms for the masses of weak bosons and fermions, and their wave function renormalizations are the same form as the corresponding one in the SM.
Those for $\delta m_{h_i}^2$ and $\delta Z_{h_i}^{}$ ($h_1 = h$ and $h_2 = H$) are given by
\begin{align}
  \delta m_{h_i}^2 & = \Pi_{h_i h_i}^\textrm{}(m_{h_i}^2),  \,\,\,
  \delta Z_{h_i}
  = - \frac{d}{dp^2}\Pi_{h_i h_i}^\textrm{1PI}(p^2)
  \big|_{p^2=m_{h_i}^2}.    \label{abc1}
\end{align} 
Those for mixing parameters of the CP-even Higgs bosons are given by
\begin{align}
  \delta C_{h}^{} & = \frac{1}{2(m_H^2 - m_h^2)}\left[
   \Pi_{Hh}^\textrm{1PI}(m_h^2) - \Pi_{Hh}^\textrm{1PI}(m_H^2)
    \right],\label{delta_Ch} \\ 
  \delta \alpha & = \frac{1}{2(m_H^2 - m_h^2)}\left[
    \Pi_{Hh}^\textrm{}(m_h^2) + \Pi_{Hh}^\textrm{}(m_H^2)    \right]. \label{delta_a}
\end{align}
Finally, we give the explicit forms of $\delta \lambda_{hhh}^{}$ and $\delta \lambda_{Hhh}$ which appear in the renormalized $hhh$ and $Hhh$
couplings given in Eqs.~(\ref{hhh-hsm2}) and \eqref{bHhh-hsm2}, respectively:
\begin{align}
  \delta \lambda_{hhh}^{} &=
  -(\lambda_{hhh}-s_\alpha^2 \mu_S^{})\frac{\delta v}{v}
  - \frac{c_\alpha^3}{2v}\delta m_h^2
  + F_\alpha^\textrm{HSM}\delta \alpha + \delta M, \\
  \delta \lambda_{Hhh} &=
  -(\lambda_{Hhh} + 3c_\alpha^{}s_\alpha^2\mu_S^{})\frac{\delta v}{v}
  - \frac{s_\alpha c_\alpha^2}{2v}(2\delta m_h^2 + \delta m_H^2)
  + G_\alpha^\textrm{HSM}\delta\alpha + \delta M', 
\end{align}
where 
\begin{align}
    F_\alpha^\textrm{HSM} & = \frac{3s_\alpha c_\alpha^2}{2v}m_h^2
  + v\lambda_{\Phi S}^{} s_\alpha^{} (s_\alpha^2 -2c_{\alpha}^{2})
  + 3 s_\alpha^2 c_\alpha^{} \mu_S^{}, \\  
  G_\alpha^\textrm{HSM} &=\frac{c_\alpha^{}}{2v}
  (2s_\alpha^2 -c_\alpha^2)(2m_h^2 + m_H^2)
  - \frac{v}{4}\lambda_{\Phi S}^{}(c_\alpha - 9 c_{3\alpha}^{})
  +3 \mu_S^{} s_\alpha^{}(s_\alpha^2 -2 c_\alpha^2). 
\end{align}
We note that $\delta M$ and $\delta M'$ are linear combinations of the counter terms $\delta\mu_S^{}$ and $\delta\lambda_{\Phi S}$~\cite{HSM-KKY2}. 
Their explicit forms are given as follows:
\begin{align}
\delta M &= -\frac{s_\alpha^2}{16\pi^2}\Big[\sum_f\frac{2N_c^fm_f^2}{v}\lambda_{\Phi S}c_\alpha -\frac{2c_\alpha^3}{v^3}(2m_W^4+m_Z^4)
-\frac{3}{v}\lambda_{\Phi S}c_\alpha(2m_W^2+m_Z^2)\notag\\
&+\frac{m_h^2}{4v}\lambda_{\Phi S}(11c_\alpha+c_{3\alpha})
+\frac{m_H^2}{v}\lambda_{\Phi S}c_\alpha s_\alpha^2
+4v\lambda_{\Phi S}(3\lambda_S + \lambda_{\Phi S}) c_\alpha
-36\mu_S\lambda_S s_\alpha
\Big]\Delta_{\text{div}},  \\
\delta M' &= \frac{s_\alpha}{16\pi^2}\Big[\sum_f\frac{N_c^fm_f^2}{v}\lambda_{\Phi S}(1+3c_{2\alpha}) 
-\frac{2m_W^4+m_Z^4}{v^3}c_\alpha^2(c_{2\alpha}-3)\notag\\
&-\frac{3(2m_W^2+m_Z^2)}{2v}\lambda_{\Phi S}(1+3c_{2\alpha})
+\frac{3m_h^2}{2v}\lambda_{\Phi S}c_\alpha^2(3 + c_{2\alpha})
-\frac{3m_H^2}{v}\lambda_{\Phi S} s_\alpha^4\notag\\
&+2v\lambda_{\Phi S}(3\lambda_S + \lambda_{\Phi S})(1+3c_{2\alpha})
-108\mu_S\lambda_S c_\alpha s_\alpha
\Big]\Delta_{\text{div}}, 
\end{align}
where $\Delta_{\text{div}}$ expresses the UV divergent part of the loop integral and $N_c^f$ is the color factor; i.e., $N_c^f=3~(1)$ for $f$ being quarks (leptons).

In the KOSY scheme, $\delta \alpha$, $\delta m_h^2$, and $\delta m_H^2$ are given by the same way as those given in Eqs.~(\ref{abc1}), (\ref{delta_Ch}) and (\ref{delta_a}), but 
we should use the scalar two-point functions defined in Eq.~\eqref{Pi_replace_SM}, 
where each $\delta T_{ij}\big|_{\text{KOSY}}$ is given by
\begin{align}
  \delta T_{hh}^{}\big|_{\text{KOSY}} &=
  - \frac{c_\alpha^2}{v}\left(s_\alpha^{} T_H^\textrm{1PI}
  + c_\alpha^{} T_h^\textrm{1PI} \right), \\
  \delta T_{HH}^{}\big|_{\text{KOSY}} &=
  - \frac{s_\alpha^2}{v}\left(s_\alpha^{} T_H^\textrm{1PI}
  + c_\alpha^{} T_h^\textrm{1PI} \right),\\
    \delta T_{hH}^{}\big|_{\text{KOSY}} &= 
    -\frac{s_\alpha^{} c_\alpha^{}}{v}\left(
    s_\alpha^{} T_H^\textrm{1PI}
  + c_\alpha^{} T_h^\textrm{1PI} \right). 
\end{align}

\subsection{THDM}

We give the expressions for the counter terms appearing in Sec.~B--3. 
The explicit formulae for the relevant 1PI diagram contributions to 1-point and 2-ponint functions  are given in Refs.~\cite{THDM-KKY2}. 

Counter terms for the masses of weak bosons and fermions, and their wave function renormalizations are the same form as the corresponding one in the SM.
Those for masses of Higgs bosons $\varphi (= h,H,A,H^\pm$) and their wave function renormalizations are expressed as
\begin{align}
  \delta m_\varphi^2 & = \Pi_{\varphi \varphi}^\textrm{}(m_\varphi^2),  \,\,\,
  \delta Z_\varphi 
  = - \frac{d}{dp^2}\Pi_{\varphi \varphi}^\textrm{1PI}(p^2)
  \big|_{p^2=m_\varphi^2}. 
\end{align}
Counter terms for the mixing parameters for the CP-odd scalar bosons and those for the singly-charged scalar bosons are
given by
\begin{align}
  \delta C_{A}^{} & = -\frac{1}{2m_A^2}\left[
    \Pi_{AG^0}^\textrm{1PI}(m_A^2) - \Pi_{AG^0}^\textrm{1PI}(0) \right], \\
  \delta C_{H^{\pm}}  &=
  -\frac{1}{2m_A^2}\left[
  \Pi_{AG^0}^\textrm{1PI}(m_A^2) + \Pi_{AG^0}^\textrm{1PI}(0)
  - \frac{2m_A^2}{m_{H^\pm}^2}     \Pi_{H^+G^-}^\textrm{1PI}(0) 
  \right], \\
  \delta \beta & = -\frac{1}{2m_A^2}\left[
    \Pi_{AG^0}^\textrm{}(m_A^2) + \Pi_{AG^0}^\textrm{}(0) 
    \right]. 
\end{align}
We note that $\delta C_{h}$ and $\delta\alpha$ take the same form as given in Eq.~\eqref{delta_Ch} and \eqref{delta_a}, respectively.
In the THDMs, $\delta \lambda_{hhh}^{}$ and $\delta \lambda_{Hhh}$ are expressed as 
\begin{align}
\delta \lambda_{hhh}^{} &=
  - \lambda_{hhh}\frac{\delta v}{v}
  - \frac{    c_{3\alpha-\beta} + 3c_{\alpha+\beta}}{4v s_{2\beta}^{}}\delta m_h^2
  + F_\alpha^\textrm{THDM}\delta\alpha 
  + F_\beta^{}\delta\beta +
  \frac{c_{\beta - \alpha}^2 c_{\alpha + \beta}^{}}{vs_{2\beta}^{}}\delta M^2,\\
  \delta \lambda_{Hhh} &=
    - \lambda_{Hhh}\frac{\delta v}{v}
  - \frac{s_{2\alpha}^{}c_{\beta-\alpha}^{}}{2v s_{2\beta}^{}}(2\delta m_h^2
  + \delta m_H^2 ) 
 +G_\alpha^\textrm{THDM}\delta \alpha + G_\beta \delta \beta
 + \frac{3c_{\beta - \alpha}}{2v}\left(\frac{s_{2\alpha}^{}}{s_{2\beta}^{}}
 -\frac{1}{3}\right)
  \delta M^2 ,   
\end{align}
where 
\begin{align}
     F_\alpha^\textrm{THDM} &=\frac{c_{\beta -\alpha}}{2v}\left[
    3\frac{s_{2\alpha}}{s_{2\beta}}(m_h^2 - M^2) +M^2
    \right], \\
  G_\alpha^\textrm{THDM} &=
  \frac{s_{\beta - \alpha}^{}}{c_{\beta - \alpha}^{}}\lambda_{Hhh}^{}
  - \frac{c_{2\alpha}^{}c_{\beta-\alpha}}{v s_{2\beta}^{}}
  (2m_h^2 + m_H^2 -3M^2), 
    \end{align}
and $F_\beta$ and $G_\beta$ are given in Eqs.~\eqref{F_beta} and \eqref{G_beta}, respectively. 
The expression for $\delta M^2$ is given by 
\begin{align}
  \frac{\delta M^2}{M^2} & =
  \frac{1}{16\pi^2v^2}\Big[
    2\sum_f N_c^f m_f^2 \zeta_f^2 + 4M^2 -2 m_{H^\pm}^2 - m_A^2 
     \frac{s_{2\alpha}}{s_{2\beta}} (m_H^2 - m_h^2) -3(2m_W^2 + m_Z^2)\Big]\Delta_\textrm{div}, 
  \end{align}
where $\zeta_f$ are given in Tab.~\ref{yukawa_tab}.

Similar to the case in the HSM, in the KOSY scheme, 
scalar two-point functions $\Pi_{ij}^{}$ are defined in Eq.~(\ref{Pi_replace_SM}), where each of the 
counter term of the tadpole is given by 
\begin{align}
  \delta T_{hh}^{}\big|_{\text{KOSY}} &=
  \frac{1}{vs_\beta c_\beta}\left[
  -s_\alpha c_\alpha c_{\beta-\alpha}^{} T_H^\textrm{1PI}
  + (s_\alpha^3 s_\beta^{} - c_\alpha^3 c_\beta^{})T_h^\textrm{1PI} \right],\\
  \delta T_{HH}^{}\big|_{\text{KOSY}} &=
  \frac{1}{vs_\beta c_\beta} \left[
  - (s_\beta^{} c_\alpha^3 + c_\beta^{} s_\alpha^3)T_H^\textrm{1PI}
  + s_\alpha^{} c_\alpha^{} s_{\beta - \alpha}^{} T_h^\textrm{1PI}  
  \right], \\ 
    \delta T_{hH}^{}\big|_{\text{KOSY}} &= 
    \frac{s_\alpha c_\alpha}{vs_\beta^{} c_\beta^{}}
    \left(s_{\beta-\alpha}^{} T_H^\textrm{1PI} -c_{\beta-\alpha}^{} T_H^\textrm{1PI}\right), \\
  \delta T_{AA}^{}\big|_{\text{KOSY}} & =  \delta T_{H^+ H^-}\big|_{\text{KOSY}} =
  \frac{1}{v s_\beta c_\beta}\left(
  - s_\beta^{} c_\beta^{} T_H^\textrm{1PI}
  + (s_\beta^2 - c_\beta^2) T_h^\textrm{1PI}\right), \\
  \delta T_{AG}^{}\big|_{\text{KOSY}} & = \delta T_{H^+G^-}^{}\big|_{\text{KOSY}}=
  \frac{1}{v}\left( s_{\beta-\alpha}^{} T_H^\textrm{1PI}
  - c_{\beta - \alpha}^{} T_h^\textrm{1PI} \right).
\end{align}

\end{appendix}

\end{document}